\documentclass[11pt]{article}
\usepackage{fullpage,amsmath, amsfonts, amsthm,xr}
\usepackage{graphicx, algorithmic, algorithm, enumerate,bm}
\usepackage{natbib}
\usepackage[bbgreekl]{mathbbol}
\usepackage{csquotes,caption}
\usepackage{multirow}
\usepackage[toc,page]{appendix}

\usepackage[dvipsnames]{xcolor}

\usepackage{hyperref}

\begin{document}

\begin{center}
{\bf\LARGE The Jensen Effect and Functional Single Index Models: \\ Estimating the Ecological Implications of Nonlinear Reaction Norms}\\
\vspace{0.5cm}
{\sc Zi Ye\footnote{PhD Candidate, Department of Statistical Science, Cornell University, zy234@cornell.edu}, Giles Hooker\footnote{Associate Professor, Department of Statistical Science, Cornell University, gjh27@cornell.edu}, Stephen Ellner\footnote{Professor, Department of Ecology and Evolutionary Biology, Cornell University, spe2@cornell.edu}}\\
\end{center}

\begin{abstract}
This paper develops tools to characterize how species are affected by environmental variability, based on a functional single index model relating a response such as growth rate to environmental conditions. In ecology, the curvature of such responses are used, via Jensen's inequality, to determine whether environmental variability is harmful or beneficial, and differing nonlinear responses to environmental variability can contribute to the coexistence of competing species.

Here, we address estimation and inference for these models with observational data on individual responses to environmental conditions. Because nonparametric estimation of the curvature (second derivative) in a nonparametric functional single index model requires unrealistic sample sizes, we instead focus on directly estimating the effect of the nonlinearity, by comparing the average response to a variable environment with the response at the expected environment, which we call the \emph{Jensen Effect}. We develop a test statistic to assess whether this effect is significantly different from zero. In doing so we re-interpret the SiZer method of \cite{chaudhuri1999sizer} by maximizing a test statistic over smoothing parameters. We show that our proposed method works well both in simulations and on real ecological data
from the long-term data set described in \cite{drake2005population}.
\end{abstract}

\section{Introduction}
In natural ecosystems, environmental conditions are highly variable over time and space
\citep[e.g.,][]{vasseur-mccann-2007} and many
classical questions in ecology and evolution are therefore concerned with the potential consequences of
this variation. Two important topics have been how species' traits and life histories evolve so that
species can persist in environments that may be favorable at some times and unfavorable at others
(see \cite{cohen1966optimizing} and \cite{koons2008evolution}). Differing nonlinear responses to
environmental variability can contribute to maintaining the biodiversity of competing species,
allowing them to coexist stably (see for example \cite{hutchinson1961paradox, chesson1981environmental,
ellner1987alternate, chesson1994multispecies, chesson2000mechanisms, chesson2000general}). Nonlinear
responses to environmental conditions are also important for forecasting responses to climate change,
as environmental variability can either increase population growth rate (\cite{drake2005population,koons2009life})
or decrease it (\cite{lewontin1969population}), depending on the shape of the norm of reaction between environmental
variables and the components of population growth rate (survival, growth, and reproduction).

The goal of this paper is to develop methods for determining the effect of environmental variability on some component of population growth. Within mathematical biology, this is a result of the curvature of the model; from Jensen's inequality convex functions result in higher growth under a variable environment, than when the environment is held constant at its mean, and {\em vice versa} for concave functions. However, common statistical models for species growth make parametric assumptions that pre-determine these effects. For example, exponential models will always return convex results, while Michaelis-Menten saturation effects produce concave relationships.  Recent statistical research in semi-parametric or nonparametric methods inspires us to understand the effect of environmental variation via
nonparametric models that do not impose these assumptions. Specifically, we consider spline-based methods (see \cite{wood2000modelling}, \cite{ramsay2006functional}, \cite{ramsay2009functional} and \cite{ruppert2003semiparametric}) to predict nonlinear responses under environmental fluctuation.

The nonparametric model considered here is
\begin{align}
\bm{G} = g\left(\bm{E}\right) + \epsilon,\label{model}
\end{align}
where $\bm{G}$ and $\bm{E}$ are the growth (or future size) and environment of an organism. The
function $g$ is the link function to be estimated, and $\epsilon$ is random error.
We assume that the environmental $\bm{E}$ is described by a climate history, such as temperature or precipitation observed at a fine time-scale, such as daily resolution. Following \cite{teller2016linking}, these are thought of as functional covariates, leading to representation of $\bm{E}$ as a functional linear term
\begin{align}
\bm{E} = \int X\left(t\right)\beta\left(t\right)\mathrm{d}t,  \label{env}
\end{align}
where $\beta\left(t\right)$ is the coefficient function to be estimated, and $X\left(t\right)$ is the observed climate history.


Combining $\left(\ref{model}\right)$ and $\left(\ref{env}\right)$, a functional model for observed species growth is
\begin{align}
Y = g\left(\int X\left(t\right)\beta\left(t\right)\mathrm{d}t\right) + \epsilon. \label{FSIM}
\end{align}
This is the Functional Single Index model introduced in \cite{chen2011single} and \cite{ma2016estimation}. The functional single index model is an extension of the single index model to a functional covariate via $\int X\left(t\right)\beta\left(t\right)\mathrm{d}t$. This model allows a nonlinear relationship between response $Y$ and covariate function $X\left(t\right)$. In addition, it improves stability in estimating the link function $g$ through imposing smoothness on $\beta\left(t\right)$.

With estimates of $g$ and $\bm{E}$, we wish to assess the impact of environmental variability. Within mathematical biology, this is done by an analysis of $g''$: an always positive second derivative corresponds to $g$ being convex and increased growth under environmental variability due to Jensen's inequality: $E g(\bm{E}) > g(E \bm{E})$. \cite{ye2018local} investigates estimating $g''$ in a functional single index model using a local quadratic estimate for $g$ and demonstrates convergence rates but finds that the resulting estimators are impractical for finite sample size. A brief demonstration of this same effect using our estimators is given in Appendix \ref{sec:smallsimulation}.

The methods in this paper bypass estimating curvature and instead we estimate the consequences of environmental variation directly. That is, we define the quantity
\[
\delta = g\left[\mathrm{E}\left(\bm{E}\right)\right] - \mathrm{E}\left[g\left(\bm{E}\right)\right]
\]
directly and conduct inference about its sign.  We have titled this the ``Jensen Effect'' as being the result of Jensen's inequality. However,  we note that $g$ need not be strictly convex for $\delta$ to be positive (or conversely for it to be negative) and observe that the ecological interest lies in $\delta$ rather than $g''$.  Indeed, in our real-world examples in Section \ref{sec:real} we observe that $g$ is often not estimated to have the same curvature over the whole range; nonetheless the interaction of curvature and the distribution of covariates does produce a consistent Jensen Effect.

We argue that this target is also better suited for inference: it avoids estimating derivatives and, by averaging, we expect to gain stability relative to pointwise estimates of $g$. By contrast, direct inference about curvature requires a test of $\min_s g''(s) > 0$ or $\max_s g''(s) < 0$, which is substantially less stable. While bounding curvature does allow statements about consequences of variability to be independent of the distribution of $\bm{E}$, we observe that using a non-parametric estimate of $g$ already constrains the range of the argument $s$ at which we can make statements about $g''(s)$ and our tests can be employed using any given alternative distribution for $\bm{E}$ that is of interest.

In order to conduct inference, we take inspiration from the SiZer method of \cite{chaudhuri1999sizer}. Our estimates $\hat{g}_{\lambda}(\cdot)$ and $\hat{\beta}_{\lambda}(\cdot)$ both employ smoothing parameters. Rather than choosing these parameters, we instead examine the resulting $\hat{\delta}_{\lambda}$ as a function of $\lambda$ and use $\max_{\lambda} |\hat{\delta}_{\lambda}|$ as a test statistic.  We can assess the significance of this statistic by treating $\hat{\delta}_{\lambda}$ as a Gaussian process over $\lambda$ and simulating from a null distribution in which $E \hat{\delta}_{\lambda} = 0$. This avoids the need to account for the selection of $\lambda$ as well as allowing us to detect relationships that may not be significant at smoothing parameters chosen by GCV or other criteria.


\subsection{Related Literature: Single Index Models}

There is a large literature on  single index models covering both applied methodology and theoretical properties. The link function $g$
and the coefficient vector $\bm{\beta}$ have been estimated by three different methods. (1) The most widely used is the Projection Pursuit Regression (PPR)
approach introduced in \cite{hardle1993optimal}. This method is a nested estimation procedure, with the link function $g$ estimated by local polynomial approximation and the coefficient function by minimizing the MSE. Theoretical properties were studied in \cite{hardle1993optimal}
and \cite{ichimura1993semiparametric}. (2) The Average Derivative approach was introduced in \cite{hristache2001direct}. (3) \cite{li1991sliced} introduced the Sliced Inverse Regression method, which considered the estimation of the coefficient vector as a dimension-reduction problem.

In contrast, there are few studies of the functional single index model. A counterpart to the Projection Pursuit Regression was introduced in \cite{chen2011single}, where the coefficient function $\beta$ was approximated by a spline basis and the coefficient vector was estimated.
In addition, a convergence rate was found for this method. \cite{ma2016estimation} used two spline bases to approximate the coefficient function and the link function, respectively, and derived some asymptotic properties of the resulting estimate. Methods that use single index models as an additive term for function-on-scalar regression have been developed and studied in \citet{li2017functional}.

\subsection{Related Literature: SiZer}

The SiZer (SIgnificant ZERo crossings of derivative) method that we adopt to assess significance was introduced by \cite{chaudhuri1999sizer} to assess the significance of peaks and other features in nonparametric smooths while bypassing the selection of  smoothing parameters. \cite{chaudhuri1999sizer} observe that smoothing parameters can have a substantial impact on the features observed, and rather than base inference on a selected smoothing parameter, they examine a range of reasonable smooths.

Specifically, SiZer is particularly concerned with the assessment of sign changes of derivatives. Local modes, in particular, can be represented by sign changes in a first derivative. To assess the significance of these \citet{chaudhuri1999sizer} obtains a nonparametric smooth $\hat{f}(x;\lambda)$ of a relationship over $x$ indexed by smoothing parameters $\lambda$. The method then constructs the $t$-statistic $t(x;\lambda) = \hat{f}^{(k)}(x;\lambda)/\mbox{sd}\left(\hat{f}^{(k)}(x;\lambda)\right)$. $\hat{f}^{(k)}(x;\lambda)$ is then plotted over both $x$ and $\lambda$ with regions in which $|t(x;\lambda)|>C$ are indicated. In particular, changes from a positive value of $\hat{f}^{(1)}(x;\lambda)$ to a negative value as $x$ is changed indicate a local maximum.  \citet{chaudhuri1999sizer} present their methods in the context of local polynomial smoothing and kernel density estimation, but exactly the same tools can be employed for any nonparametric method including the smoothing splines we employ here.  See \citet{sonderegger2009using} for an example of the use of  SiZer in ecological models when searching for points of rapid ecological change to in response to environmental forcing.

The interpretation of the significance of these changes depends on the selection of the critical value $C$.  \citet{chaudhuri1999sizer} suggest a variety of choices, including using pointwise significance levels (in which a number of false positives may be expected when searching over both $x$ and $\lambda$), using a critical value that approximately controls the familywise error rate in $x$ for each $\lambda$ and obtaining a uniform bound by finding a critical value of $\max_{x,\lambda} \left|t(x;\lambda)\right|$ via a bootstrap. \citet{marron2005sizer} develops approximations for Gaussian processes that reduce the computational burden while improving coverage probabilities.

We borrow from these ideas, and particularly repurpose the search over ``scale-space'' to examine the Jensen Effect over a range of smoothing parameters. In contrast to SiZer, which examines features over a range of $x$, the Jensen Effect averages over covariates. Similar to \citet{marron2005sizer}, we select a significance threshold designed to control the familywise error rate over smoothing parameters, however we select this from an explicit Gaussian approximation which can be efficiently simulated.

7
 smoothing parameter selection by comparing the estimates of a curve over a range of smoothing parameters. Our test statistic is inspired by the SiZer  method. Instead of trying to select an optimal smoothing parameter for estimation and inference, we examine  estimates over a range of smoothing parameters and conduct inference based on maximizing a test statistic over that range.


In the remainder of this paper we provide details of our estimation procedure for a functional Single Index Model in Section \ref{sec:estimation}, and our assessment and test of the Jensen Effect in Section \ref{sec:Jensen}.  Simulation studies to assess the efficacy of our test are conducted in Section \ref{sec:simulations}.  Section \ref{sec:real} provides a motivating example in which we examine the response of 8 copepod species to water temperature variability in data obtained from the North Temperate Lakes Long Term Experimental Research station where we find evidence for positive adaptation to environmental variability in 5 species. Section \ref{sec:conclusion} concludes.

\section{Estimation Procedure} \label{sec:estimation}


Here we provide details of our representation of $\beta$ and $g$ and our estimation procedure given particular smoothing parameters. Appendix \ref{sec:smallsimulation} provides a demonstration of using these approaches to estimate curvature rather than the Jensen effect we examine here.

 $\beta$ and  $g$ are represented using smoothed basis expansions. Assume that $n$ independent and identically distributed data pairs $\left(X_1\left(t\right),Y_1\right), \cdots, \left(X_n\left(t\right),Y_n\right)$ are observed where $X_j(t)$ is a real-valued function on [0,1]. 
The Functional Single Index model is
\begin{align*}
Y = g\left(\int X\left(t\right)\beta\left(t\right)\mathrm{d}t\right) + \epsilon
\end{align*}
where the coefficient function $\beta\left(t\right)$ is, like X(t), defined on the interval $\left[0,1\right]$.
$\epsilon$ is assumed to be Gaussian random error.
%
%
This integral may need to be evaluated numerically, depending on the representations used for $X_i$ and $\beta$,
and we assume that this is done up to ignorable error throughout the calculations below.
To ensure identifiability of the model, we require that $\left\| \beta\right\| = \int \beta\left(t\right)^2\mathrm{d}t = 1$.

We use a $K_1$-dimensional B-spline basis for the link function $g$. For any $s$ in the
range of possible $\int X\beta$ values, the link function $g$ can be written as
\begin{align*}
g\left(s\right) = \bm{\phi}^{\top}\left(s\right)\bm{d},
\end{align*}
where $\bm{d}$ is a $K_1$-dimensional column coefficient vector.

We use a $K_2$-dimensional basis for the coefficient function $\beta$, such that
\begin{align*}
\beta\left(t\right) = \bm{\psi}^{\top}\left(t\right)\bm{c},
\end{align*}
where $\bm{c}$ is a $K_2$-dimensional column coefficient vector, and $t \in \left[0,1\right]$.

The coefficient vectors $\bm{c}$ and $\bm{d}$ are estimated by minimizing a penalized sum of squares
\begin{align}
2 \text{PLS} \doteq & \sum\limits_{i=1}^n \left(Y_i - g_i\right)^2 + \lambda_g \int \left(g^{\left(2\right)}\left(s\right)\right)^2\mathrm{d}s + \lambda_{\beta} \int \left(\beta^{\left(2\right)}\left(t\right)\right)^2\mathrm{d}t \label{PLS}\\
=& \sum\limits_{i=1}^n \left\{Y_i - \bm{\phi}^{\top}\left[\left(\int X_i\bm{\psi}^{\top}\right)\bm{c}\right]\bm{d}\right\}^2 + \lambda_g \bm{d}^{\top}\mathbb{P}_g\bm{d} +\lambda_{\beta} \bm{c}^{\top}\mathbb{P}_{\beta}\bm{c}, \nonumber
\end{align}
where $g_i \doteq g\left(\int X_i\beta\right)$ and the penalty matrices $\left[\mathbb{P}_\beta\right]_{ij} = \int \psi_i^{(2)}(t) \psi_j^{(2)}(t) dt$ and $\left[\mathbb{P}_g\right]_{ij} = \int \phi_i^{(2)}(t) \phi_j^{(2)}(t) dt$ are available analytically for most common choices of basis expansion.

Equation $\left(\ref{PLS}\right)$ specifies a nonlinear optimization problem, which we solve numerically using
built-in optimizers in {\tt R} (see below). 
Denoting the estimated coefficients as $\hat{\bm{c}}$ and $\hat{\bm{d}}$, the estimates are
\begin{align*}
\hat{\beta}\left(t\right) = \bm{\psi}^{\top}\left(t\right)\hat{\bm{c}},
\end{align*}
and
\begin{align*}
\hat{g}_i \doteq \hat{g}\left(\int X_i\hat{\beta}\right) = \bm{\phi}^{\top}\left[\left(\int X_i\left(t\right)\bm{\psi}^{\top}\left(t\right)\mathrm{d}t\right)\hat{\bm{c}}\right]\hat{\bm{d}},
\end{align*}
where $i = 1, \cdots, n$.

\subsection{Notes on Implementation and Model Selection}
Our objective criterion \eqref{PLS} requires nonlinear numerical optimization.  In our experiments below we have used the {\tt R} function {\tt optim} with some additional modifications. The simulations reported in Section \ref{sec:simulations} used the BFGS gradient-based optimizer.
However, at large values of $\lambda$ we find that very tight convergence criteria are needed to reduce the numerical
error to below that of the estimated noise. This was mitigated with two strategies:
\begin{enumerate}
    \item We initialize our optimization with $\bm{d}$ chosen so that $\hat{g}$ is exactly linear and $\bm{c}$ is obtained from functional linear regression.
    \item We re-initialize BFGS once it converges, and run it a second time. BFGS uses a sequentially-calculated approximate Hessian, and can stop early
    due to poor estimation of this Hessian. Re-initialization resets the approximate Hessian to the identity, so that optimization is restarted with
    a steepest descent step.
    \item In our motivating data example we additionally attempted re-initializing from the solutions found at surrounding 8 combinations of $\lambda_g$ and $\lambda_\beta$ and chose the best optima among these. This was repeated until the maximum relative improvement in optima was less than 0.01. This strategy was employed in order to ensure a smooth relationship between smoothing parameters and estimated effects, but can substantially increase computational costs.
\end{enumerate}
To maintain identifiability of our model, we normalize our estimate of $\beta(t)$ within each evaluation of the objective function and multiply by $\text{sign}(\beta(0))$.

In order to represent $\hat{g}$ with a basis expansion, we need to control the range of its arguments. Throughout our estimates below, we have used the identifiability requirement that $\left\| \beta \right\| = 1$ to use a range of $[-S, \ S]$ where $S$ is the largest score for the maximum eigenvalue from a principal components decomposition of the $X_i$. If $|\int X_i(t) \beta(t) dt| > S$ for some $i$, we replace the argument with the corresponding end-point of the range and add a penalty of $|\int X_i(t) \beta(t) dt - S|$ to the objective \eqref{PLS}. In practice, while we find that this excecdance can occur during optimization, it never appears in the final result.

We find that these procedures are sufficient to provide reliable inference for the Jensen Effect. However, estimates of $g''(\cdot)$ can be highly sensitive to initial conditions and optimization strategies Appendix \ref{sec:smallsimulation} provides a brief example of the sensitivity of curvature to both noise and initial conditions. We speculate that this sensitivity is a result of a complex optimization landscape in which there are many good estimates of $g$, but these can vary substantially for $g''$.


While our assessment of statistical significance avoids selecting $\lambda$, it will be useful to have a value for visualization and
for an estimate of residual variance.  To choose $\lambda$, we define a smoother matrix $\mathbb{S}_{\lambda}$ associated with $\lambda$,
and the GCV value for selecting $\lambda$ is calculated from
\begin{align*}
\text{GCV}\left(\lambda\right) \doteq \frac{\frac{1}{n}\left\|\left(\mathbb{I}-\mathbb{S}\left(\lambda\right)\right)\bm{Y}\right\|^2}
{\left[\frac{1}{n}\text{tr}\left(\mathbb{I}-\mathbb{S}\left(\lambda\right) \right)\right]^2},
\end{align*}
where $\mathbb{I}$ is the $\left(n \times n\right)$-dimensional identity matrix. We derive $\mathbb{S}_{\lambda}$ from a Taylor expansion in \eqref{eq:smootherM} below.

\section{Jensen Effect} \label{sec:Jensen}
The ecological interest in $g''$ is in the comparison of $g\left[\mathrm{E}\left(\bm{E}\right)\right]$ and $\mathrm{E}\left[g\left(\bm{E}\right)\right]$. Because reliable estimation of $g''$ requires unrealistic sample sizes, we instead  compare these quantities directly to estimate what we call the ``Jensen Effect''.

We define a difference statistic by
\begin{align*}
\delta = \frac{1}{n}\sum\limits_{i=1}^n g\left(\int X_i\beta\right) - g\left(\int \bar{X}\beta\right),
\end{align*}
where $\bar{X} = \frac{1}{n}\sum\limits_{i=1}^n X_i$. In mathematical analyses, if the link function $g$ is convex, then $\delta > 0$ which indicates
better growth with a varying environment; otherwise, the difference $\delta < 0$ and a constant environment is better for
growth. However, this estimate still depends on the smoothing parameters $\lambda_g$ and $\lambda_{\beta}$. Inspired by the SiZer method of \citet{chaudhuri1999sizer}, we examine the
difference $\delta$ over a range of $\lambda$ values for $g$ and $\beta$, and generate
hypothesis tests using the maximum or minimum value of $\delta$ as a function of $\lambda$.

\subsection{Hypothesis Test $1$: Nonparametric Smoothing}
We begin by briefly developing our SiZer-inspired test for a standard smoothing spline (treating the environment $\bm{E}$ as known)
before developing the test for a functional single index model. Here defining $\Phi$ to be matrix of evaluations, $\Phi_{ij} = \phi_j\left(E_i\right)$ and $\mathbb{P}$ to be the second derivative penalty matrix, the standard smoothing spline estimate is
\begin{align}
\hat{g}_{\lambda} \left(e\right) = \phi\left(e\right)^{\top} \left(\Phi^{\top} \Phi + \lambda \mathbb{P}\right)^{-1} \Phi^{\top} \mathbf{Y}.
\end{align}

Define the $\left(n+1\right)$-dimensional column vector $\bm{a} \doteq
\left(\frac{1}{n},\cdots,\frac{1}{n},-1\right)^{\top}$ and the augmented set of evaluation
points $\mathbf{e} = \left(E_1,\ldots,E_n,\bar{E}\right)^{\top}$, where $E_i = \int X_i\hat{\beta}$ (at each observed environment value) and $\bar{E} = \frac{1}{n}\sum\limits_{i=1}^n \int X_i\hat{\beta}$ (averaged across all environment values), with corresponding evaluation matrix $\Phi^+$. We can write
\begin{align}
\delta_{\lambda} = \frac{1}{n} \sum \hat{g}_{\lambda}\left(E_i\right) -  \hat{g}_{\lambda}(\bar{E}) = \bm{a}^{\top} \Phi^+ \left(\Phi^{\top} \Phi + \lambda \mathbb{P}\right)^{-1} \Phi^{\top} \mathbf{Y} = \bm{u}_\lambda \bm{Y},
\end{align}
which we can standardize to obtain the t-statistic
\begin{align}
t_{\lambda} = \frac{\bm{u}_{\lambda} \bm{Y}}{\hat{\sigma} \sqrt{ \bm{u}_\lambda \bm{u}_{\lambda}^{\top}}}
\end{align}
in which $\hat{\sigma}$, the estimate of the standard deviation of the random error $\epsilon$, is obtained from the value of $\lambda$ selected by GCV (see details below).

Since the response variable $Y$ is  Gaussian, the test statistics $t_{\lambda}$ is also a Gaussian process with the covariance function
\begin{align}
\Sigma(\lambda_1,\lambda_2) = \frac{\bm{u}_{\lambda_1} \bm{u}_{\lambda_2}}{\left\| \bm{u}_{\lambda_1} \right\|\left\| \bm{u}_{\lambda_2} \right\|}
\end{align}
which involves no unknown parameters. We can thus use $\max_{\lambda} \left|t_{\lambda}\right|$ as a test statistic, obtaining critical values by simulating from the Gaussian process $\mathbb{GP}\left(\bm{0},\Sigma\right)$. Under the null hypothesis $\delta = 0$, $t_{\lambda}$ is a Gaussian process with mean $\bm{0}$.

An important consideration here is that we expect $\delta_{\lambda}$ to inherit smoothing bias, but this should result in under-estimation of the Jensen Effect because it will shrink the estimated second derivative. In analogy to SiZer, by examining $t_{\lambda}$ over the whole range of $\lambda$ we can assess this effect at various levels of smoothing; our use of $\max |t_{\lambda}|$ as a test statistic allows to maintain a conservative test.  We do still need to choose $\lambda$ by GCV in our estimate $\hat{\sigma}^2$, because we use the same $\hat{\sigma}^2$ when calculating the covariance matrix $\Sigma$. We expect  $\hat{\sigma}^2$ to be relatively insensitive to the specific $\lambda$ chosen so long as we do not over-smooth (see arguments in \cite{ruppert2003semiparametric}); maintaining a constant $\hat{\sigma}^2$ in the $t$-statistic removes the need to account for changes in $\hat{\sigma}^2$ across
$\lambda$.

We note the potential for $\hat{\delta}_{\lambda}$ to change signs over the range of $\lambda$.  For example, if the underlying $g$ is strongly convex in a very narrow region but concave more broadly we might find a positive effect at small values of $\lambda$ and a negative effect at large values as the convex portion of $g$ is smoothed over. We would regard this as good reason to examine the resulting estimates of $g$ with an eye to plausibility at both values of $\lambda$.  In our motivating data in Section \ref{sec:real}, we found a couple of examples in which $\delta_\lambda$ was declared significant at two values of  $\lambda$ where $\delta_\lambda$ had opposite signs. One of these could be dismissed easily as occuring only at one set of extreme values. The case of {\em Diacyclops Thomasi} required further investigation which we defer to Section \ref{sec:real}.

\subsection{Hypothesis Test $2$: Functional Single Index Model}
The functional single index model complicates the process described above by including
two smoothing parameters and nonlinear effects of $\hat{\beta}$, necessitating a Taylor
expansion to approximate the recipe above.  For each
pair of smoothing parameters $\left(\lambda_g,\lambda_{\beta}\right)$, we obtain an estimate of $\beta$, $g$,
and $\delta$, denoted as $\hat{\delta}\left(\lambda_g,\lambda_{\beta}\right)$.

Defining
\begin{eqnarray}
\bm{i} &=& \left(\int X_1\hat{\beta},\cdots,\int X_n\hat{\beta},\int \bar{X}\hat{\beta}\right)^{\top},\\
\bm{i}_{-1} &=& \left(\int X_1\hat{\beta},\cdots,\int X_n\hat{\beta}\right)^{\top},\\
\bm{v} &=& \left(\hat{g}\left(\int X_1\hat{\beta}\right),\cdots,\hat{g}\left(\int X_n\hat{\beta}\right),\hat{g}\left(\int \bar{X}\hat{\beta}\right)\right)^{\top},
\end{eqnarray}
the estimated difference function given $\left(\lambda_g,\lambda_{\beta}\right)$ is
\begin{align}
& \hat{\delta}\left(\lambda_g,\lambda_{\beta}\right) = \bm{a}^{\top}\bm{v} = \bm{a}^{\top}\bm{\phi}\left(\bm{i}\right)\hat{\bm{d}} \\
& \hspace{.4cm} = \bm{a}^{\top}\bm{\phi}\left(\bm{i}\right)\left(\bm{\phi}\left(\bm{i}_{-1}\right)^{\top}\bm{\phi}\left(\bm{i}_{-1}\right)+\lambda_g\mathbb{P}_g\right)^{-1}\bm{\phi}\left(\bm{i}_{-1}\right)^{\top}\bm{Y}. \nonumber
\end{align}

To construct a t-statistic to test the significance of $\delta$, an estimate of the variance of the difference function $\hat{\delta}$ is needed. The estimated difference function $\hat{\delta}\left(\lambda_g,\lambda_{\beta}\right)$ is defined on an estimate of $\hat{\bm{c}}$ and $\hat{\bm{d}}$, which are the coefficients of $\beta$ and $g$ respectively. Therefore, we need to calculate the covariance of the estimated $\hat{\bm{c}}$ and $\hat{\bm{d}}$.

Recall \eqref{PLS},  the penalized least squares criterion to be minimized,
and define the matrices of linear basis effects $\Psi_{ij} = \int X_i\left(t\right) \psi_j\left(t\right) \mathrm{d}t$ and evaluations of the link function bases and derivatives $\Phi^{\left(k\right)}_{ij} = \phi_j^{\left(k\right)}\left( \Psi_{i \cdot} \bm{c} \right)$ with $\bm{c}$ taken at its expected estimate. We derive gradients of PLS as
\begin{align}
\left(
\begin{array}{c c}
\bigtriangledown_{\bm{d}}  \\
\bigtriangledown_{\bm{c}}
\end{array}\right)
= \left(\begin{array}{c}
\Phi^{\top}\left\{\bm{Y} - \Phi\bm{d}\right\} + \lambda_g \mathbb{P}_g \bm{d} \\
 \Psi^{\top}\text{diag}\left\{\Phi^{(1)}\bm{d}\right\}\left\{\bm{Y} - \Phi^{\top}\bm{d}\right\} + \lambda_\beta \mathbb{P}_\beta \bm{c}
\end{array}\right) = \left( \begin{array}{c} \mathbb{Z}_g \\ \mathbb{Z}_\beta \end{array} \right) (\bm{Y} - \Phi \bm{d}) + \left( \begin{array}{c} \lambda_g \mathbb{P}_g \bm{d} \\ \lambda_\beta \mathbb{P}_\beta \bm{c} \end{array} \right)
\end{align}
and expected Hessian
\begin{align}
 \mathbb{H} = \left( \begin{array}{cc}  \mathbb{Z}_g^{\top} \mathbb{Z}_g + \lambda_g \mathbb{P}_g & \mathbb{Z}_g^{\top} \mathbb{Z}_\beta \\  \mathbb{Z}_\beta^{\top} \mathbb{Z}_g  &
 \mathbb{Z}_\beta^{\top}  \mathbb{Z}_\beta + \lambda_\beta \mathbb{P}_\beta \end{array} \right).
\end{align}

We can now obtain the sandwich covariance
\begin{align}  \label{eq:coefcov}
\text{cov}\left(
\begin{array}{c c}
\bm{d} \\
\bm{c}
\end{array}\right) = \hat{\sigma}^2\mathbb{H}^{-1}\left(
\begin{array}{c c}
\mathbb{Z}_g \\
\mathbb{Z}_\beta
\end{array}\right)^{\top}\left(
\begin{array}{c c}
\mathbb{Z}_g \\
\mathbb{Z}_\beta
\end{array}\right)
\mathbb{H}^{-1}
\end{align}
where we estimate $\sigma^2$ from
\begin{align}
\hat{\sigma}^2 = \frac{1}{\text{df}_{\text{res}}}\sum\limits_{i=1}^n\left[Y_i - \hat{g}\left(\int X_i\hat{\beta}\right)\right]^2
\end{align}
where, following \cite{ruppert2003semiparametric}, the residual degree of freedom is defined as
\begin{align}
\text{df}_{\text{res}} = n - 2\text{tr}\left(\mathbb{S}\right) + \text{tr}\left(\mathbb{S}\mathbb{S}^{\top}\right)
\end{align}
where
\begin{equation}
\mathbb{S} \doteq \mathbb{S}\left(\lambda_g,\lambda_{\beta}\right) = \left(
\begin{array}{c c}
\mathbb{Z}_g \\
\mathbb{Z}_\beta
\end{array}\right) \mathbb{H} \left(
\mathbb{Z}_g^{\top}, \
\mathbb{Z}_\beta^{\top}
\right)
\label{eq:smootherM}
\end{equation}
is an approximate smoother matrix in which we use the values of $(\lambda_g,\lambda_\beta)$ selected by GCV.

We now define a t-statistic for $\delta$ as a function of $\lambda$,
\begin{align}
t \doteq t\left(\lambda_g,\lambda_{\beta}\right) \doteq \frac{\hat{\delta}\left(\lambda_g,\lambda_{\beta}\right)}{\text{sd}\left[\hat{\delta}\left(\lambda_g,\lambda_{\beta}\right)\right]}.\label{tstat}
\end{align}
where $\text{sd}\left[\hat{\delta}\left(\lambda_g,\lambda_{\beta}\right)\right]$ is given by
\begin{align}
\text{sd}\left[\hat{\delta}\left(\lambda_g,\lambda_{\beta}\right)\right] = \left\{\left[\bm{a}^{\top}\bm{\phi}\left(\bm{i}\right)\right]\text{cov}\left(\hat{\bm{d}}\right) \left[\bm{a}^{\top}\bm{\phi}\left(\bm{i}\right)\right]^{\top}\right\}^{\frac{1}{2}}.
\end{align}
Defining the $n$-dimensional row vector $\bm{u}_{\lambda}$ as
\begin{align}
\bm{u}_{\lambda} = \bm{a}^{\top}\bm{\phi}\left(\bm{i}\right)\left(\bm{\phi}\left(\bm{i}_{-1}\right)^{\top}\bm{\phi}\left(\bm{i}_{-1}\right)+\lambda\mathbb{P}\right)^{-1}\bm{\phi}\left(\bm{i}_{-1}\right)^{\top},
\end{align}
the estimated covariance matrix of $\hat{\delta}_{\lambda}$ is $\hat{\sigma}^2\bm{u}_{\lambda}\bm{u}_{\lambda}^{\top}$.  $\delta_{\lambda}$ is therefore approximately a Gaussian process indexed by $\lambda$ and  we can get the estimated variance of $t_{\lambda}$ from $\frac{\bm{u}_{\lambda}\bm{u}_{\lambda}^{\top}}{\left\|\bm{u}_{\lambda}\right\|^2}$.

We want to test if $\delta \equiv 0$. Denote the number of the smoothing parameters $\lambda_g$ as $m$, we test $H_0$: $\left(\hat{\delta}_{\lambda_1},\cdots,\hat{\delta}_{\lambda_m}\right)^{\top} = \bm{0}_m$. Under $H_0$,
$ \left(t_{\lambda_1},\cdots,t_{\lambda_m}\right)^{\top} \sim \mathrm{N}\left(\bm{0}_m,\mathbb{A}_{mm}\right)$, where $\bm{0}_m$ is a $m$-dimensional column vector, and the covariance matrix $\mathbb{A}$ is $\left(m \times m\right)$-dimensional with the $\left(i,j\right)$ term equals to $\frac{\bm{u}_{\lambda_i}\bm{u}_{\lambda_j}^{\top}}{\left\|\bm{u}_{\lambda_i}\right\|\left\|\bm{u}_{\lambda_i}\right\|}$. The test statistic that we examine is $T = \max\left\{t_{\lambda_1},\cdots,t_{\lambda_m}\right\}$.

In order to obtain a critical value for this statistic, we repeatedly simulate $t_{\lambda}$ from $\mathrm{N}\left(\bm{0}_m,\mathbb{A}_{mm}\right)$ and obtain a distribution for $\max_{\lambda} \left|t_{\lambda}\right|$.

\section{Simulation Study} \label{sec:simulations}
In this section, we use simulated data to explore the power of our test for both single index and functional single index models. Computation time for these models depends on numerous properties of the model: the number of smoothing parameter values to try, the nonlinearities of the underlying $g(\cdot)$, the size of the error variance and the data set size.  Because differing settings could result in substantially different computing times for exactly the same problem, we have not given exact timings here, but note that our single index models run within a few minutes per simulation, while functional versions can require ten to thirty minutes. Our real-world examples, with larger data sets and a finer mesh of smoothing parameters required more than half an hour on a recently purchased laptop.

\subsection{Single Index Model} \label{subsec:sim.sim}
We test for a Jensen Effect by calculating the difference function $\delta$ over a range of smoothing parameters. If the link function $g$ is convex, the $\delta$ function will be positive for most of $\lambda$ values, although it may have high variance at low $\lambda$ and high bias at high $\lambda$. For each simulation, we conduct the hypothesis test introduced in previous section.

Our simulation study starts with the single index model with $p = 5$ covariates generated independently and uniformly on
$\left[-0.5, 0.5\right]$, and the coefficient $\bm{\beta} = \frac{1}{\sqrt{p}} \bm{1}_p$ so that $\left\| \bm{\beta} \right\| = 1$.

\begin{figure}[ht]
\centering
\begin{tabular}{cc}
\includegraphics[width =0.35\textwidth,height = 0.3\textheight]{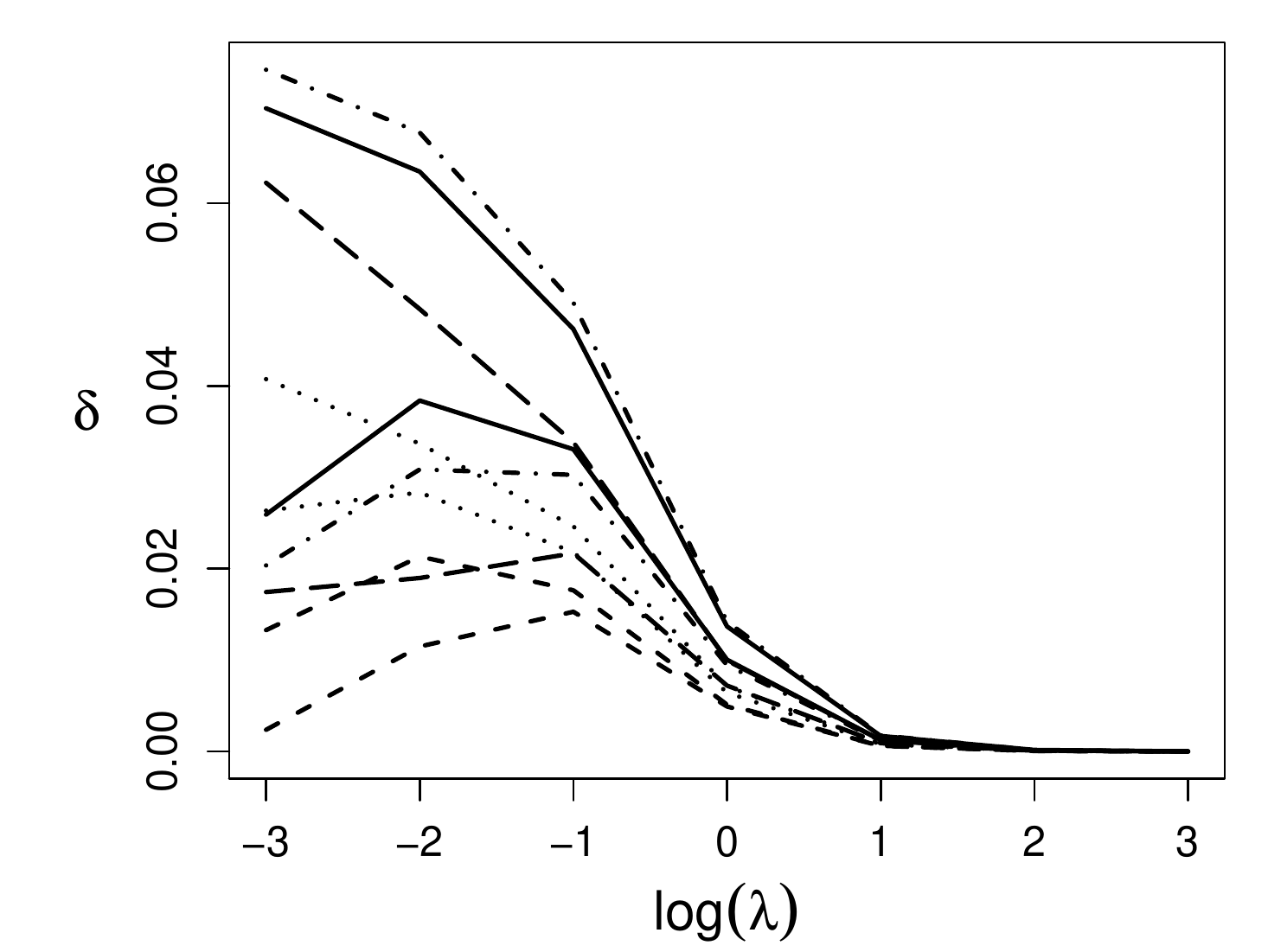}
\includegraphics[width =0.35\textwidth,height = 0.3\textheight]{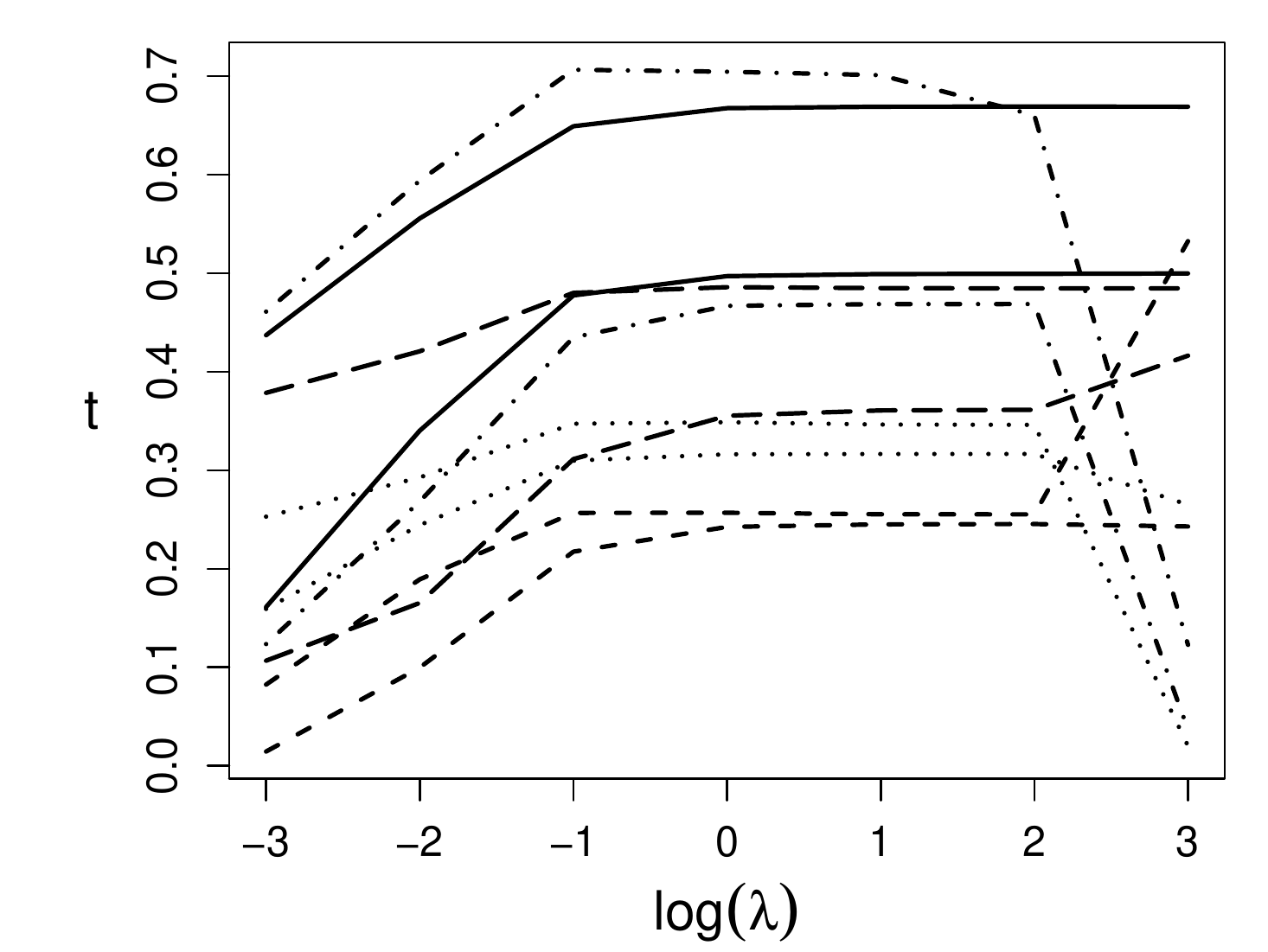}
\end{tabular}
\caption{Left: a sample of $\delta_{\lambda}$ as a function of $\lambda$ in a single index model with link function $g\left(s\right) = e^{s}$. Right: the corresponding $t_{\lambda}$ functions.}
\label{fig:3}
 \end{figure}

To illustrate the Jensen Effect, we choose three different link functions, (1) $g\left(s\right) = e^s$, (2) $g\left(s\right) = -s^2$, (3) $g\left(s\right) = s$. We represented $g$ by a $25$-dimensional quintic B-spline basis. For each link function, we simulated $1000$ data sets of size $100$, with error standard deviation $0.1$. We obtained critical values for our test by simulating $5000$ normal samples from the null distribution. Figure \ref{fig:3} presents a sample of $\delta_{\lambda}$ and $t_{\lambda}$ functions functions versus $\log(\lambda)$ for $g\left(s\right) = e^{s}$; plots  for the other link functions are in  Appendix \ref{appendix:sim}. The rejection rates for these functions are: $99.2\%$, $99.3\%$ and $5.7\%$ respectively.

\subsection{Functional Single Index Model} \label{subsec:fsim.sim}
To define a distribution for the functional covariates, we use a $25$-dimensional Fourier basis $\bm{\psi}\left(t\right)$, where $t \in \left[0,1\right]$.
The covariate functions $X\left(t\right)$ are generated as
\begin{align}
X\left(t\right) = \sum\limits_{i=1}^{25} \xi_i\psi_i\left(t\right),
\end{align}
where $\xi_i \sim \mathrm{N}\left(0,e^{-\left(i-1\right)/12}\right)$. The coefficient function is
\begin{align}
\beta\left(t\right) = \bm{c}^{\top}\bm{\psi}\left(t\right),
\end{align}
where $\bm{c} = \left(0,1,1,0.5,0,\cdots,0\right)^{\top}$.

Again we used the three link functions $g\left(s\right) = e^2$, $g\left(s\right) = -s^2$, $g\left(s\right) = s$.
We represented $g$ by a $25$-dimensional quintic B-spline basis.
For each link function, we generated $1000$ simulated data sets of size $100$ with error
standard deviation $0.1$, and for each such data set we generated $5000$ normal samples from
the null distribution to obtain critical values.

A plot of the $\delta_{\lambda}$ and $t_{\lambda}$ functions for $g\left(s\right) = e^{s}$ is presented in Figure \ref{fig:4}. We have placed equivalent plots for $g\left(s\right) = -s^2$ and $g\left(s\right) = s$ in  Appendix \ref{appendix:fsim}. The rejection rates for the three link functions were $100\%$, $100\%$ and $7.3\%$, showing very good power with a reasonable sample size and close to nominal rate when the null hypothesis is true (no curvature).

\begin{figure}[ht]
\centering
\begin{tabular}{cc}
\includegraphics[width =0.35\textwidth,height = 0.3\textheight]{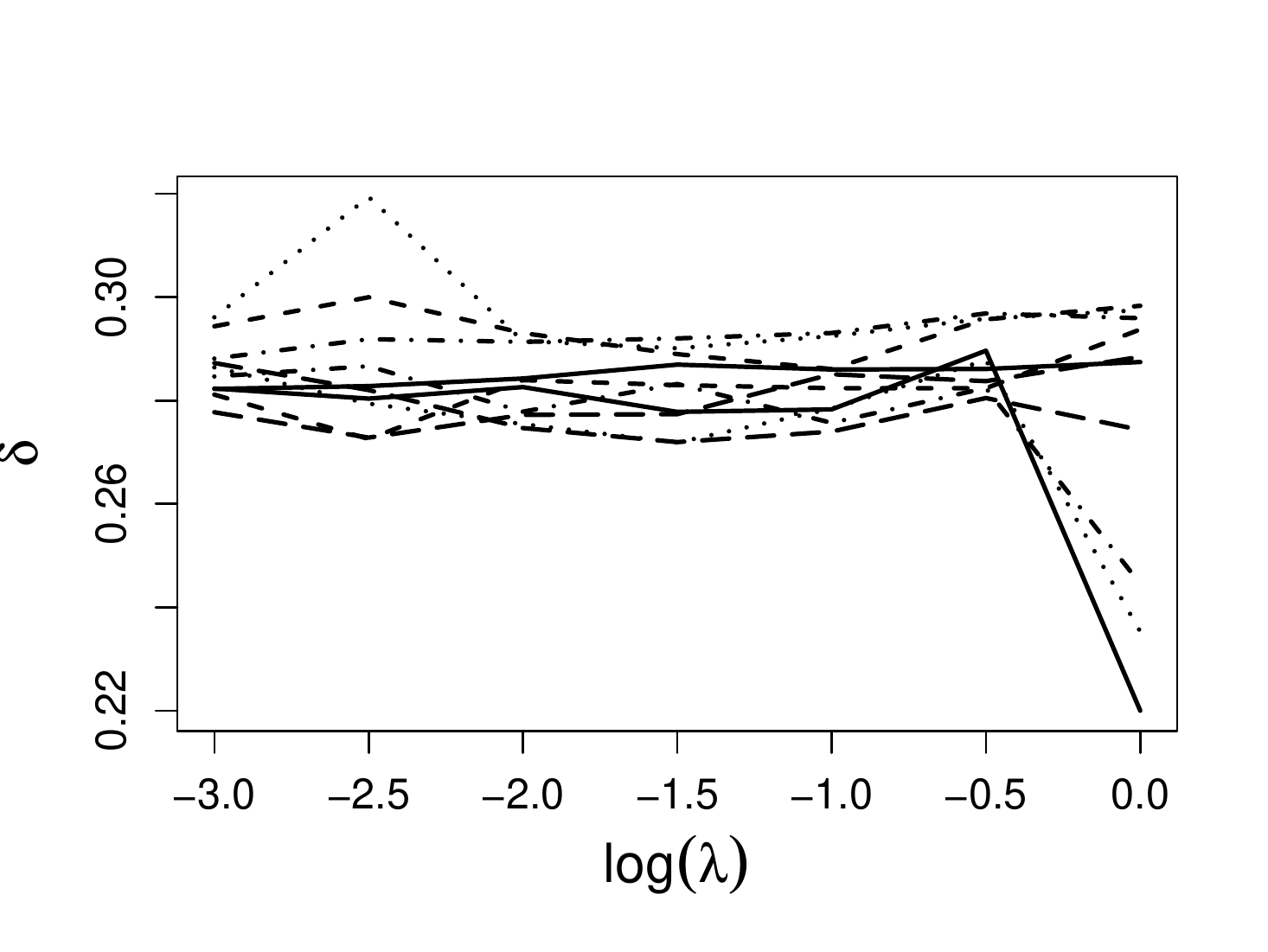}
\includegraphics[width =0.35\textwidth,height = 0.3\textheight]{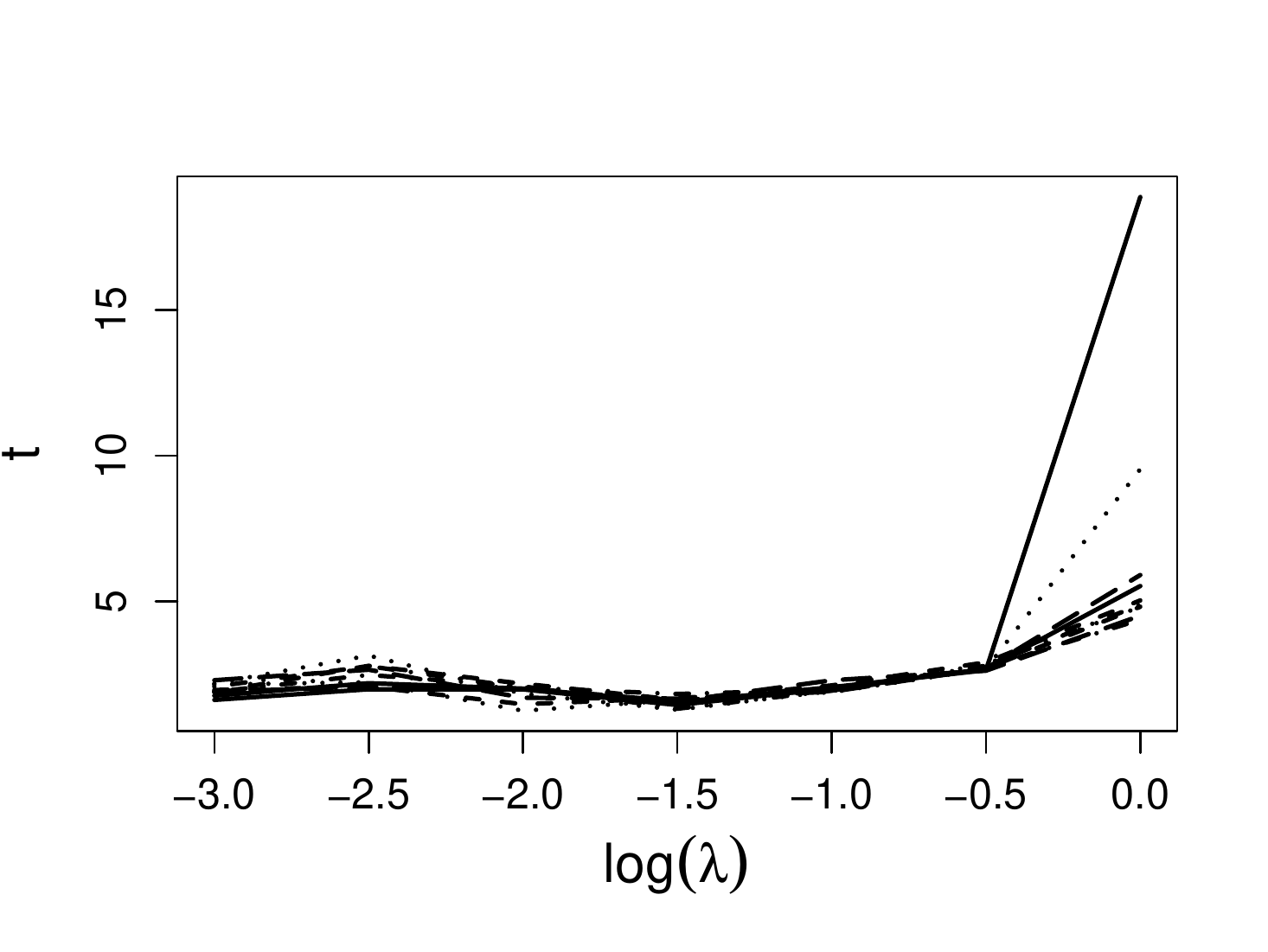}
\end{tabular}
\caption{Left: a sample of $\delta_{\lambda}$ as a function of $\lambda$ in a functional single index model with link function $g\left(s\right) = e^{s}$. Right: the corresponding $t_{\lambda}$ functions.}
\label{fig:4}
 \end{figure}

\subsection{Power Analysis} \label{sec:power}
To investigate the power of our test in more detail we consider a series of increasingly nonlinear
link functions
\begin{align}
g\left(s\right) = s + \eta e^{-s},\label{FSIM_power}
\end{align}
with $ 0 \le \eta \le 1.2$ for the single index model and $0 \le \eta \le 0.8$ for the
functional single index model. As $\eta$ increases, $g$ becomes strongly convex. For each $\eta$, we generate $1000$ simulated data sets and again used $5000$ normal samples under the null distribution to obtain critical values.  Due to the computational overhead associated with searching over another smoothing parameter, we used only 200 simulations for the functional single index model. We replicated each simulation changing $n$ to 200 and changing $\sigma$ to 0.2 to test for the expected loss of power with increasing noise and gain of power with sample size.

\begin{figure}[ht]
\centering
\begin{tabular}{cc}
\includegraphics[width =0.35\textwidth,height = 0.3\textheight]{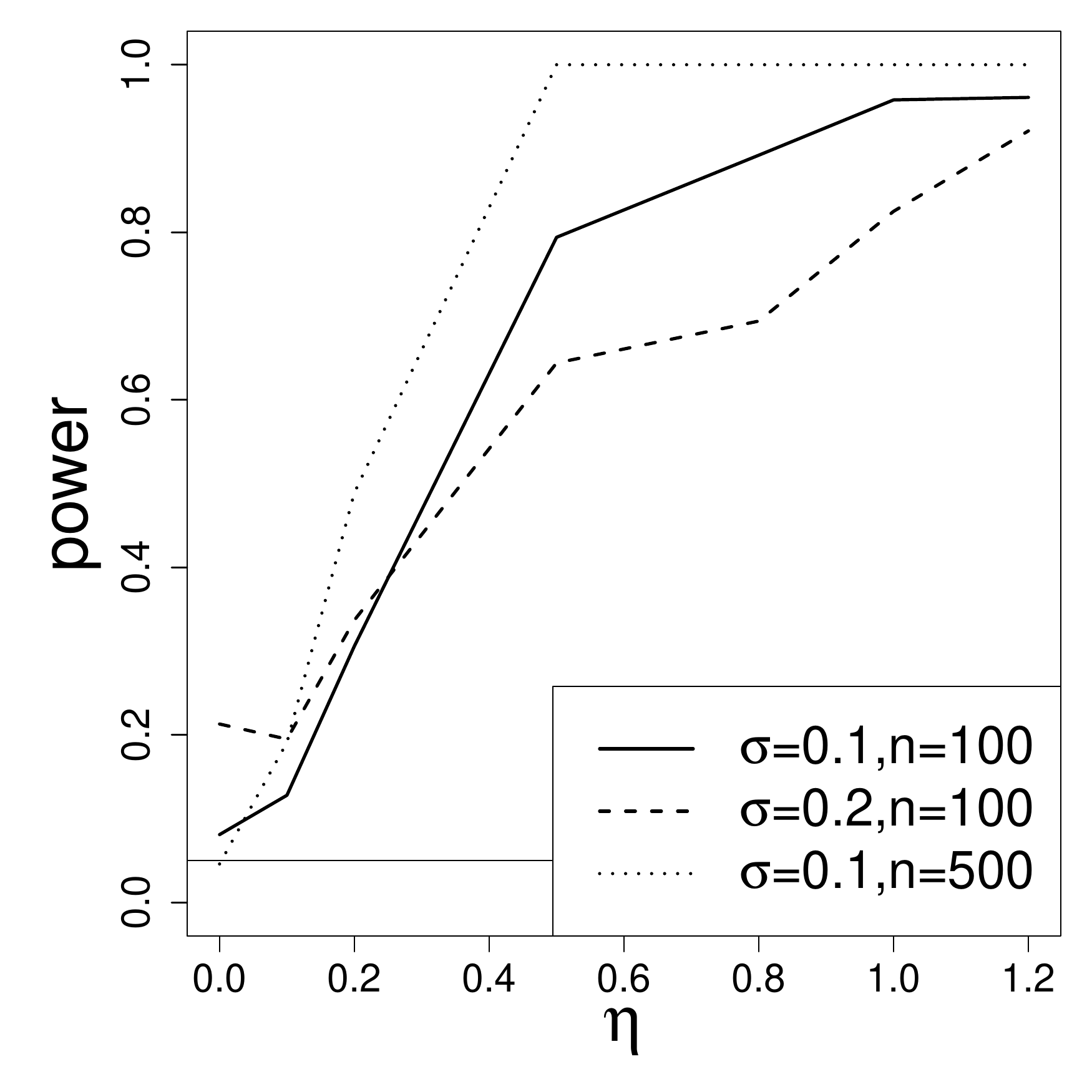} 
\includegraphics[width =0.35\textwidth,height = 0.3\textheight]{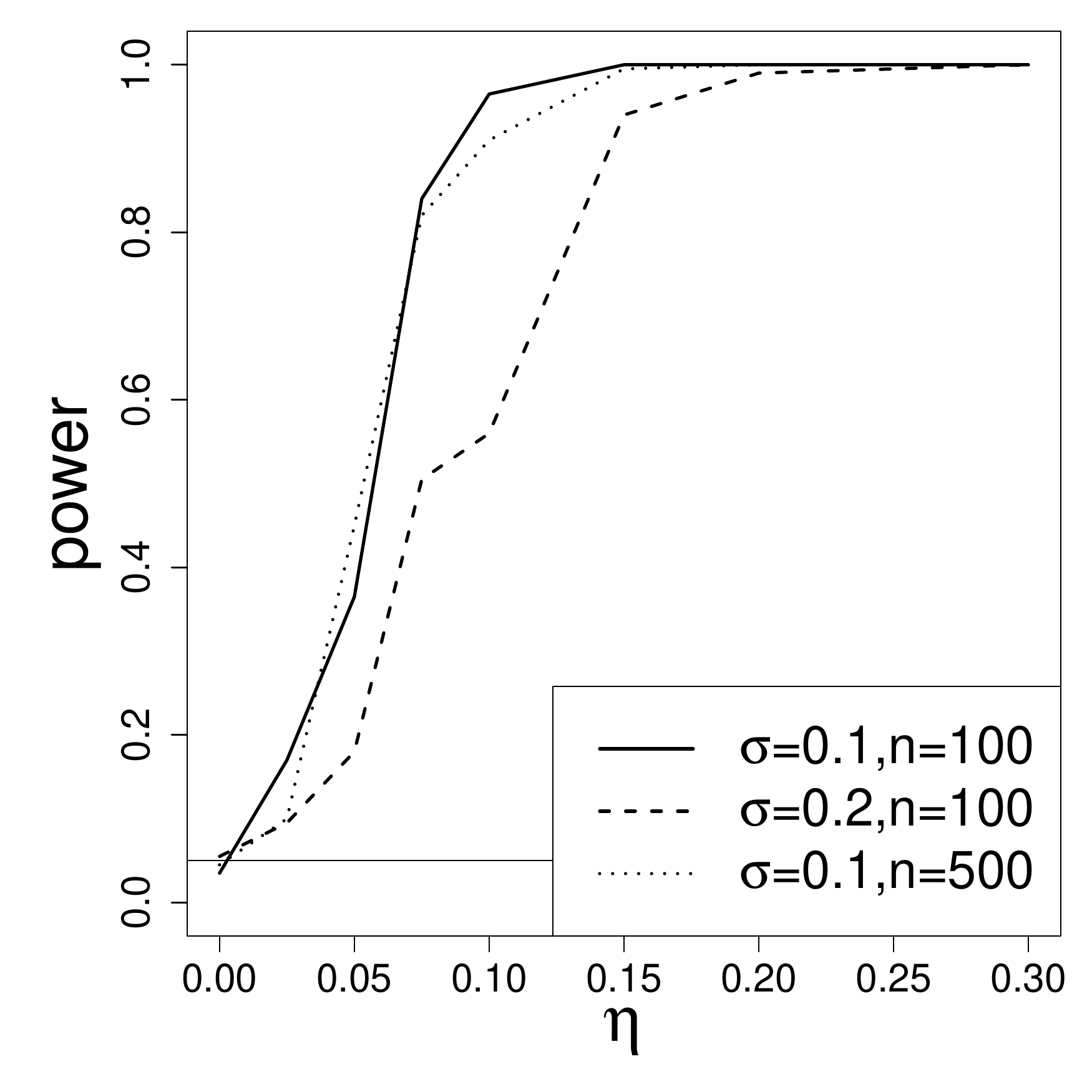}
\end{tabular}
\caption{The power function of the Jensen Effect test plotted against $\eta$ in the link function  $g\left(s\right) = s + \eta e^{-s}$ at sample sizes $n \in \{100,200\}$ and residuals standard errors $\sigma \in \{0.1,0.2\}$. A single index model using the simulation settings from Section \ref{subsec:sim.sim} produced the left plot, while the right-hand plot is obtained using a functional single index model with settings described in Section \ref{subsec:fsim.sim}.}
\label{fig:power}
 \end{figure}

Figure \ref{fig:power} presents the rejection rate plotted against $\eta$. We observe a sharp increase as $\eta$ increases, as expected: as the link function $g$ becomes more and more convex, the rejection rate will converge to $1$. The expected patterns of decreasing power with increasing $\sigma$ and increasing power with $n$ are observed, although a smaller simulation size makes this less clear in the functional single index model.

We also used this experiment as an opportunity to verify that our estimates of residual standard error, $\sigma$ performed well. See Appendix \ref{app:sigest} for further details.

 \section{Application to real ecological data} \label{sec:real}
        To demonstrate application of our tests, we analyze the North Temperate Lakes LTER: Zooplankton - Trout Lake Area data set\footnote{Zooplankton records from \\ \href{https://portal.edirepository.org/nis/mapbrowse?scope=knb-lter-ntl\&identifier=37\&revision=29}{https://portal.edirepository.org/nis/mapbrowse?scope=knb-lter-ntl\&identifier=37\&revision=29} \\ water temperature from \\ \href{https://portal.edirepository.org/nis/mapbrowse?scope=knb-lter-ntl\&identifier=129}{https://portal.edirepository.org/nis/mapbrowse?scope=knb-lter-ntl\&identifier=129}}.
An earlier version of these data were analyzed by \cite{drake2005population} to examine the temperature-dependence of copepod populations. In our data set, the density of the populations of eight species of copepods and rotifers, along with water temperature, were recorded from $1981$ to $2015$ in $8$ different lakes. Our choice of species was determined based on the number and length of observations available and differs from those studied in \cite{drake2005population}.
Note that the growth response in these data is not the growth (in size) of an individual, but the growth (in numbers) of a population, but our functional single index model is still appropriate for this setting.

The values recorded in the original data set are:
\begin{enumerate}
\item $\tt{d}$: species' density at a specific time and lake.
\item $\tt{t}$: record of water temperature collected as the same time as $\tt{d}$.
\end{enumerate}
 Both measurements were recorded on irregular time points among different years and lakes, so in order to obtain functional covariates, we preprocessed the temperature data by fitting a smoothing spline; see details in Appendix \ref{appendix:real}.

Our response is change in density between successive observed time points normalized by the time change:
\[
Y_i = (\delta_{s_{i+1}} - \delta_{s_i})/(s_{i+1} - s_i)
\]
recorded so long as $s_{i+1} - s_i < 100$, where $s_i$ is the time (in days) of the $i^{th}$ sample.

At the time of each observed response, we used  temperature values over the $60$ preceding days as the climate history covariate $X\left(t\right)$. For each species, we fit a penalized spline functional single index model for the growth
in population density as a function of temperature history in each lake; details these procedures are given in Appendix \ref{appendix:real}.   We represented  $\beta$  with 12 order-6 B-splines covering the 60 day history and re-interpolated the smoothed $X(t)$ processes onto this basis.  We used a $25$-dimensional cubic B-spline basis to represent $g$ because we wanted a linear function to fall in the span of our basis and set its range to be $\pm \max \int \phi_j(t) X_i(t) dt$ to ensure that the single index values fell within it. We searched over values of $\log_{10}(\lambda)$ in the range $[-6,2]$ for $g$ and $[-2,6]$ for $\beta$. These were chosen to cover the range of values selected by GCV for most data sets while avoiding spurious significance due to optimization errors.

The Jensen Effect $\delta$ was estimated to be positive at all smoothing parameters in 6 out of the 8 species; in {\em Keratella Earlinae} only one extreme combination of smoothing parameters produced a negative $\delta$. However, statistically significant values of $\delta$ were only found for 5 out of these 7 species ({\em Kellicottia Longispina}, {\em Keratella Cochlearis}, {\em Keratella Earlinae}, {\em Polyarthra Remata},  and {\em Polyarthra Vulgaris}). Nonetheless, we conclude from this that a majority of the copepod species in this data set are evolved to take advantage of environmental variability. Examining estimates of $\beta(t)$, we find some of these are  undersmoothed at GCV values, but they tend to represent gradients corresponding to either warming or cooling water temperatures, likely associated with seasonal abundance trends. The use of water temperature as a sole covariate means that its effects are conflated with other environmental variables that change seasonally, such as the availability of nutrient sources. We thus cannot conclude a causal relationship, but note that the same analysis can be undertaken while accounting for other covariates if when they are available.

\begin{figure}
\centering
\includegraphics[height=0.45\textwidth,angle=270]{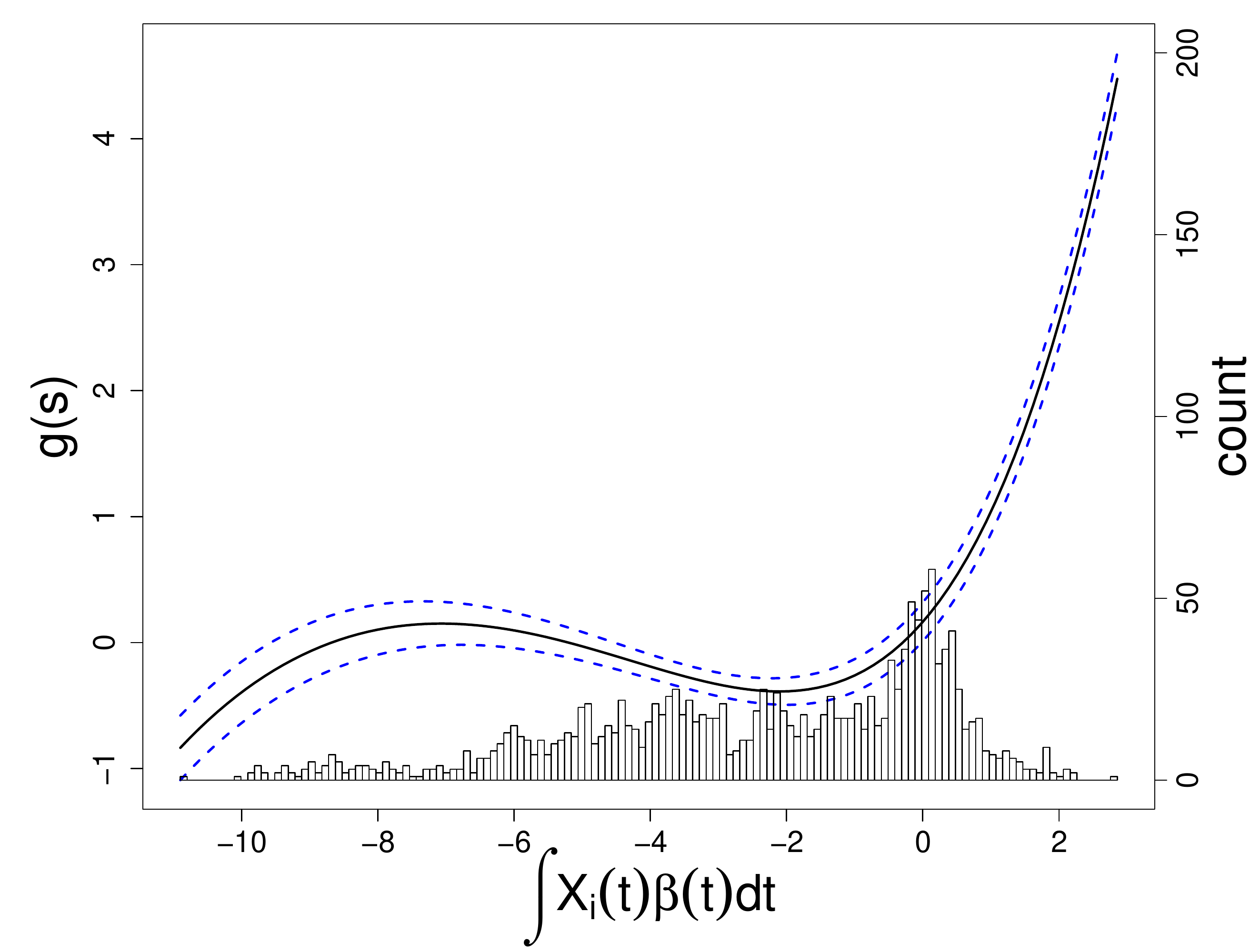}
\includegraphics[height=0.45\textwidth,angle=270]{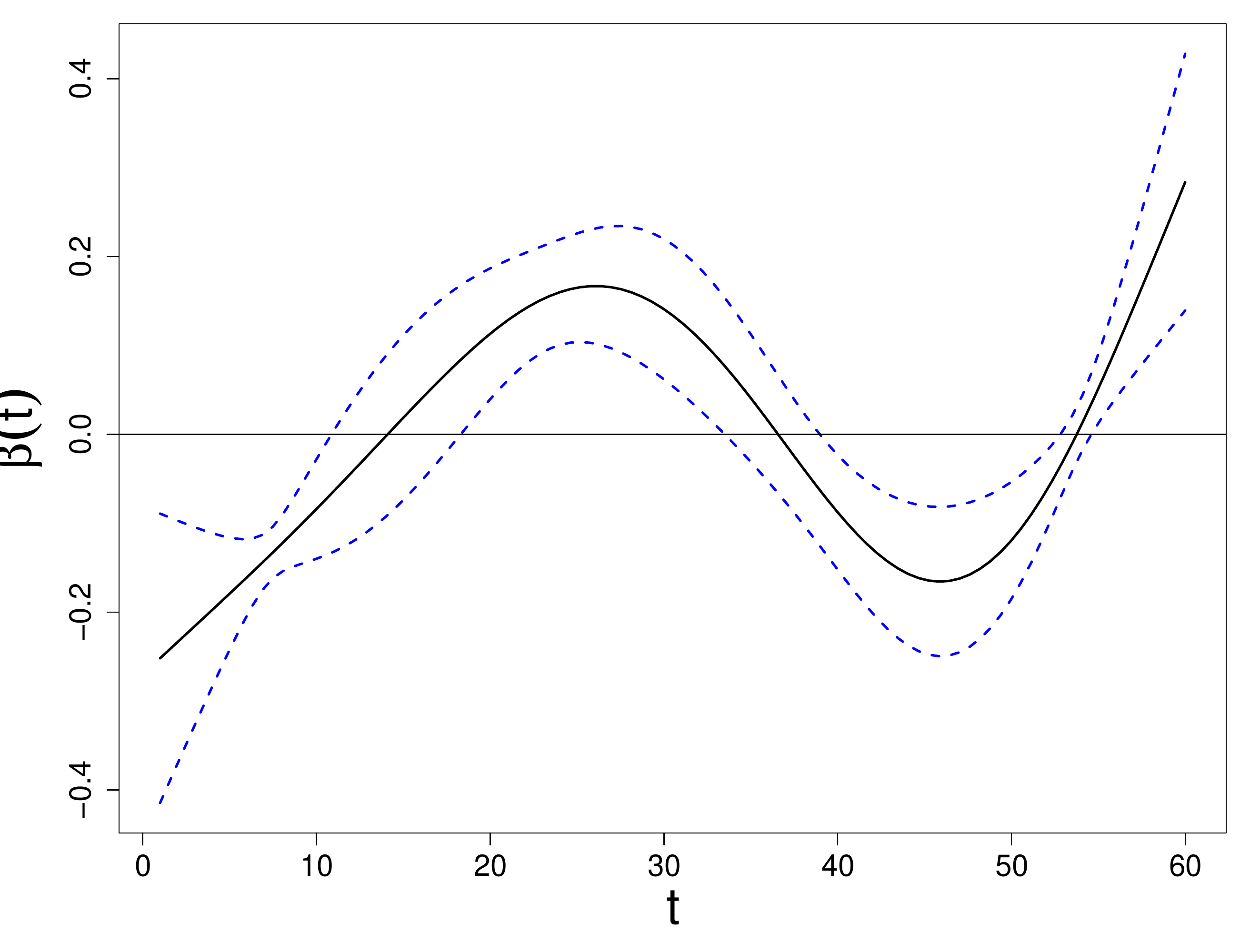}
\includegraphics[height=0.45\textwidth,angle=270]{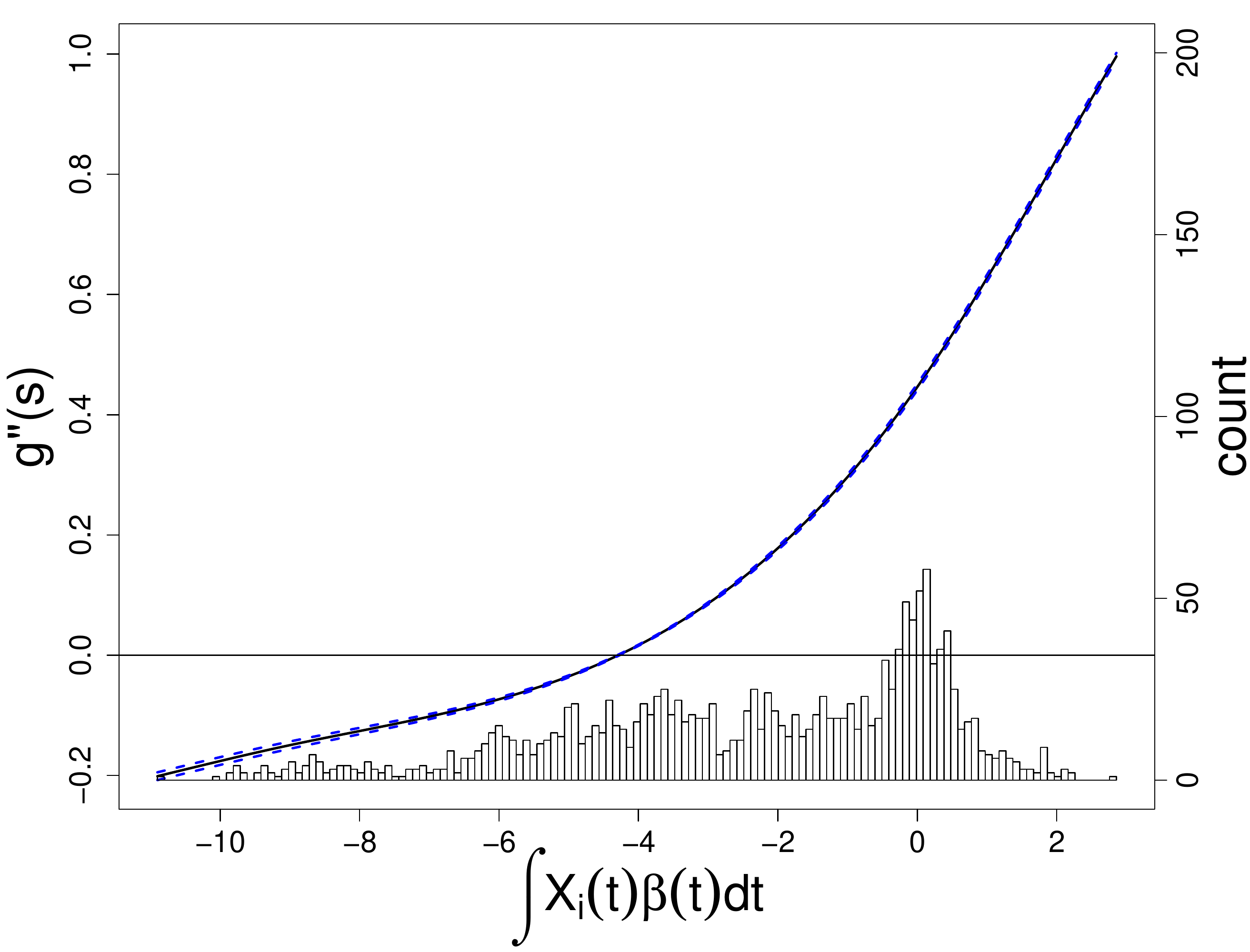}
\includegraphics[height=0.45\textwidth,angle=270]{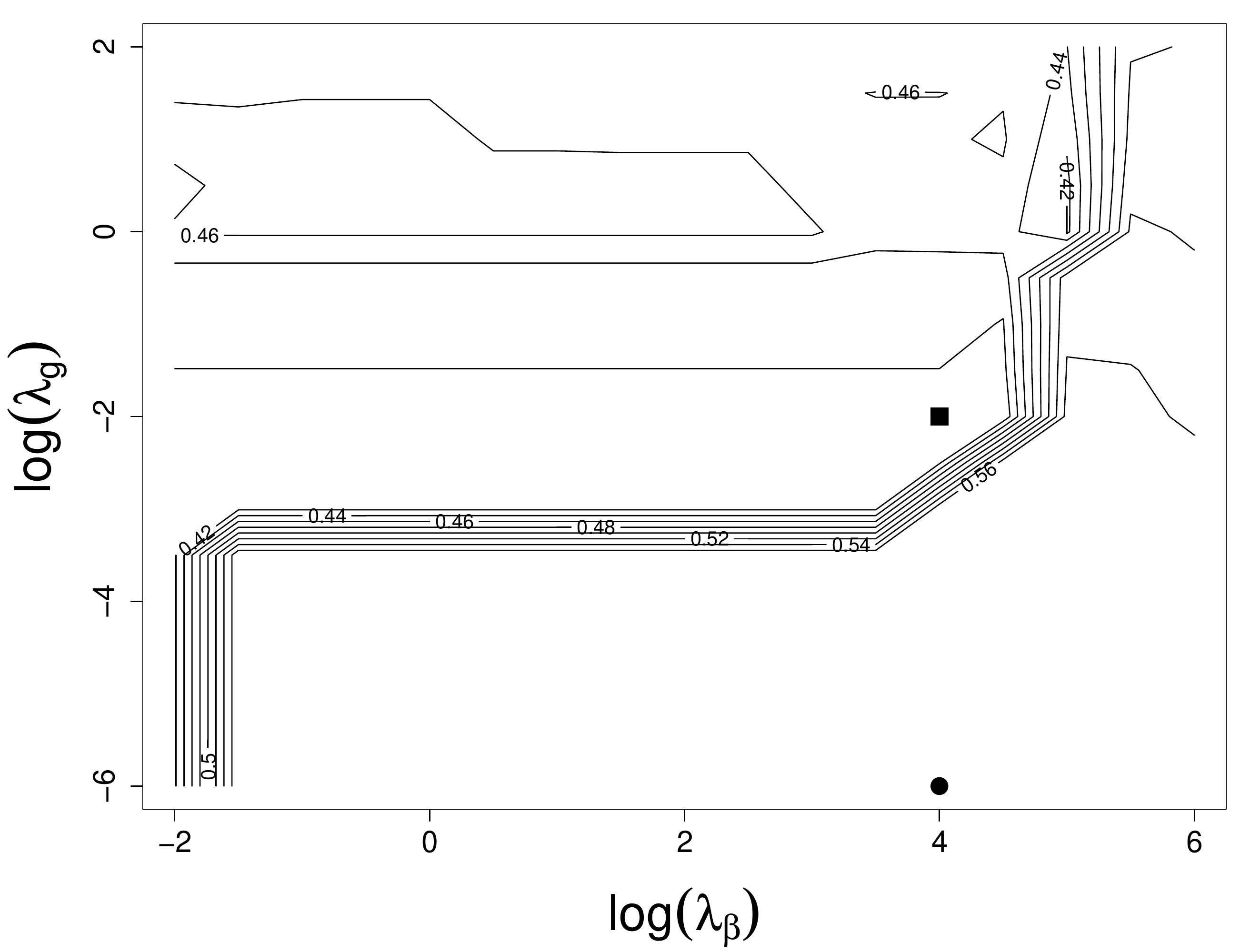} \\
\includegraphics[height=0.45\textwidth,angle=270]{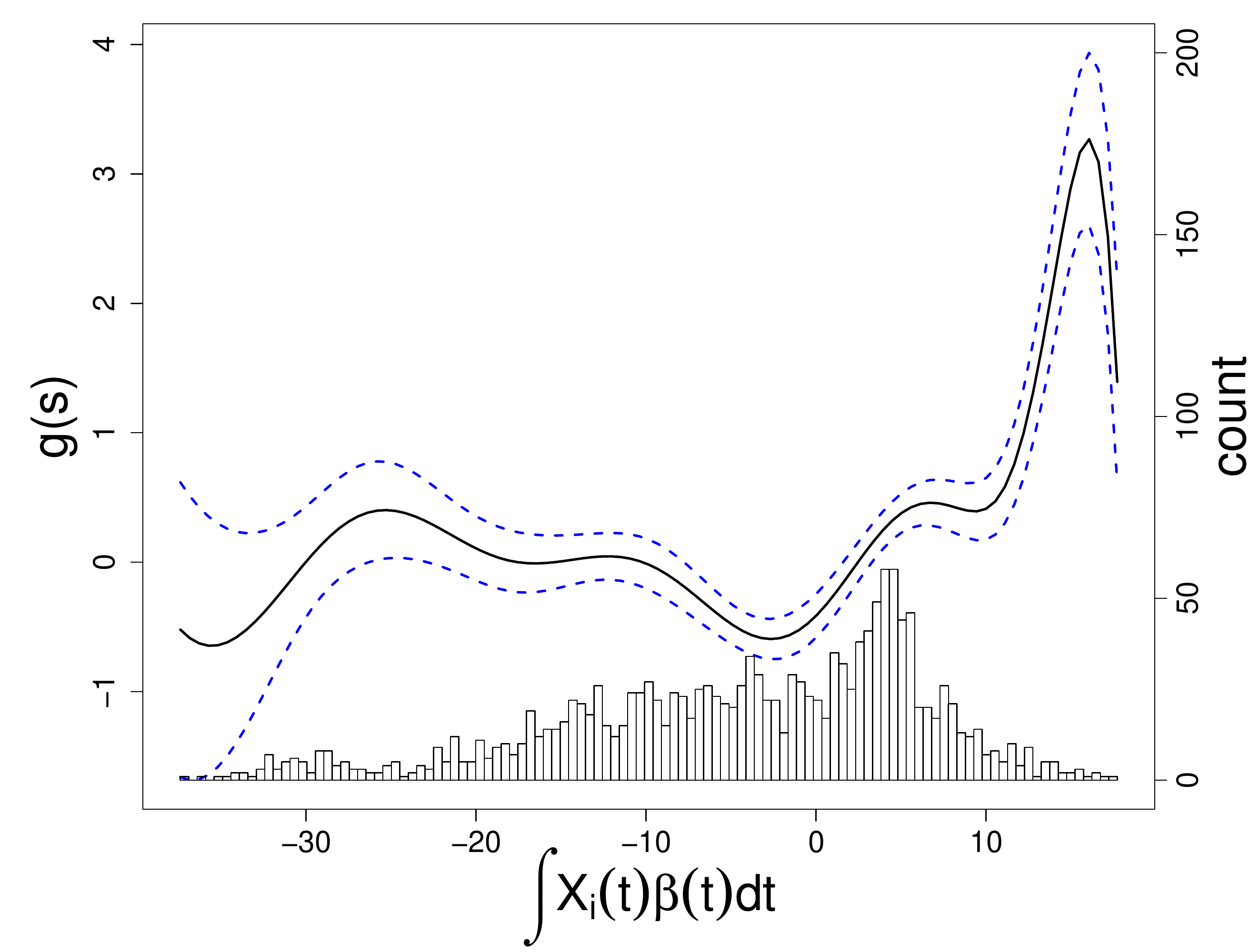}
\includegraphics[height=0.45\textwidth,angle=270]{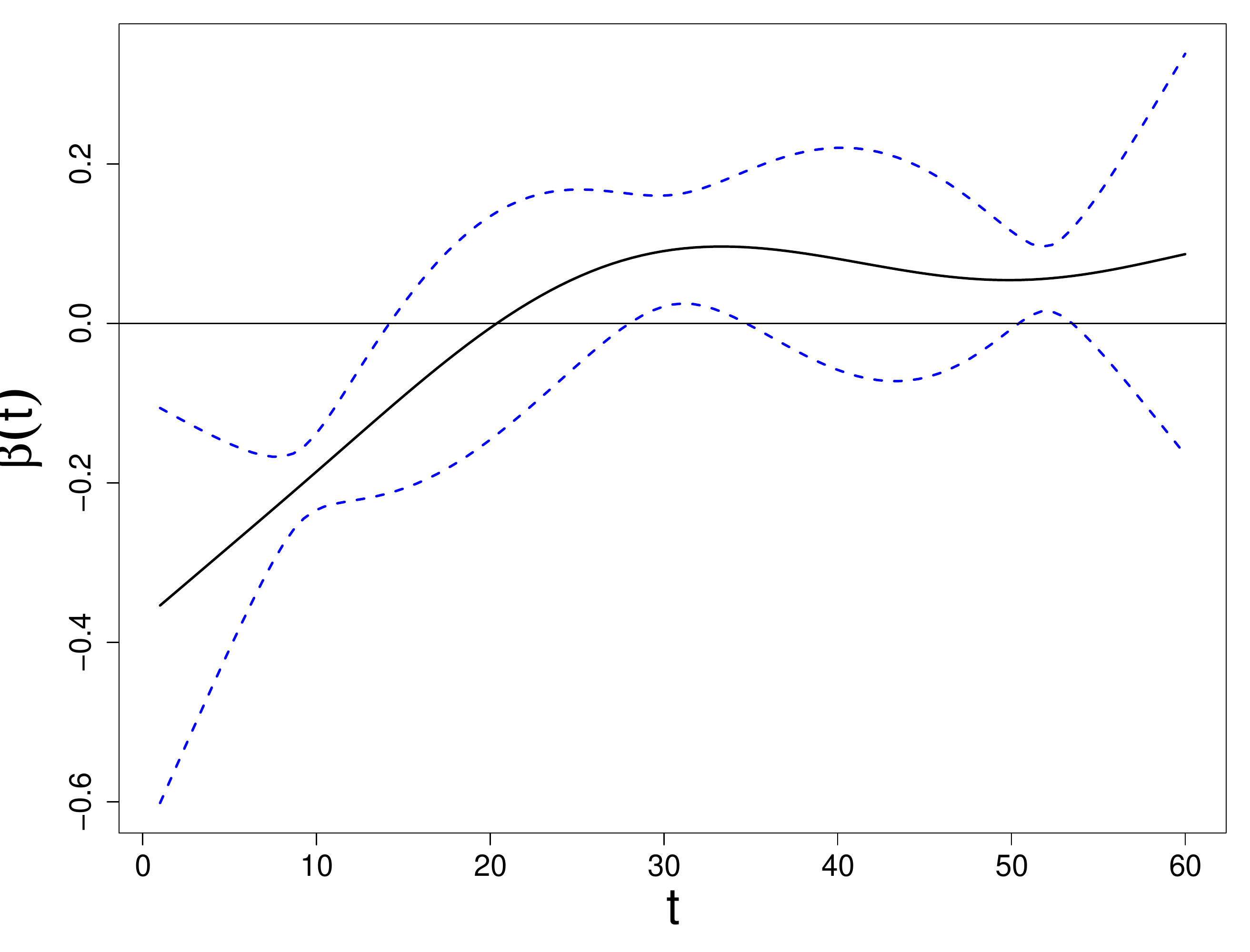}
\caption{Diagnostic plots for \emph{Polyarthra Vulgaris}, $n=1577$. The estimated $g(s)$ (top left) along with $\beta(t)$ (top right), and $g''(s)$ (middle left) along with pointwise confidence intervals.  Histograms give the distribution of estimated $\int X_i(t) \hat{\beta}(t) dt$. All plots are given at the values which minimize GCV. Middle right: $\delta$ as a function of both $\lambda_g$ and $\lambda_{\beta}$. Regions where $t_{\lambda_g,\lambda_\beta}$ exceed the critical value are indicated by a white background (the whole plot in this case) and a black square gives the minimizing value of GCV; black circle indicates values at which $\delta$ is maximized. Bottom row: $g$ and $\beta$ at the smoothing parameter values that maximize $\delta$. }  \label{fig:63005_ex}
 \end{figure}

Figure \ref{fig:63005_ex} provides a canonical set of plots for the species \emph{Polyarthra vulgaris}; equivalent plots for the remaining species are presented in Appendix \ref{appendix:real}. The top four panels provide the estimated $g(s)$, $g''(s)$ and $\beta(t)$ estimated at the smoothing parameters selected by GCV as well as a contour plot of $\delta$ over the smoothing parameters with a shaded area that indicates values at which the effect was found not to be significant. In this case, $\delta$ is positive and significant at all smoothing parameters. In order to explore the shape of the response further, we also plot $g(s)$ and $\beta(t)$ at the smoothing parameters that result in the maximum value of $\delta$. Note that maximizing $\delta$ here results in particularly large edge effects which should be treated with caution.

\begin{figure}
\centering
\includegraphics[height=0.45\textwidth,angle=270]{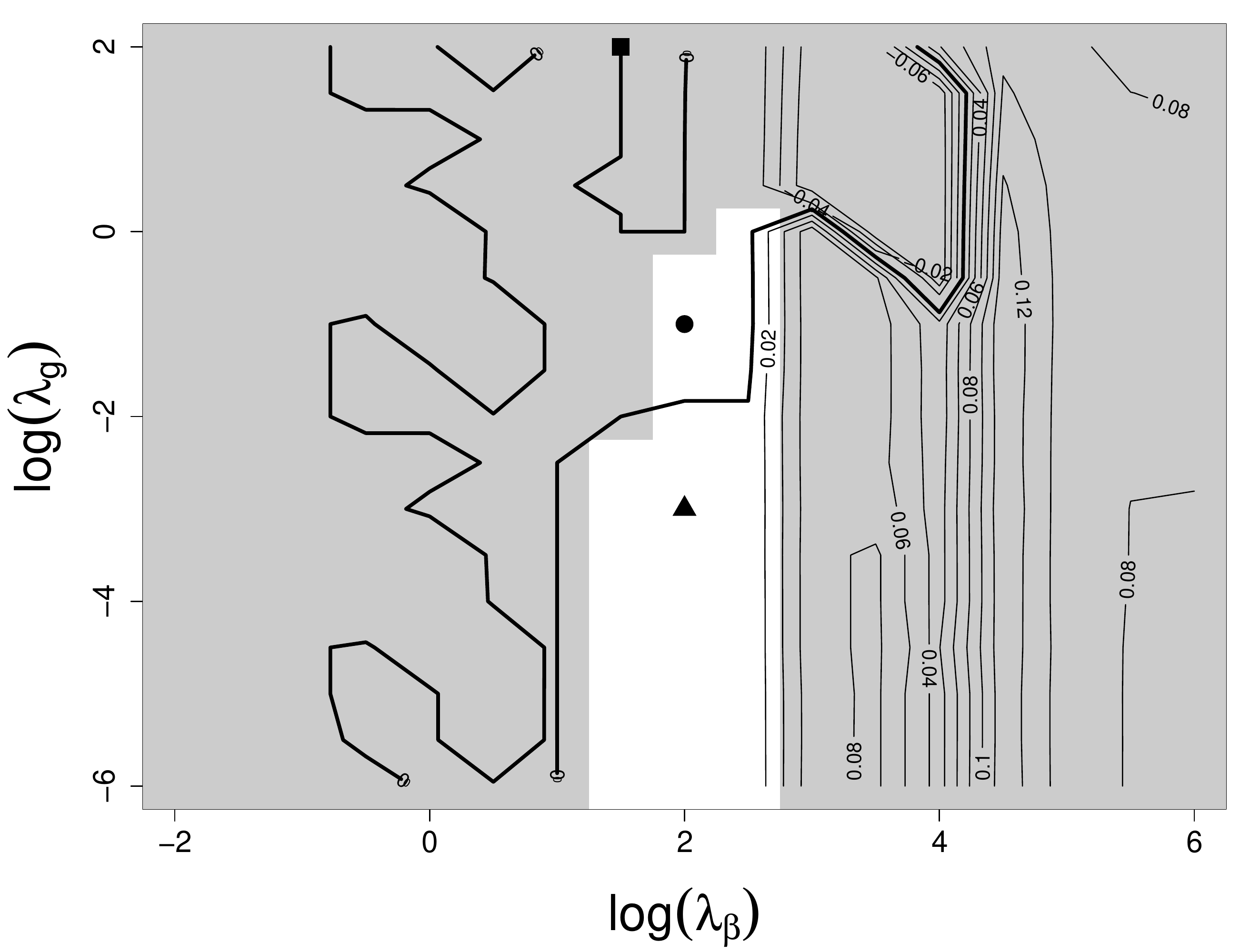}
\includegraphics[height=0.45\textwidth,angle=270]{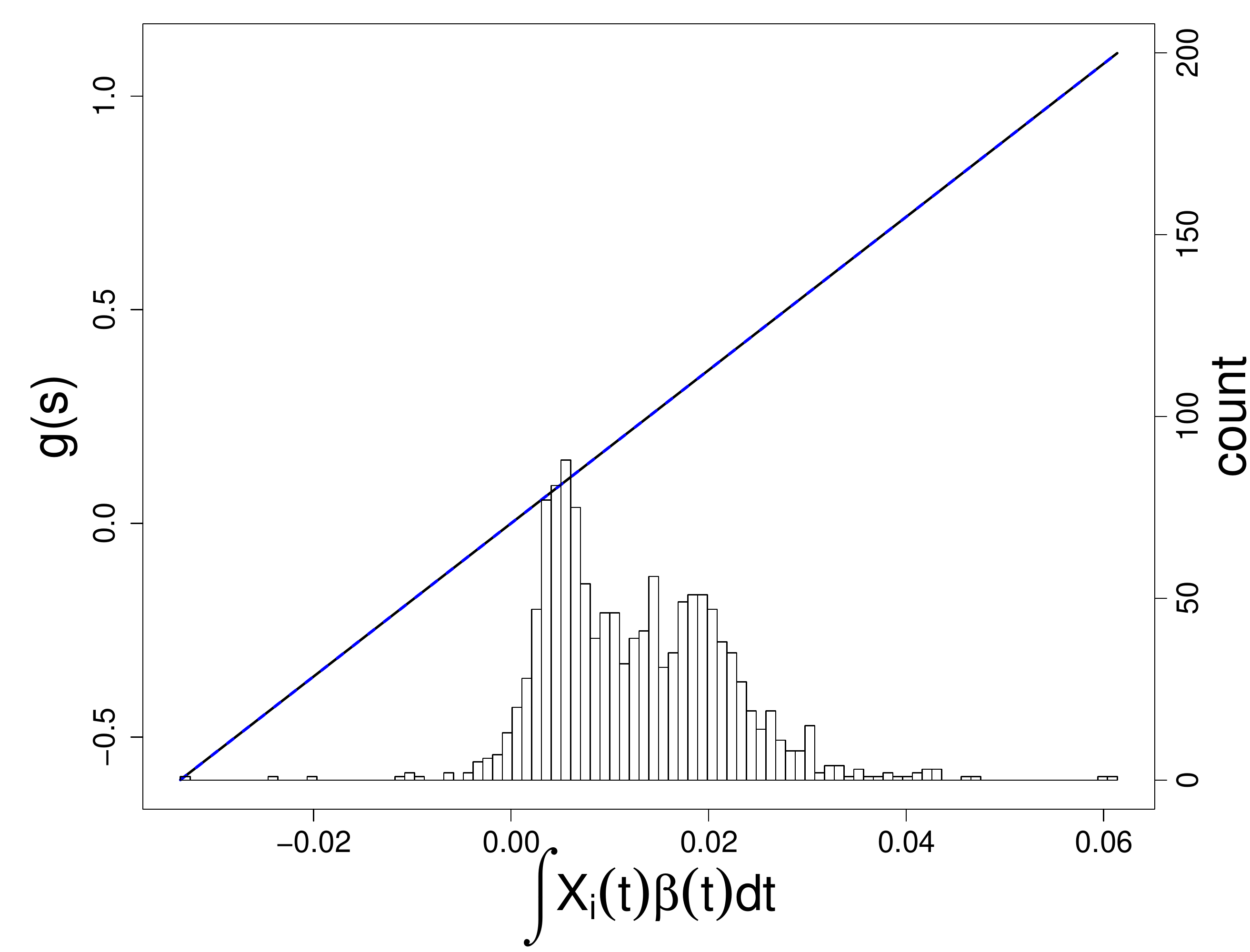} \\
\includegraphics[height=0.45\textwidth,angle=270]{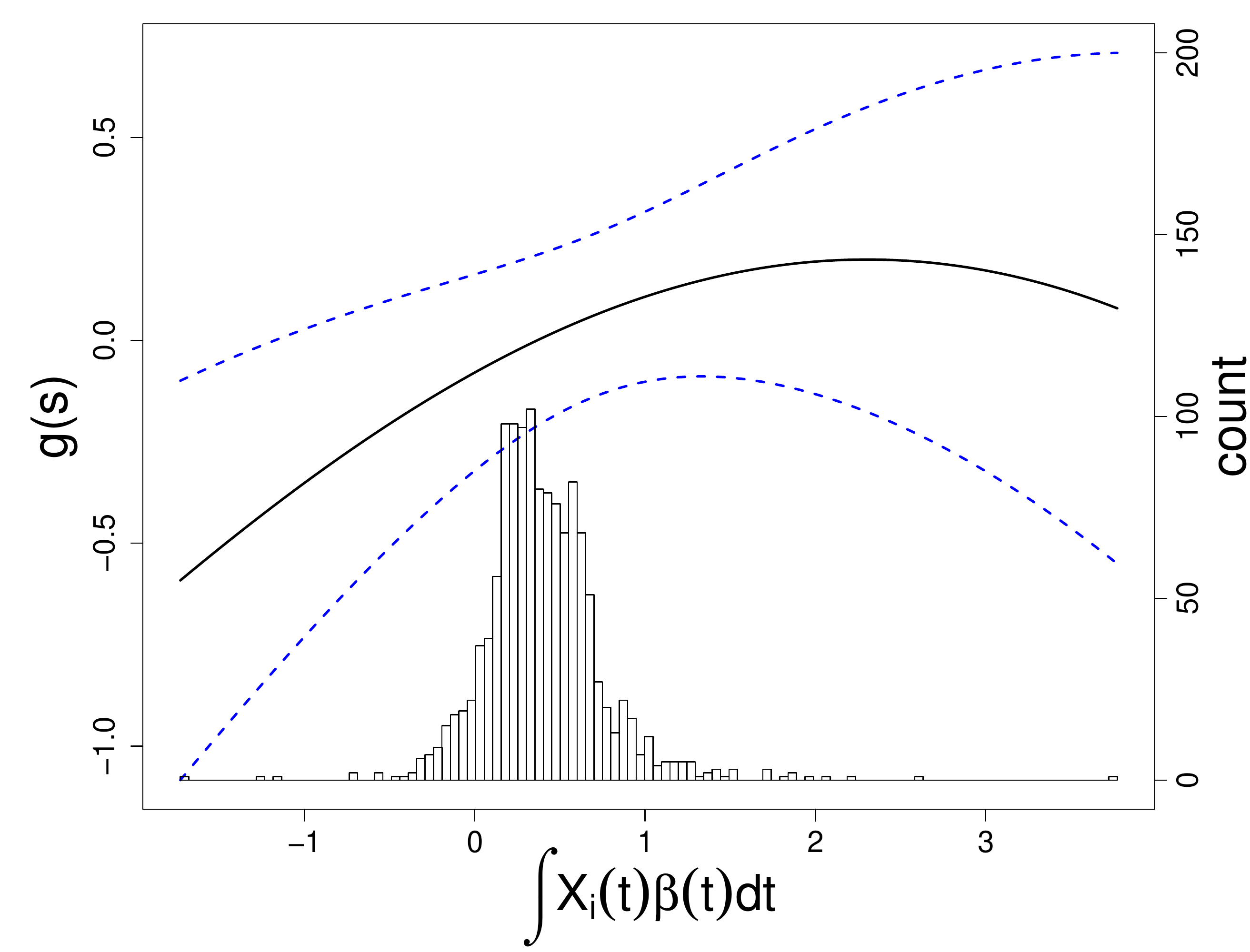}
\includegraphics[height=0.45\textwidth,angle=270]{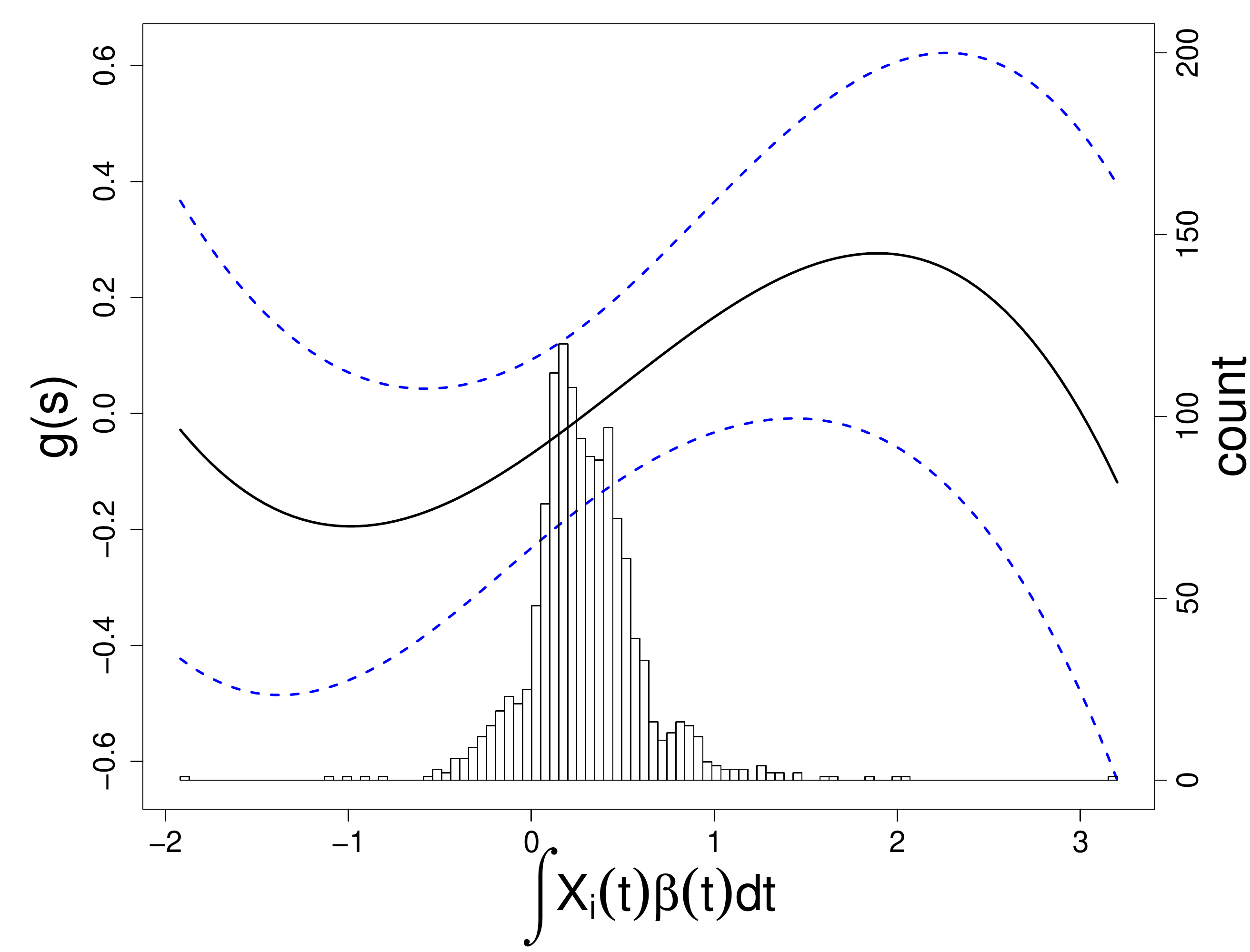}
\caption{Comparison of estimated reaction norm $g$ at different smoothing parameters for {\em Diacyclops thomasi}. Top left: contour of $\delta$ as a function of smoothing parameters with values minimizing GCV (solid square), maximizing $\delta$ (triangle) and minimizing $\delta$ (circle).  The corresponding plots of $g$ are given in the top right (minimizing GCV), bottom left (minimizing $\delta$) and bottom right (maximizing $\delta$). We judge the positive values of $\delta$ to be due to large edge effects.  }
\label{fig:20302_ex}
\end{figure}

One species, {\em Diacyclops Thomasi}, was an exception to the general pattern in our analyses, producing areas in the ($\lambda_g, \lambda_{\beta}$ plant where the estimated
$\delta$ has large positive values and other areas where it has large negative values.  Figure \ref{fig:20302_ex} provides  plots of $g(s)$ at 3 smoothing parameter values, as well as $\delta(\lambda_g,\lambda_\beta)$ along with indicators of where it is significant in order to visualize the effects of smoothing parameters. We note that $\delta$ is not significant at the values of the smoothing parameters selected by GCV (black square), but it is declared to be significant in both the negative (middle of the plot given by the black circle) and positive (bottom indicated by the triangle) directions. Figure \ref{fig:20302_ex} provides plots of $g$  at each of these two points for comparison. The positive estimate is associated with an inflection of $g$ at the low end of the range of $\int X_i(t)\hat{\beta}(t)dt$ values, and is likely the result of undersmoothing. We therefore conclude, tentatively, that {\em Diacyclops Thomasi} apparently differs from the other species in being harmed by temperature variability, but we feel that further experiments are warranted in this case. We note that this is the only species in which our signal is ambiguous (as opposed to inconclusive). In {\em Keratella Earlinae} (Figure \ref{61804}) a single significant negative value is found at the bottom corner of the contour plot, which we feel can be dismissed; see an equivalent but positive corner effect in {\em Kellicottia Longispina} in Figure \ref{61702}.

\section{Conclusion} \label{sec:conclusion}
Environmental variability is ubiquitous, and project to change significantly over the next century as one aspect of global climate change. Projecting how biological populations will respond to climate change therefore requires methods to estimate how they will respond to changes in the variance of conditions, not just the mean.

With the goal of estimating the net effect of environmental variability on components of population growth rate, we first attempted to estimate the curvature of the link function $g$ in a Functional Single Index model. In our penalized spline based method, we found that unrealistic sample sizes were needed to obtain accurate estimates. So instead, we  directly investigated the effect of environmental variability by comparing the expected response (averaged across the environmental variation) to the response at the expected environment. We termed this the ``Jensen Effect'' since it describes the effect of Jensen's Inequality, but we note that it operates far more broadly than on just convex or concave functions. Inspired by the SiZer method, our test for a nonzero Jensen Effect is based on maximizing a test statistic across a wide range of smoothing parameters encompassing all plausible values, thus avoiding the need for smoothing parameter selection. We have shown that our proposed procedures work well, on both simulated and real data.

There are multiple potential extensions of this methodology. We have used observed data as representative of the covariates of interest, to define the average and distribution of environmental variability. However, the test can be conducted for any assumed distribution of covariates, and it may be of interest to describe regions of single index values in which the estimated response function produces a Jensen Effect. A way to achieve this is to plot the $a$ and $b$ for which $E(g(S)) - g(ES)$ is significantly different from 0, when our procedure is applied under the assumption that $S \sim U[a,b]$.

It may also be of interest to ask about Jensen Effects on different scales of measurement. For example, using size (rather than change in size) as a response in our empirical application to copepod growth rate results in heteroskedasticity in the response, which can be ameliorated by a log transform. We would then employ the model $E \log(Y) = g(\int X(t) \beta(t) dt)$ and the Jensen effect of interest is $E \exp (g(\int X(t) \beta(t) dt)) - \exp(g( E \int X(t) \beta(t) dt))$. Smoothing $\hat{g}_{\lambda}$ biases the Jensen effect towards being positive, and our method would need modifications to take account of this.

The same challenge arises in extending our approach to other response structures, such as survival or count data, which also use nonlinear link functions for fitting within a generalized linear models framework. Any of the standard GLM links could be modified by placing a nonparametric $g(s)$ within the GLM link. But the Jensen Effect then applies to the composition of the GLM link with $g(s)$, so again there would need to be a way of accounting for the bias introduced by smoothing.

\bibliographystyle{chicago}
\bibliography{ecological}

\clearpage
\appendix

\section{A Simulated Demonstration} \label{sec:smallsimulation}
We present here a brief simulation study to observe the accuracy of curvature estimates. The covariate function $X\left(t\right)$ was
generated based on a Fourier basis
\begin{align*}
X_i\left(t\right) = \mu\left(t\right) + \sum\limits_{k=1}^4\xi_{ik}\eta_k\left(t\right), \quad \quad i = 1, \cdots, n,
\end{align*}
where $\mu\left(t\right) = t$, $\eta_1\left(t\right) = \frac{1}{\sqrt{2}}\sin\left(2\pi t\right)$, $\eta_2\left(t\right) = \frac{1}{\sqrt{2}}\cos\left(2\pi t\right)$, $\eta_3\left(t\right) = \frac{1}{\sqrt{2}}\sin\left(4\pi t\right)$, $\eta_4\left(t\right) = \frac{1}{\sqrt{2}}\cos\left(4\pi t\right)$, and $\xi_{ik}$ are independent $\mathrm{N}\left(0,\gamma_k\right)$ with $\gamma_1 = 1$, $\gamma_2 = \frac{1}{2}$, $\gamma_3 = \frac{1}{4}$, $\gamma_4 = \frac{1}{8}$. The coefficient function is
\begin{align*}
\beta\left(t\right) = \sqrt{2}\left[\frac{1}{\sqrt{12}}\eta_1\left(t\right)+\frac{1}{\sqrt{12}}\eta_2\left(t\right)+\frac{1}{\sqrt{6}}\eta_3\left(t\right)+\frac{1}{\sqrt{6}}\eta_4\left(t\right)\right].
\end{align*}
We observe that the coefficients for $\beta$ satisfy $\left\|\bm{c}\right\| = 1$, under an orthonormal basis. The random errors
$\epsilon_i$ are simulated as i.i.d. Gaussian noise with mean $0$ and $\text{var}\left(\epsilon\right) = 0.1\text{var}\left[g\left(\int X\beta \right)\right]$.\\
We selected the sample size as $n = 100$ and examined three link functions:
\begin{enumerate}
\item $g\left(s\right) = e^{-s}$.
\item $g\left(s\right) = -s^2$.
\item $g\left(s\right) = s$.
\end{enumerate}
To measure the performance of our estimators we define the MSE of the estimated $\beta$ and $g^{\left(k\right)}$ to be
\begin{align*}
\text{RSE} = \left[\int \left(\hat{\beta}\left(t\right) - \beta\left(t\right)\right)^2\mathrm{d}t\right]^{\frac{1}{2}},
\end{align*}
and
\begin{align*}
\text{RASE(k)} = \left\{\frac{1}{n}\sum\limits_{i=1}^n\left[\hat{Y}^{\left(k\right)}_i - g^{\left(k\right)}\left(\int X_i\left(t\right)\beta\left(t\right)\mathrm{d}t\right)\right]^2\right\}^{\frac{1}{2}},
\end{align*}
where $\hat{Y}^{\left(k\right)}_i = \hat{g}^{\left(k\right)}\left(\int X_i\left(t\right)\hat{\beta}\left(t\right)\mathrm{d}t\right)$ for $k = 0,1,\cdots$.

Of particular concern in the results (Table \ref{table1})  is the substantial discrepancy between estimates from different initial conditions.
\cite{ye2018local} similarly observed that second derivative estimates were highly sensitive to the effort placed into optimization.

\begin {table}[H]
\begin{center}
\begin{tabular}{ccc|cc|cc}
\hline
\multirow{2}{*}{} & \multicolumn{2}{c}{\textbf{g1}} &  \multicolumn{2}{c}{\textbf{g2}} & \multicolumn{2}{c}{\textbf{g3}} \\
\cline{2-7}
\textbf{Initial} & True & equal &  True & equal & True & equal \\
\hline
\textbf{RSE} & 1.1213  & 0.6417 &  0.5385 & 0.6980 & 0.7608 & 0.7024  \\
\textbf{RASE(1)} & 0.0921  & 0.0800  & 0.0490  & 0.0730  & 0.0706 & 0.0764
\\
\textbf{RASE(2)} & 5.2517 & 3.0516  & 2.9079 & 4.1328 & 5.4393 & 1.2170 \\[1ex] \hline
\end{tabular}
\end{center}
\caption {Simulation results with $(\lambda_g,\lambda_\beta)$ selected by GCV. Values in the Table are averages over 100 simulations.}
\label{table1}
\end{table}

\begin{figure}
\begin{center}
  \includegraphics[width=15cm]{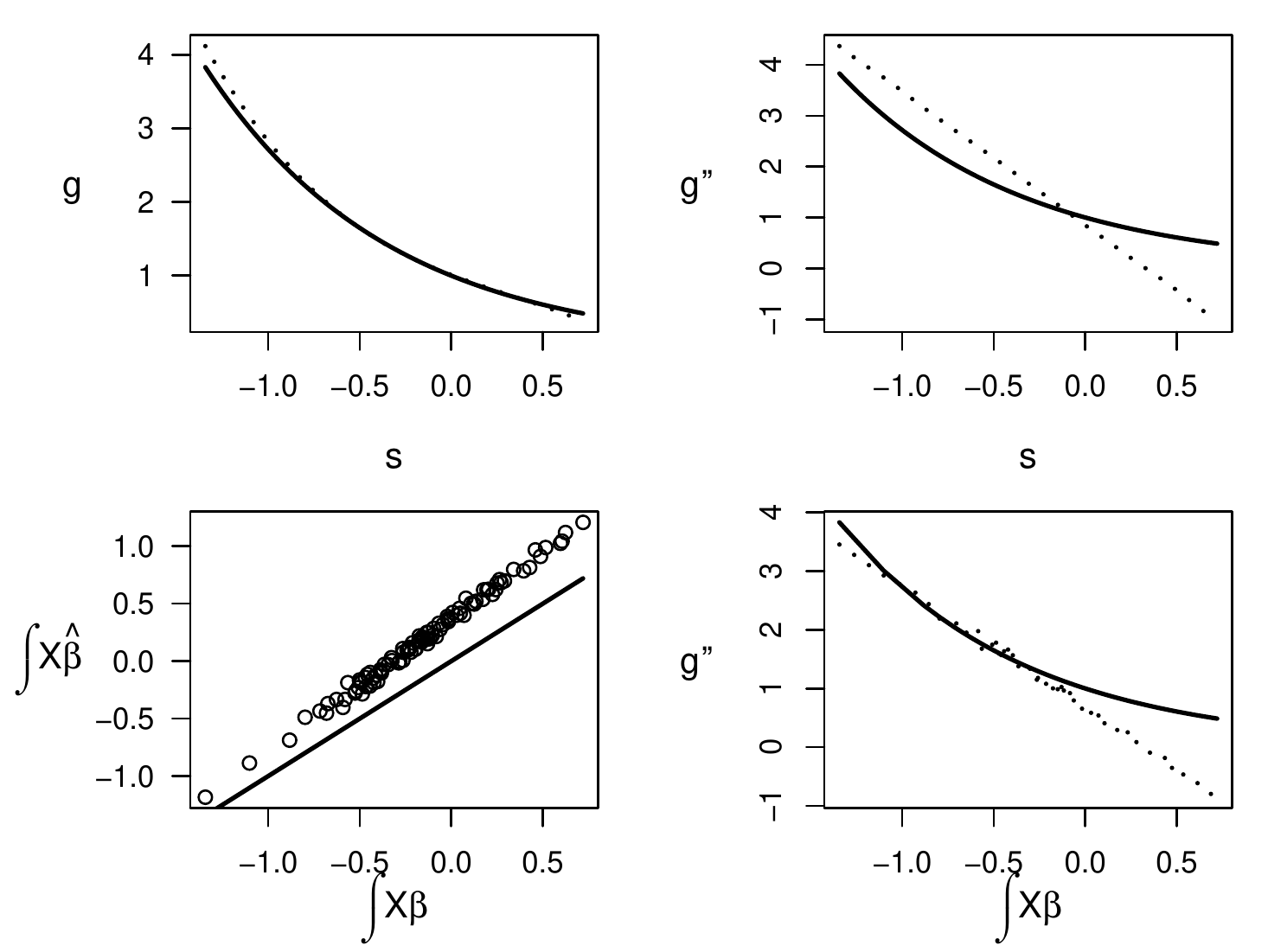}
  \caption{Example estimate for the link function $g\left(s\right) = e^{-s}$. Top-left and right panels plot $g$ and $g''$ over $1000$ equally-spaced grid points between the minimum and maximum of $\int X\left(t\right)\hat{\beta}\left(t\right)\mathrm{d}t$.
  Dots are estimated values, and the solid curves are the truth. The bottom-left panel plots $\int X\left(t\right)\hat{\beta}\left(t\right)\mathrm{d}t$ versus $\int X\left(t\right)\beta\left(t\right)\mathrm{d}t$ (circles); the solid line is the 1:1 line. The bottom-right panel presents $g''$ (black) and $\hat{g}''$ (red) evaluated at $\int X\left(t\right)\beta\left(t\right)\mathrm{d}t$ and $\int X\left(t\right)\hat{\beta}\left(t\right)\mathrm{d}t$ respectively but plotted against the true argument.}\label{fig:1}
 \end{center}
\end{figure}

\begin{figure}[ht]
\centering
\begin{tabular}{cc}
\includegraphics[width =0.35\textwidth,height = 0.3\textheight]{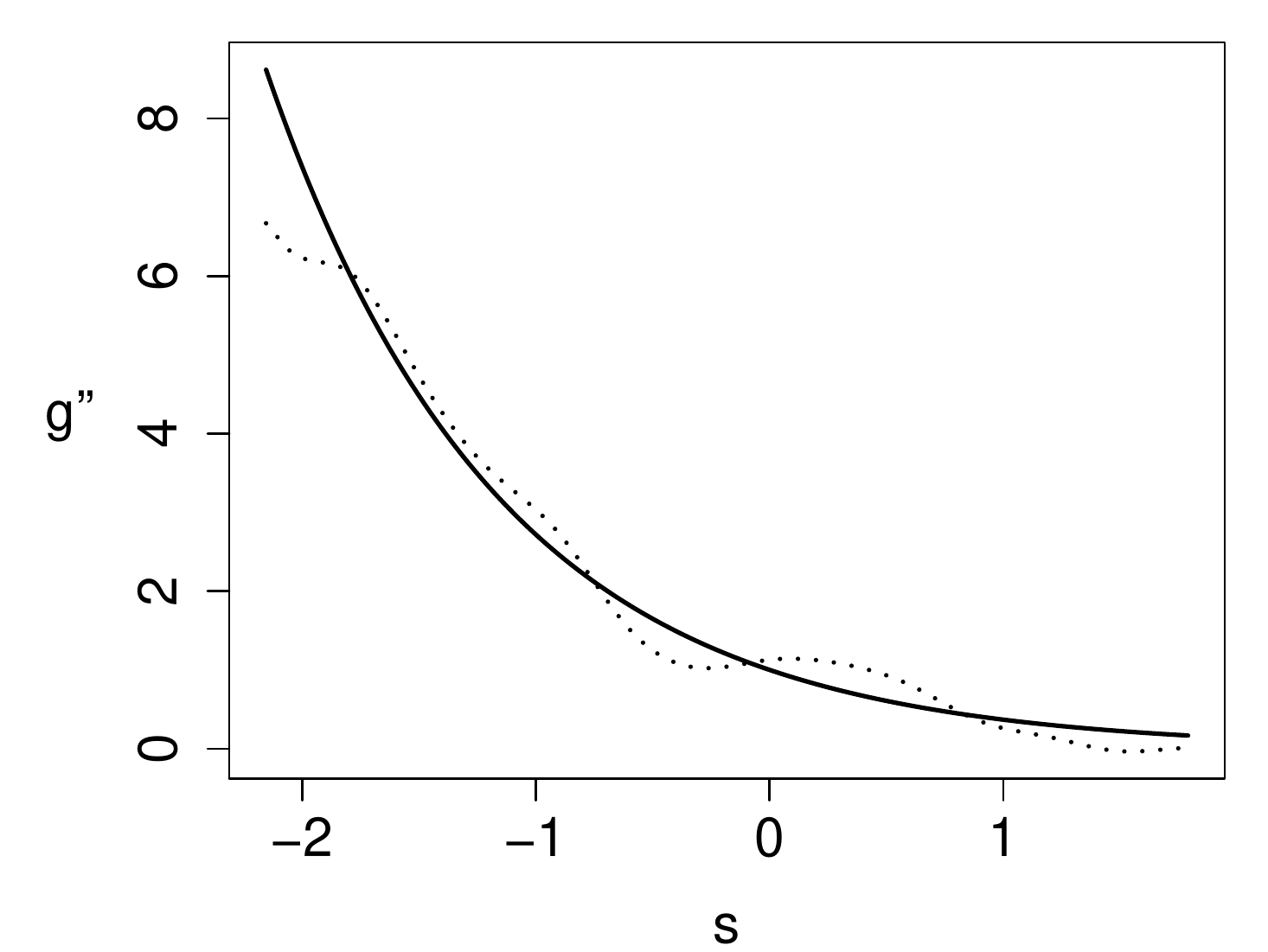} 
\includegraphics[width =0.35\textwidth,height = 0.3\textheight]{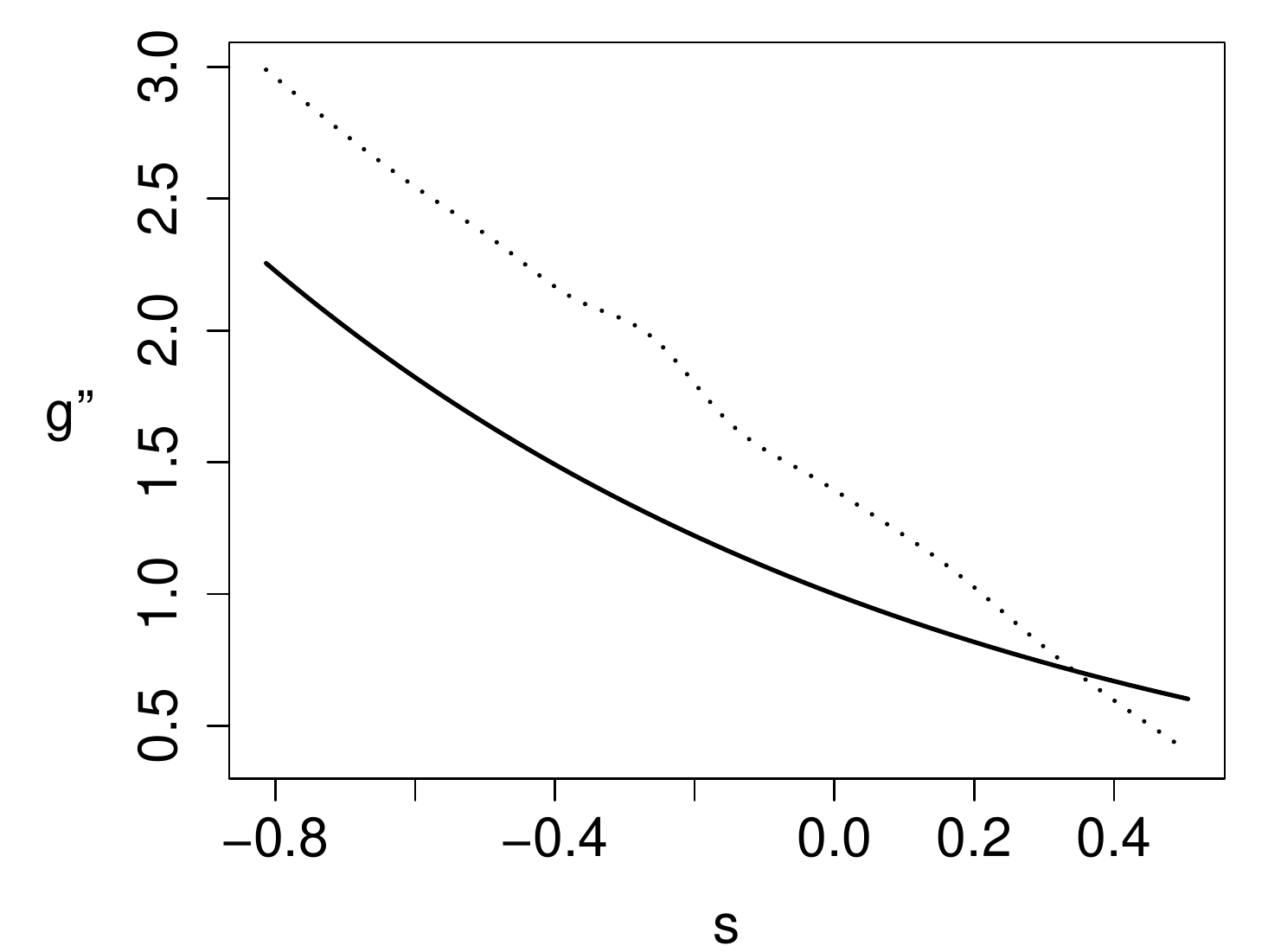}
\end{tabular}
\caption{Estimates for $g''$ from different initial conditions. Left, using the known true $g$ as the initial condition. Right, starting from equal values of the coefficients. Note that this example was chosen for illustrative purposes and does not use the same data as Figure \ref{fig:1}; the domains of the function are different due to differences in the estimate $\int X_i(t)\beta(t) dt$}
\label{fig:2}
 \end{figure}

\begin{figure}[tbp]
\begin{center}
  \includegraphics[width=15cm]{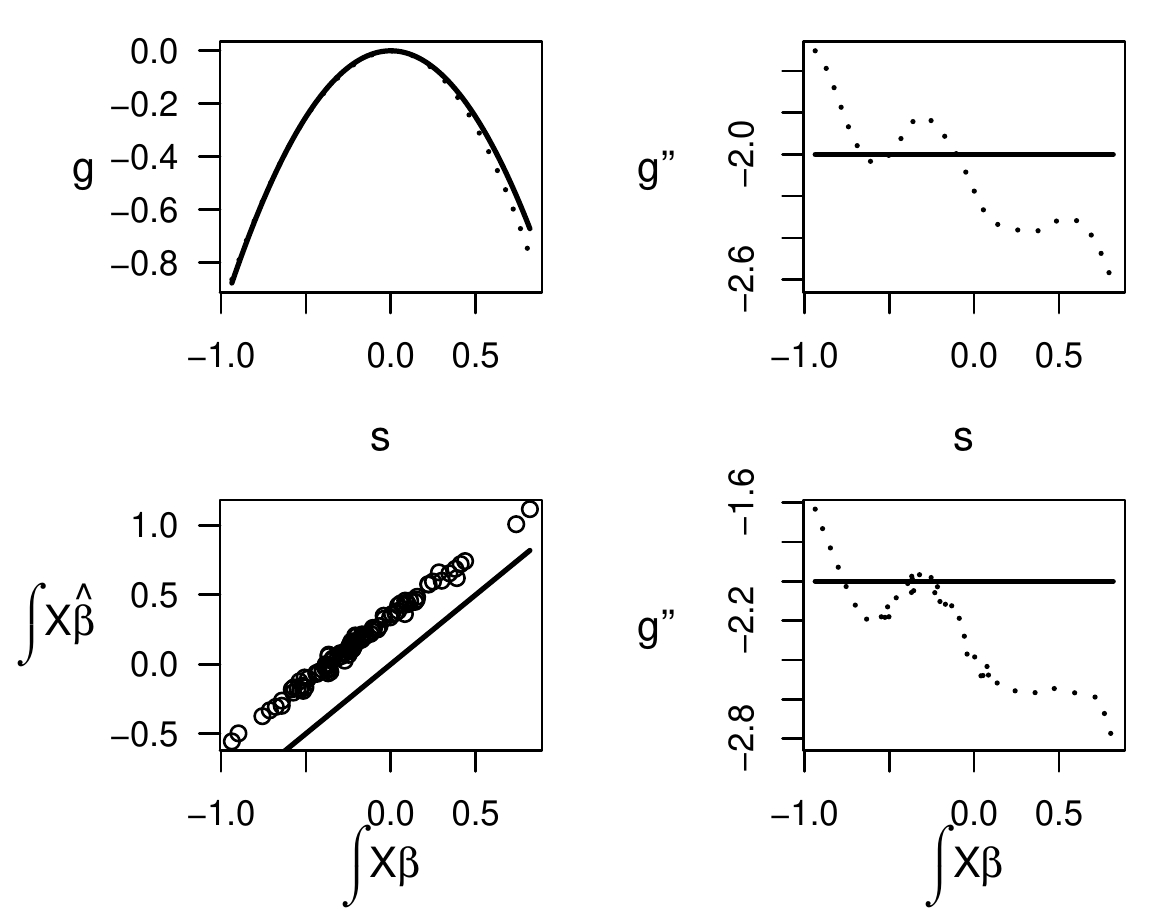}\\
  \caption{The link function is $g\left(s\right) = -s^2$. The top-left and right panel are the plots of $g$ and $g"$ over $1000$ equally-spaced grid points, while the lower and upper bound are the minimum and maximum of $\int X\left(t\right)\hat{\beta}\left(t\right)\mathrm{d}t$. The bottom-right panel is the plot of $g"$ over the true $\int X\left(t\right)\beta\left(t\right)\mathrm{d}t$. The generating model is indicated by solid lines, while dashed lines give the estimated curve. The bottom-left panel is the plot of $\int X\left(t\right)\hat{\beta}\left(t\right)\mathrm{d}t$ versus $\int X\left(t\right)\beta\left(t\right)\mathrm{d}t$, with the $y=x$ fit indicated by the solid line.} \label{fig:curvQuad}
 \end{center}
\end{figure}

\begin{figure}[tbp]
\begin{center}
  \includegraphics[width=15cm]{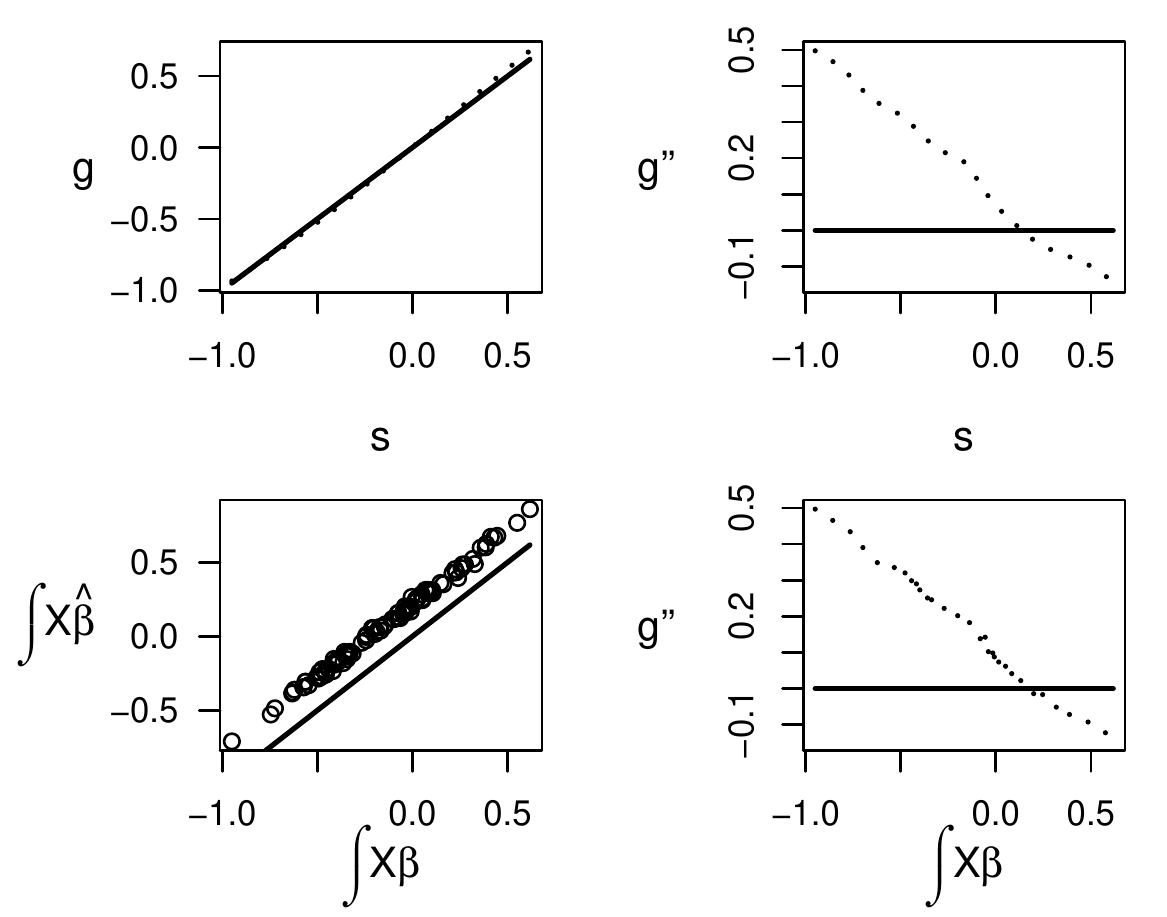}\\
  \caption{The link function is $g\left(s\right) = s$. The top-left and right panel are the plots of $g$ and $g"$ over $1000$ equally-spaced grid points, while the lower and upper bound are the minimum and maximum of $\int X\left(t\right)\hat{\beta}\left(t\right)\mathrm{d}t$. The bottom-right panel is the plot of $g"$ over the true $\int X\left(t\right)\beta\left(t\right)\mathrm{d}t$. The generating model is indicated by solid lines, while dashed lines give the estimated curve. The bottom-left panel is the plot of $\int X\left(t\right)\hat{\beta}\left(t\right)\mathrm{d}t$ versus $\int X\left(t\right)\beta\left(t\right)\mathrm{d}t$, with the $y=x$ fit indicated by the solid line.} \label{fig:curvLin}
 \end{center}
\end{figure}

\noindent The plots in Figure \ref{fig:1} provide an example of our results. The estimate of the link function nearly overlaps the true curve,
indicating that our estimate of the link function is quite accurate. However, for the second derivative, the estimate deviates from the true curve, becoming negative towards the right-hand limit. This reduced accuracy is also evident in the results in Table \ref{table1}. These plots indicate that our estimate of the curvature is not good enough to use as a basic for decisions on the convexity of $g$. In addition, the performance of the estimators varies a lot
from different initial values. In Figure \ref{fig:2} we see that different initial conditions can lead to either over- or under-fitting $g"$.
Further examples are provided in Figures \ref{fig:curvQuad} and \ref{fig:curvLin} for quadratic and linear generating $g$ respectively.

\clearpage

\clearpage
\newpage
\section{Diagnostic Plots for the Jensen Effect: Single Index Model} \label{appendix:sim}

\begin{figure}[tbp]
\centering
\begin{tabular}{cc}
\includegraphics[width=0.35\textwidth,height=0.3\textheight]{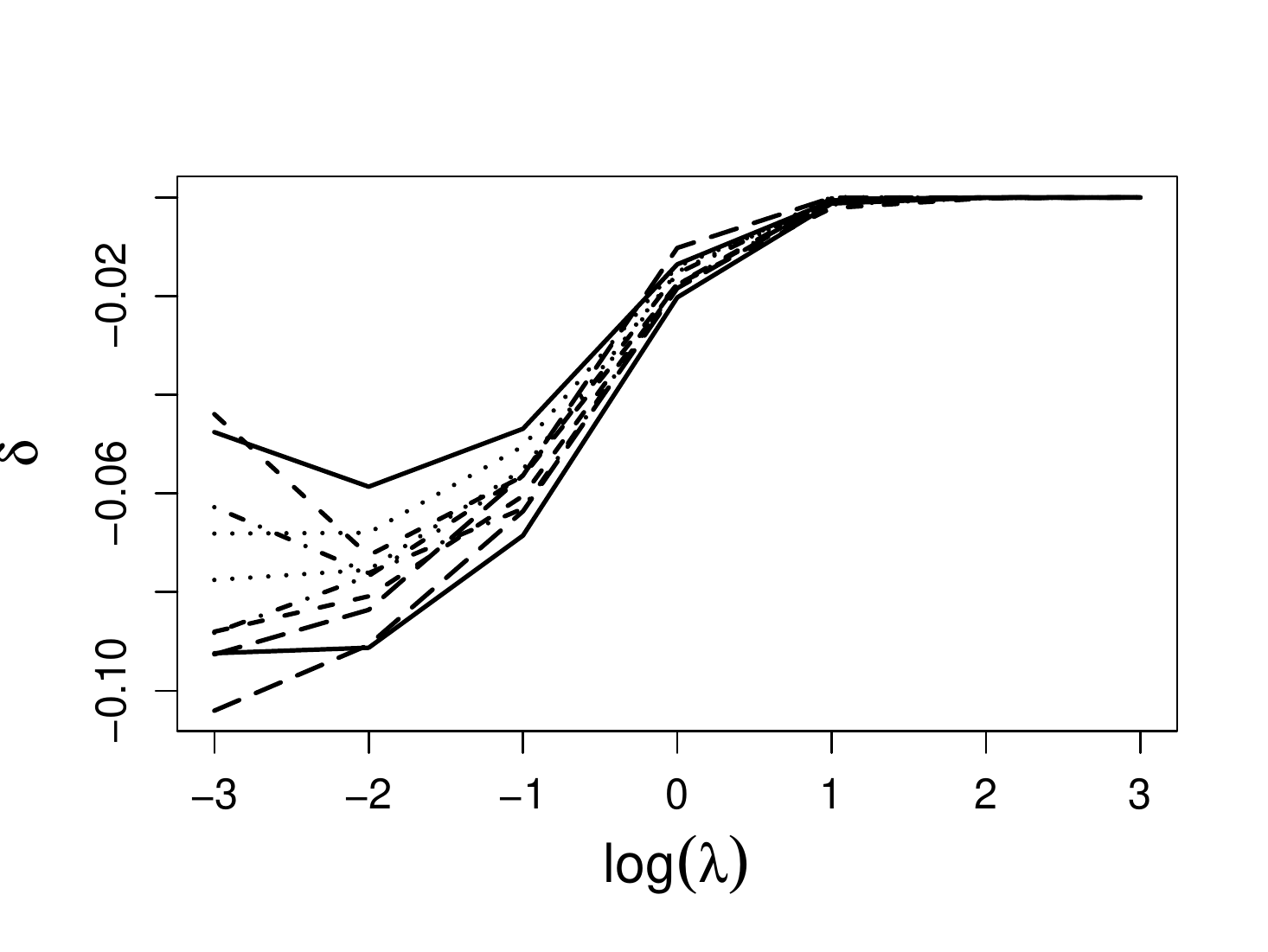} 
\includegraphics[width =0.35\textwidth,height=0.3\textheight]{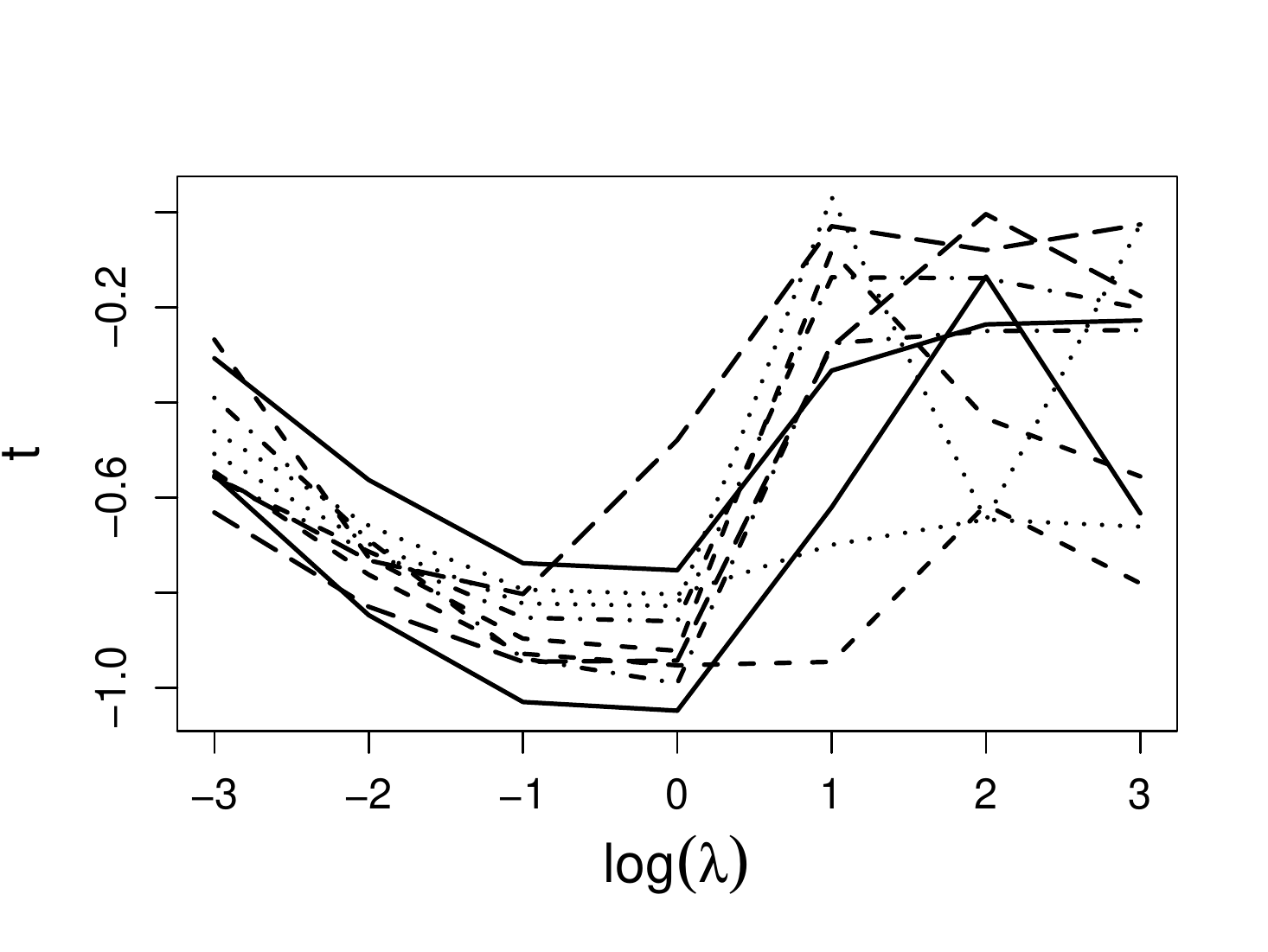}
\end{tabular}
\caption{Left: a sample of $\delta_{\lambda}$ as a function of $\lambda$ in a single index model with link function $g\left(s\right) = -s^2$. Right: the corresponding $t_{\lambda}$ functions.}
\label{fig:8}
 \end{figure}

\begin{figure}[H]
\centering
\begin{tabular}{cc}
\includegraphics[width =0.35\textwidth,height=0.3\textheight]{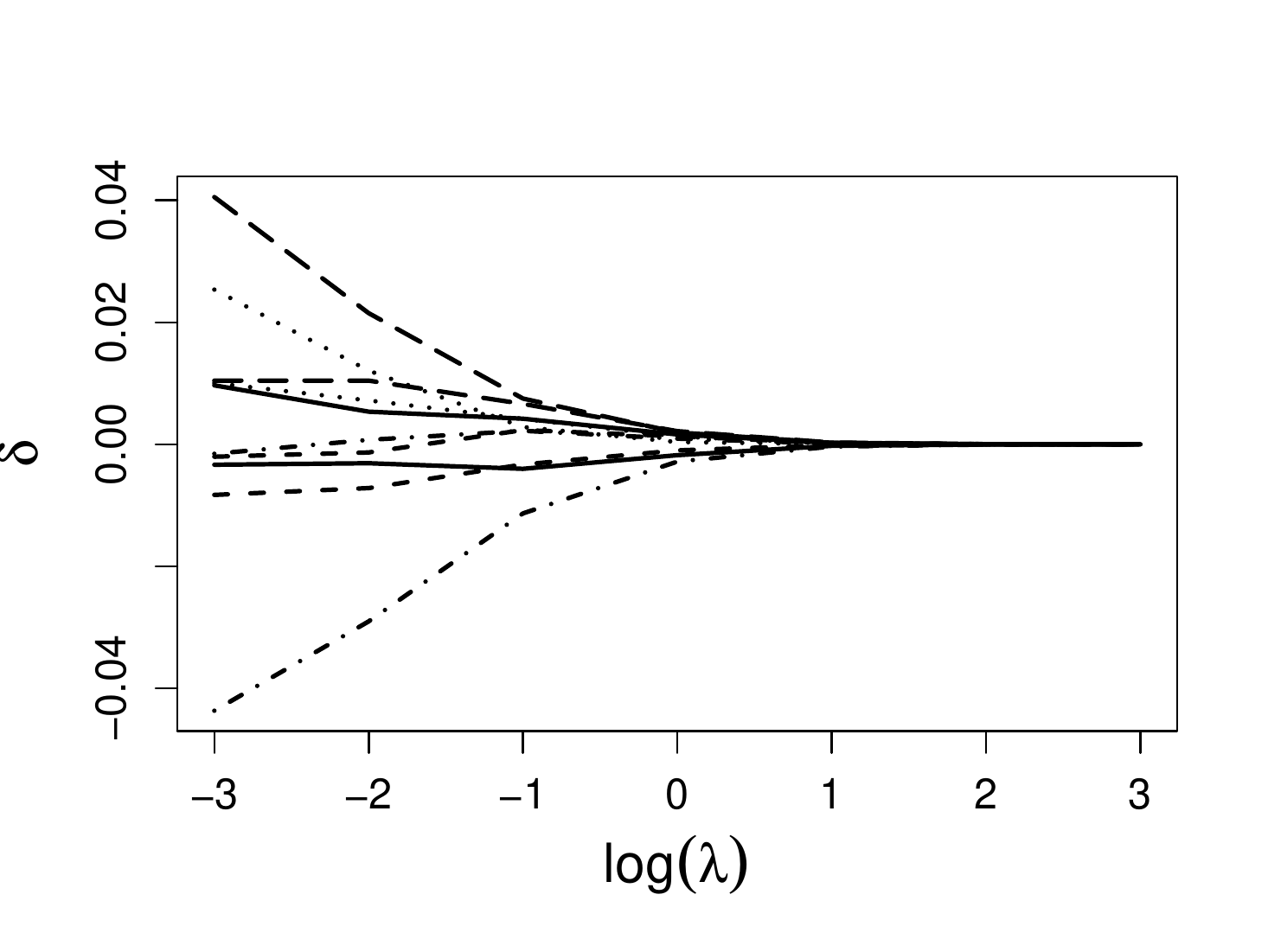} 
\includegraphics[width =0.35\textwidth,height=0.3\textheight]{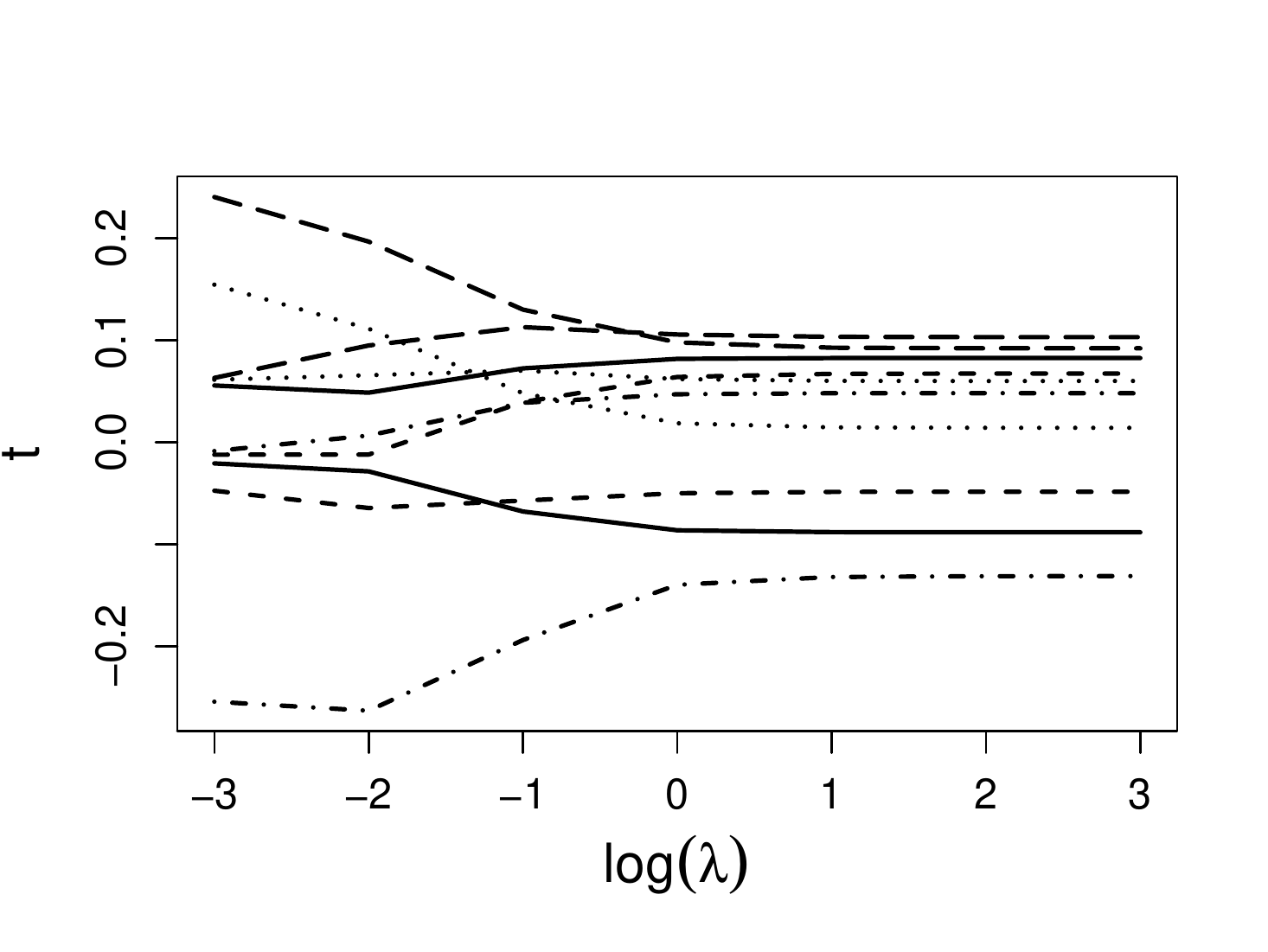}
\end{tabular}
\caption{Left: a sample of $\delta_{\lambda}$ as a function of $\lambda$ in a single index model with link function $g\left(s\right) = s$. Right: the corresponding $t_{\lambda}$ functions.}
\label{fig:9}
 \end{figure}

Figures \ref{fig:8} and \ref{fig:9} give example $\delta$ functions using a single index model and the corresponding $t$ functions for links $g(s) = -s^2$ and $g(s) = s$ respectively.

\section{Diagnostic Plots for the Functional Single Index Model} \label{appendix:fsim}
\begin{figure}[H]
\centering
\begin{tabular}{cc}
\includegraphics[width =0.35\textwidth,height = 0.3\textheight]{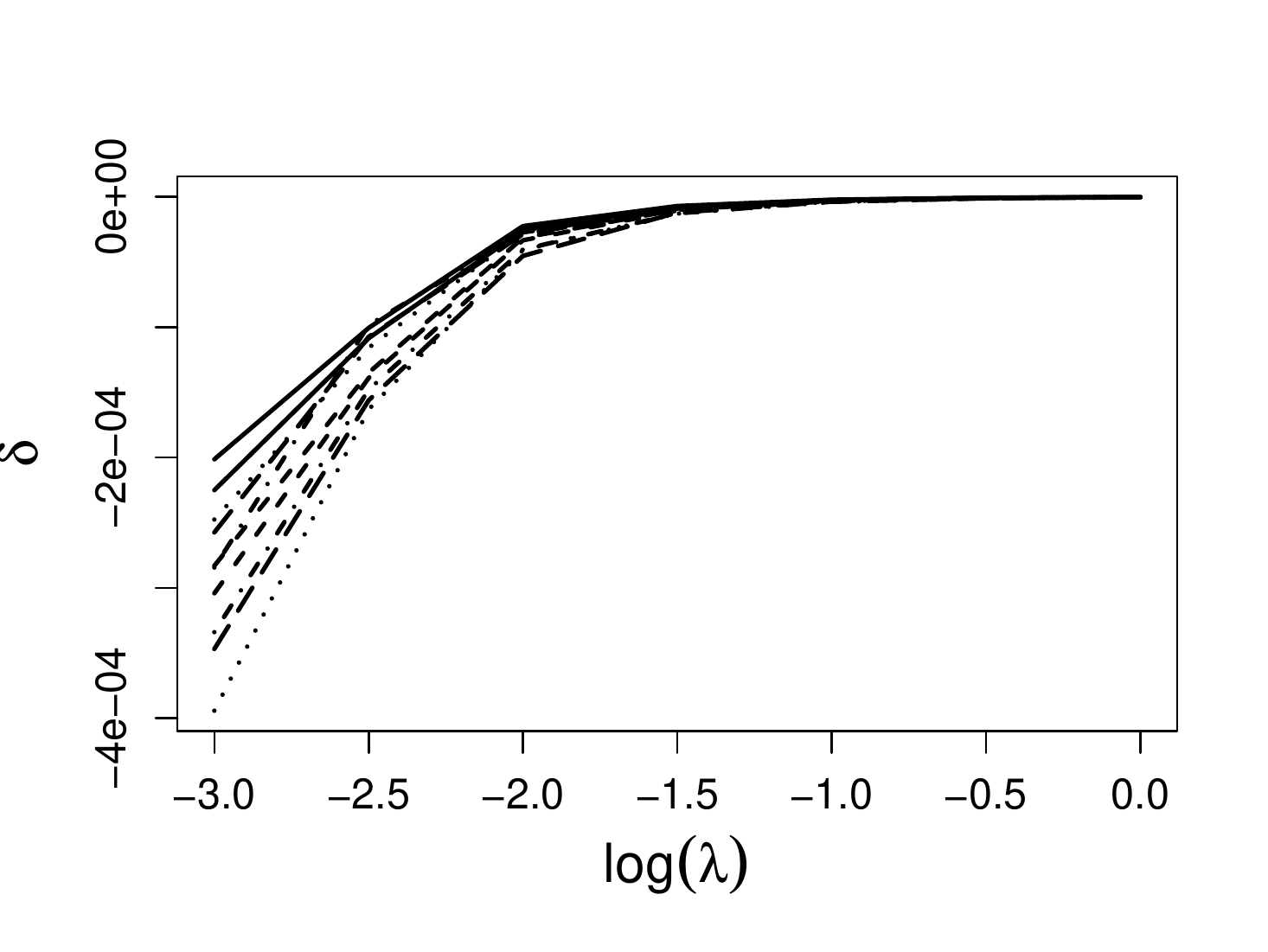} 
\includegraphics[width =0.35\textwidth,height = 0.3\textheight]{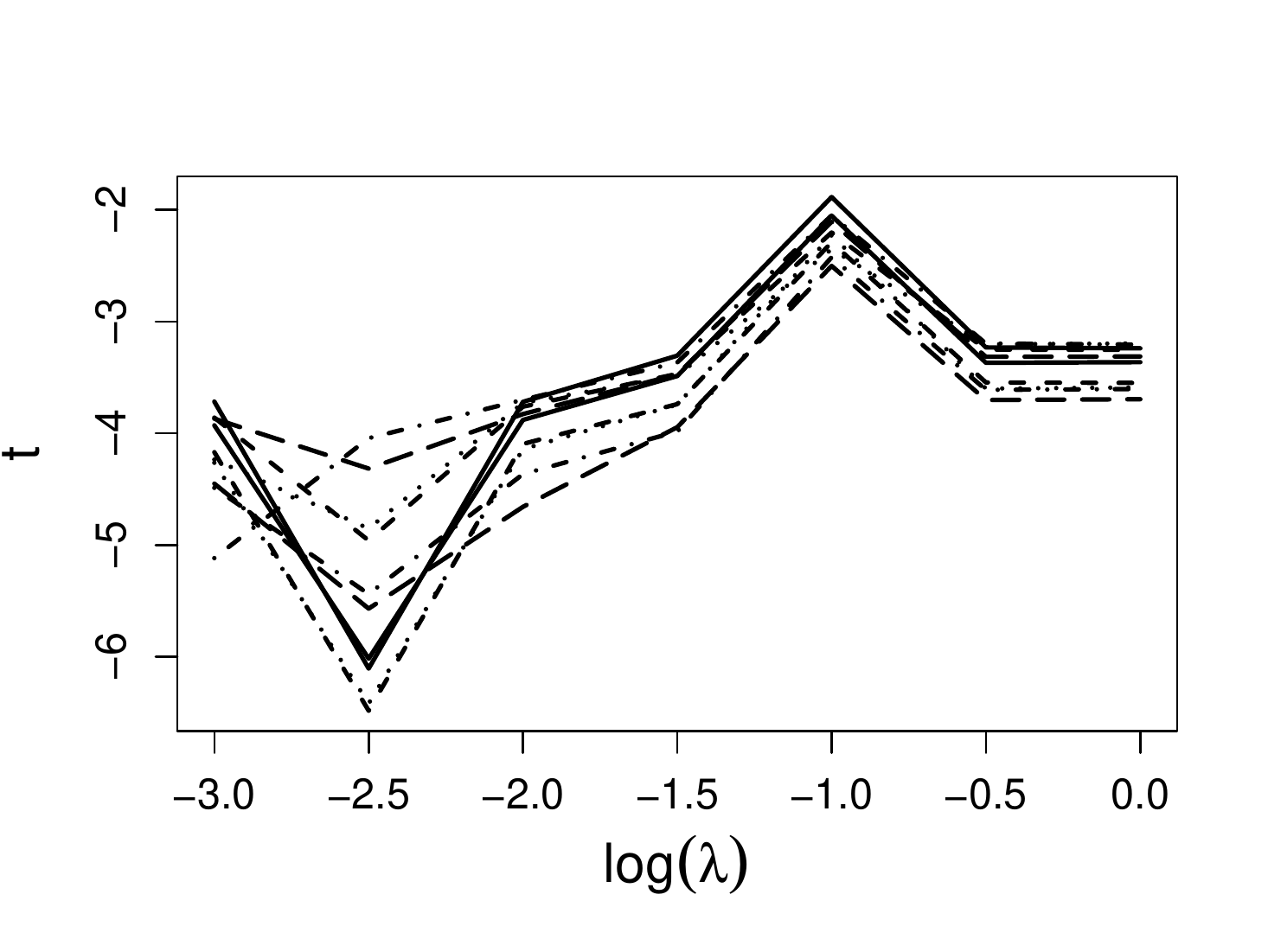}
\end{tabular}
\caption{Left: a sample of $\delta_{\lambda}$ as a function of $\lambda$ in a functional single index model with link function $g\left(s\right) = -s^2$. Right: the corresponding $t_{\lambda}$ functions.}
\label{fig:10}
 \end{figure}

\begin{figure}[H]
\centering
\begin{tabular}{cc}
\includegraphics[width =0.35\textwidth,height = 0.3\textheight]{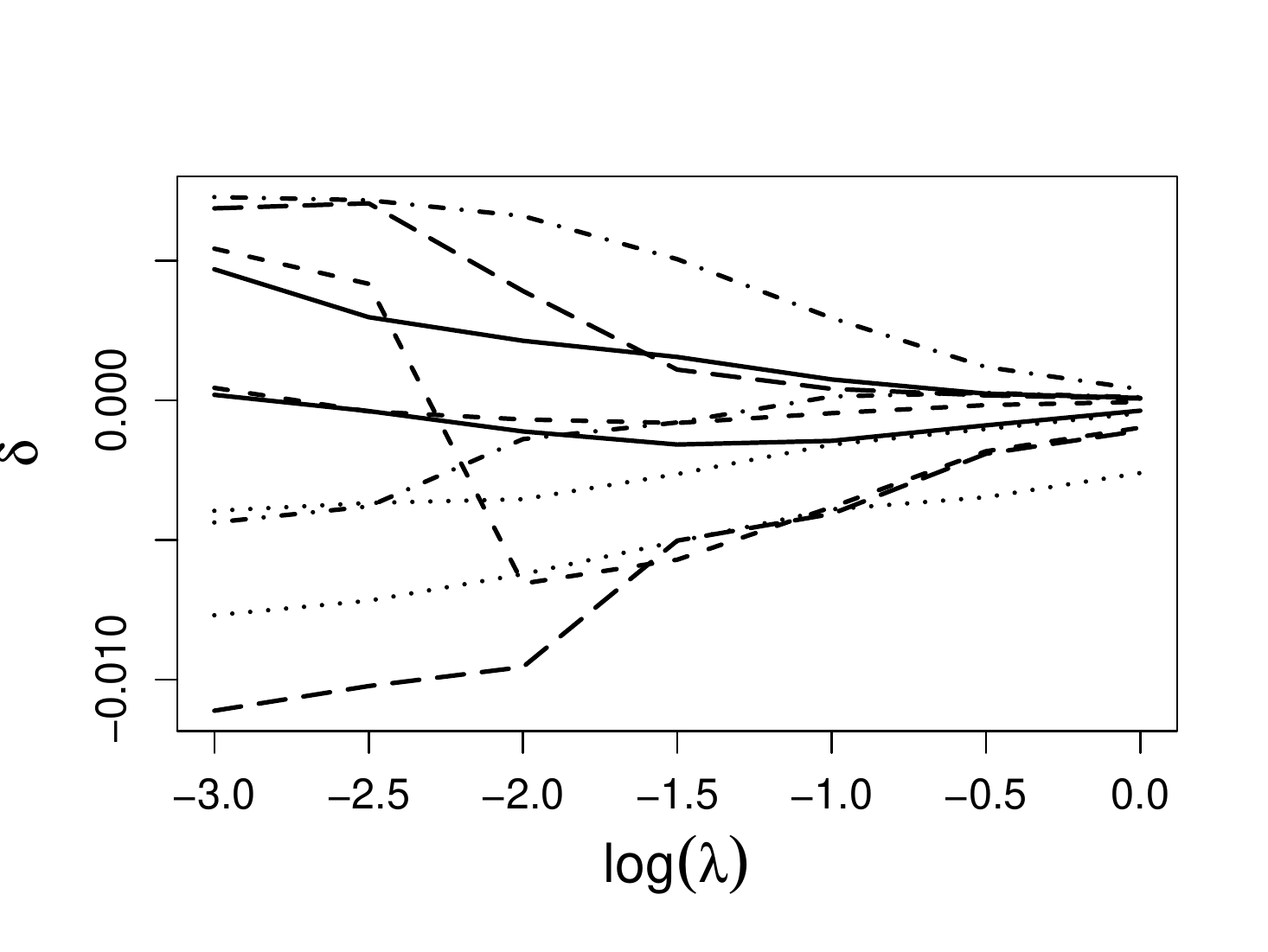} 
\includegraphics[width =0.35\textwidth,height = 0.3\textheight]{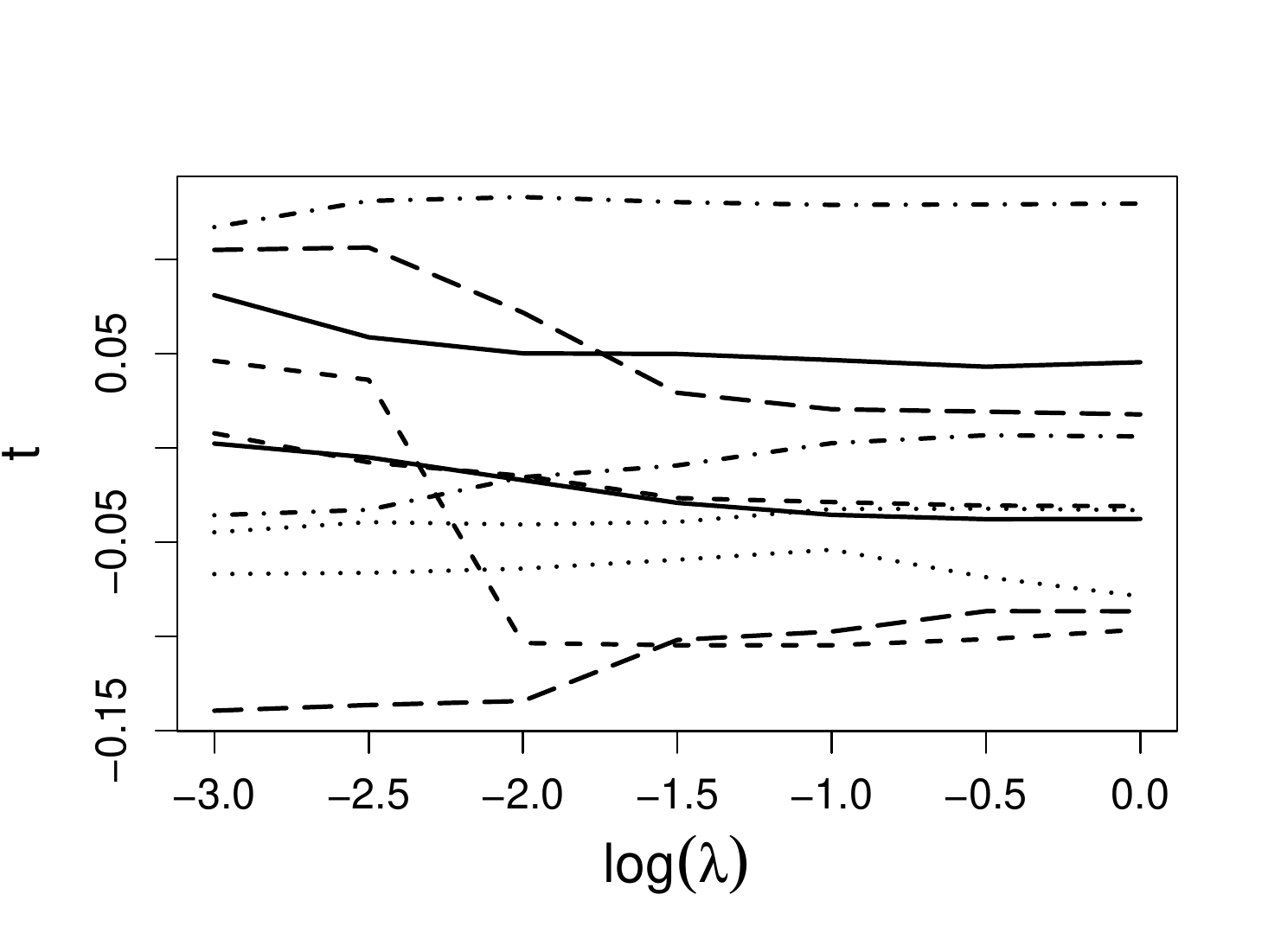}
\end{tabular}
\caption{Left: a sample of $\delta_{\lambda}$ as a function of $\lambda$ in a functional single index model with link function $g\left(s\right) = s$. Right: the corresponding $t_{\lambda}$ functions.}
\label{fig:11}
 \end{figure}

Figures \ref{fig:10} and \ref{fig:11} give example $\delta$ functions using a single index model and the corresponding $t$ functions for links $g(s) = -s^2$ and $g(s) = s$ respectively.

\section{On Estimates of Residual Variance}  \label{app:sigest}

\begin{figure}[H]
    \centering
    \includegraphics[width=0.4\textwidth]{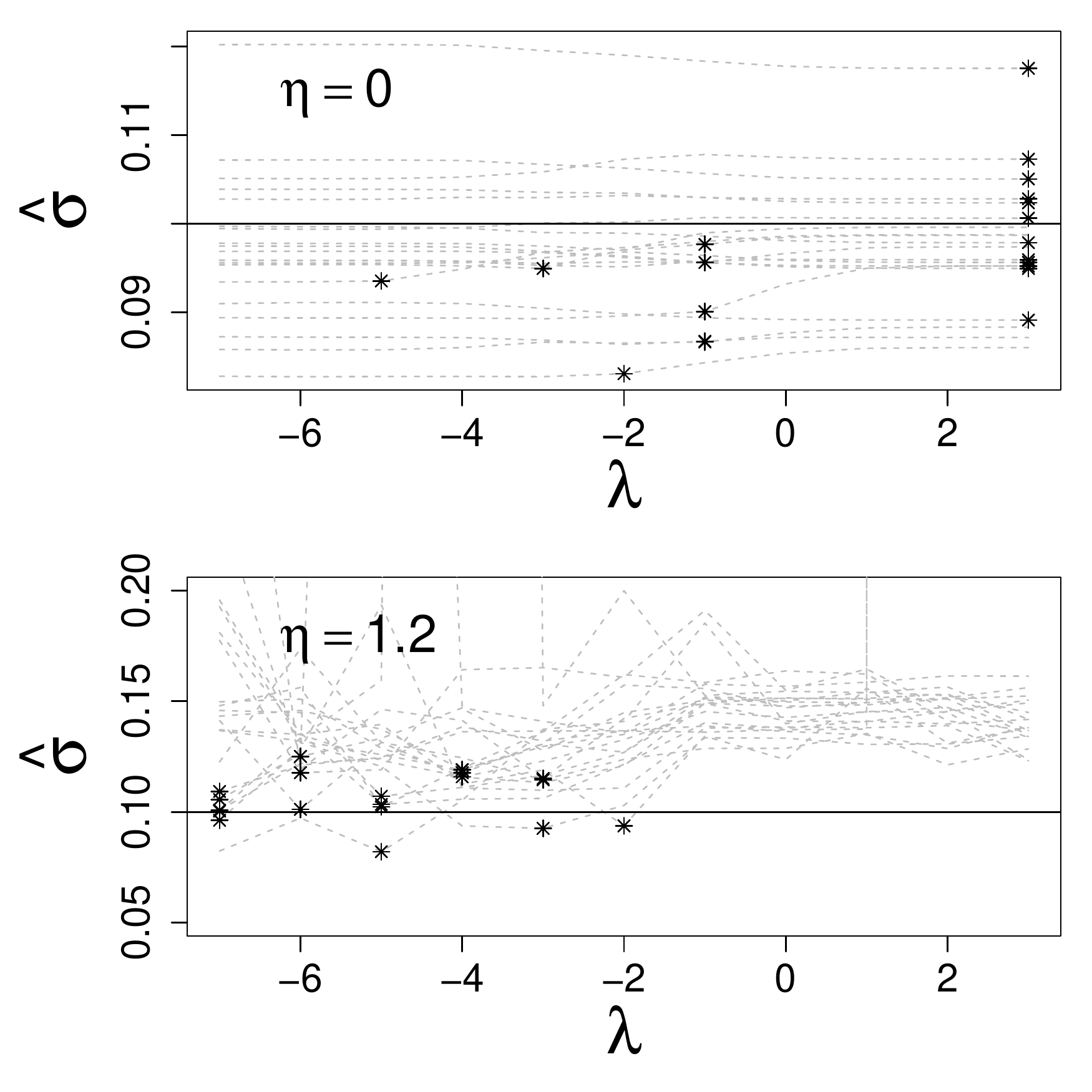}
    \includegraphics[width=0.4\textwidth]{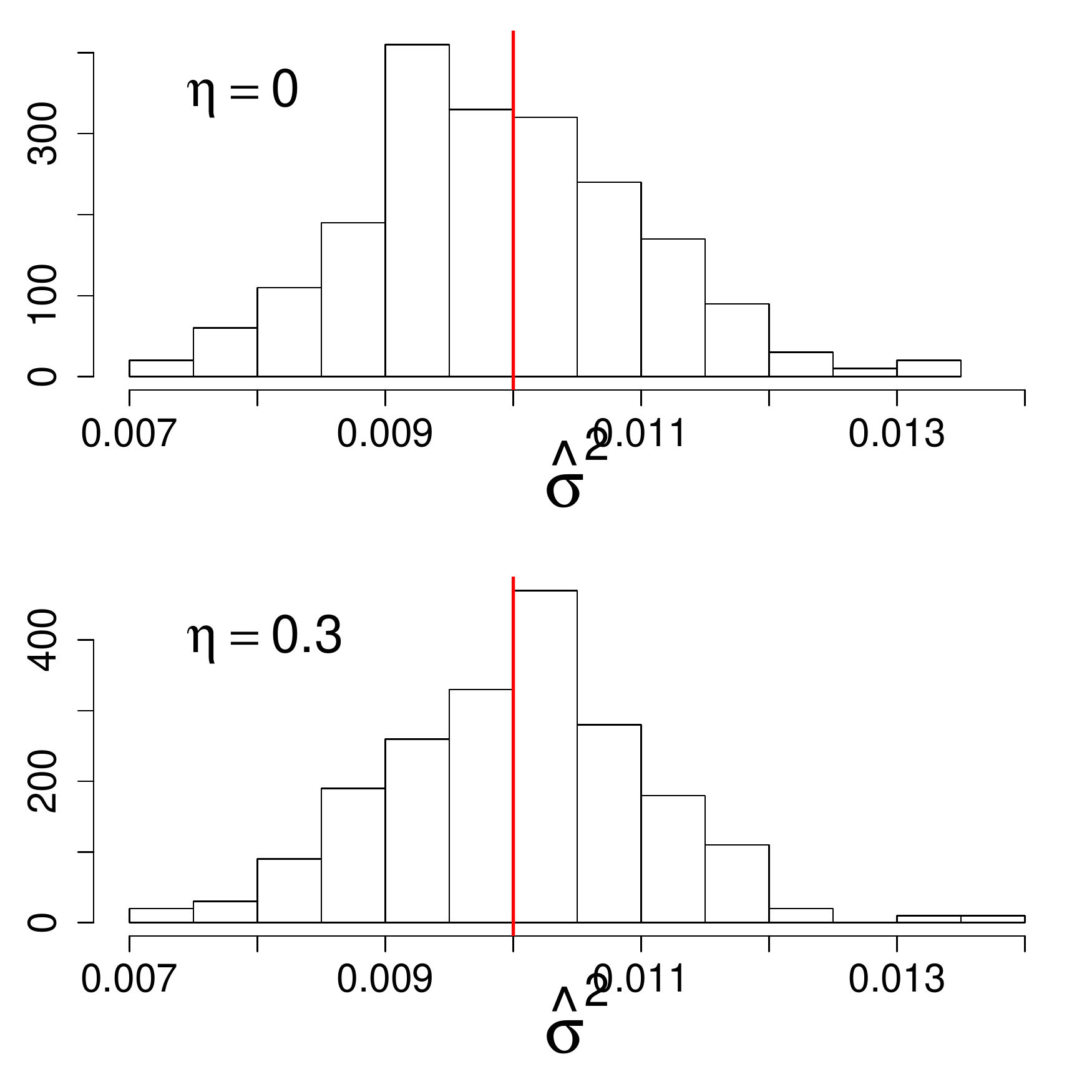}
    \caption{Estimates of residual standard error. Left: for single index models, $\hat{\sigma}_{\lambda}$ as a function of $\lambda$ for the first 10 simulations from a linear relationship (top) and maximally curved (bottom). Right: histograms estimates of $\sigma$ from linear relationships (top) and maximally curved relationships for functional single index models.}
    \label{fig:sigest}
\end{figure}

We use the power simulations in Section \ref{sec:power} to confirm our expectation that we can select $\hat{sigma}$ based on residual squared error at smoothing values selected by GCV. In Figure \ref{fig:sigest} we examine both single index models and functional single index models. For single index models we plot $\hat{\sigma}_{\lambda}$ as a function of $\lambda$ the first 10 simulations and indicate the value of GCV with an asterisk for each. When the data is generated from a linear relationship, these curves are nearly flat, when we use the maximum value of $\eta$ there are noticeable changes in $\hat{\sigma}_{\lambda}$ but each curve has an extended flat area around the correct value which is reliably estimated by GCV.  We note that GCV appropriately chooses large smoothing parameters when the relationship is linear and there is little bias, but selects lower values for simulations with higher curvature. It is less easy to make these plots for functional single index models, and we instead provide a histogram of estimates at the GCV value for both linear and maximally curved relationships where we see that in both cases our estimates focus on approximately the correct values.

\clearpage
\newpage
\section{Plots for the copepod data} \label{appendix:real}

\begin{figure}[H]
\centering
\includegraphics[width=0.85\textwidth]{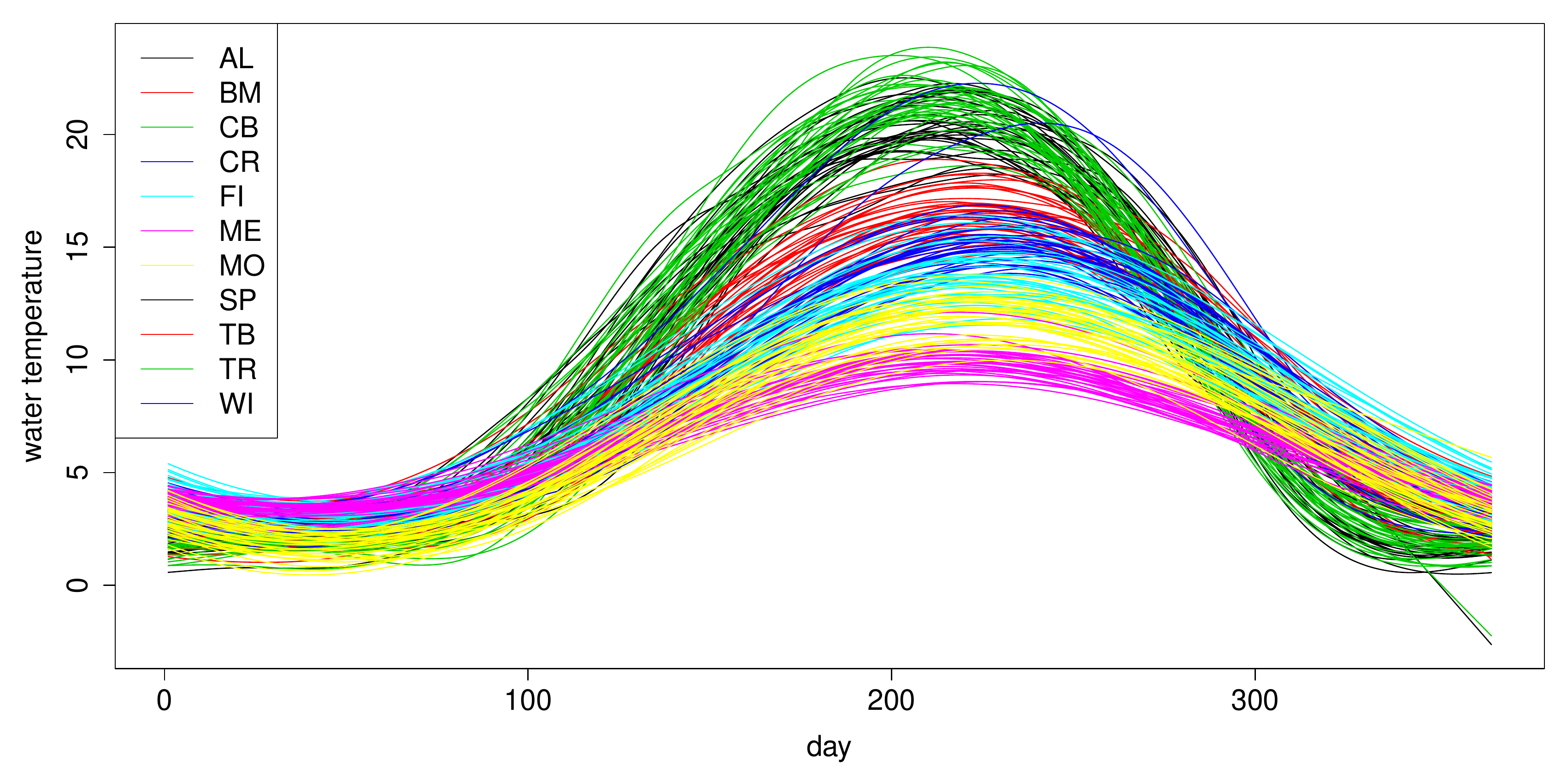}
\caption{Water temperature profiles over 7 experimental lakes and 35 years. Copepod populations were modeled as a function of the previous 60 days water temperature. } \label{fig:wtemp}
\end{figure}

In order to obtain a functional covariate, we smoothed water temperature readings in each lake employing B-splines with 21 knots per year and a second derivative penalty with penalty parameter selected by GCV. These were then evaluated on each of the 60 days prior to a population observation and re-projected onto 12 order 6 Bsplines.  Yearly temperature curves for each lake are given in Figure \ref{fig:wtemp}.

\begin{figure}[H]
\centering
\includegraphics[height=0.45\textwidth,angle=270]{g_20302_diff-eps-converted-to.pdf} 
\includegraphics[height=0.45\textwidth,angle=270]{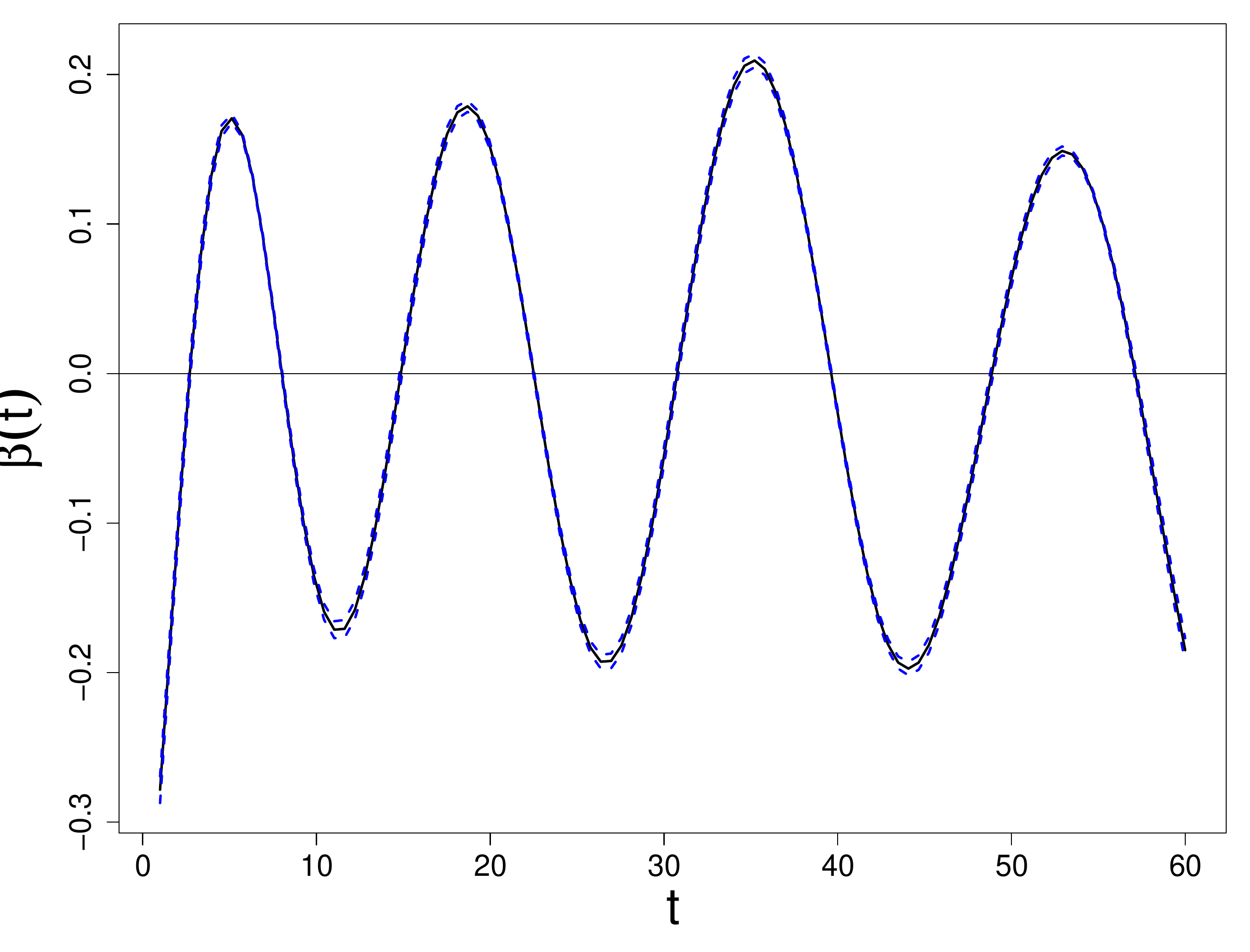}
\includegraphics[height=0.45\textwidth,angle=270]{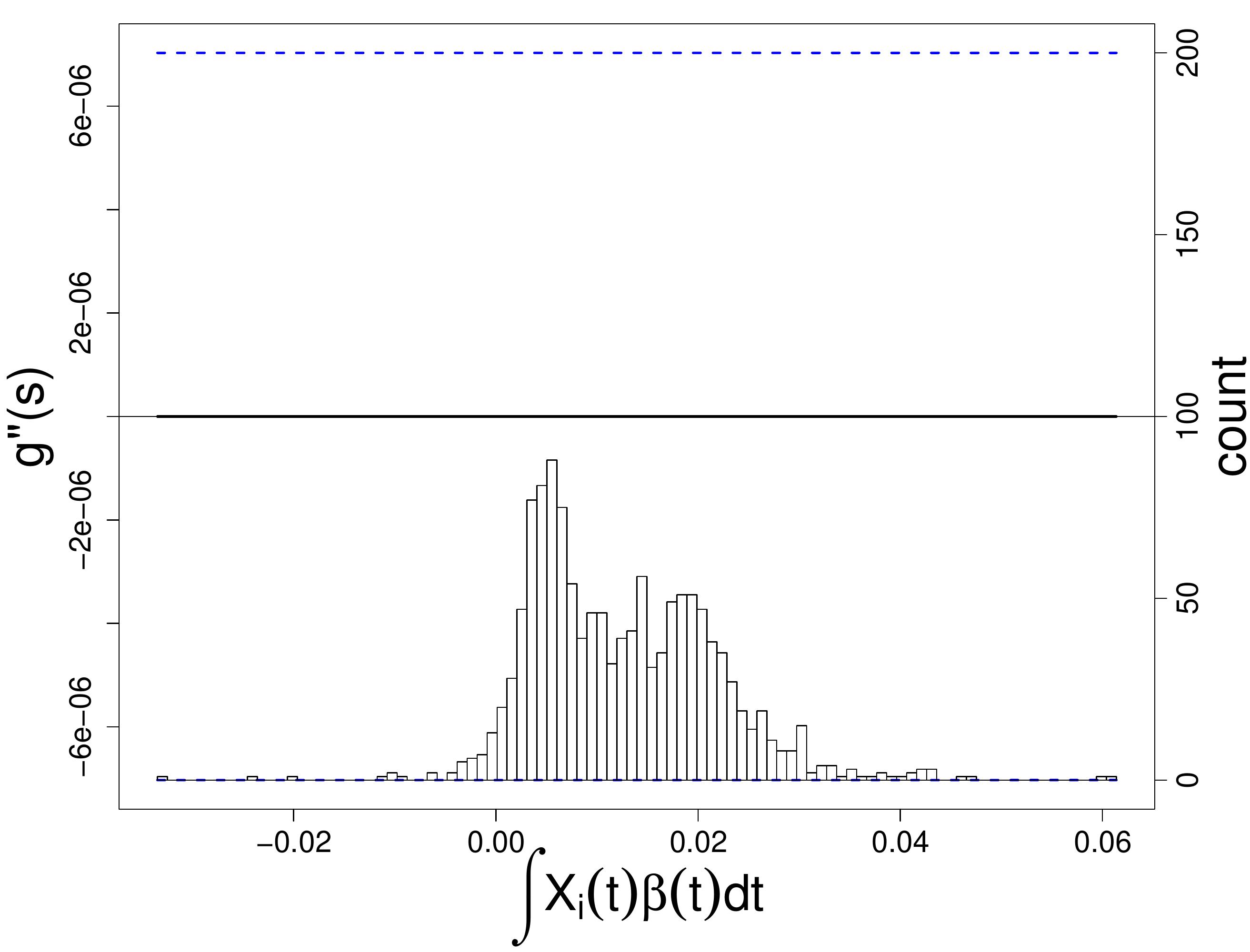}
\includegraphics[height=0.45\textwidth,angle=270]{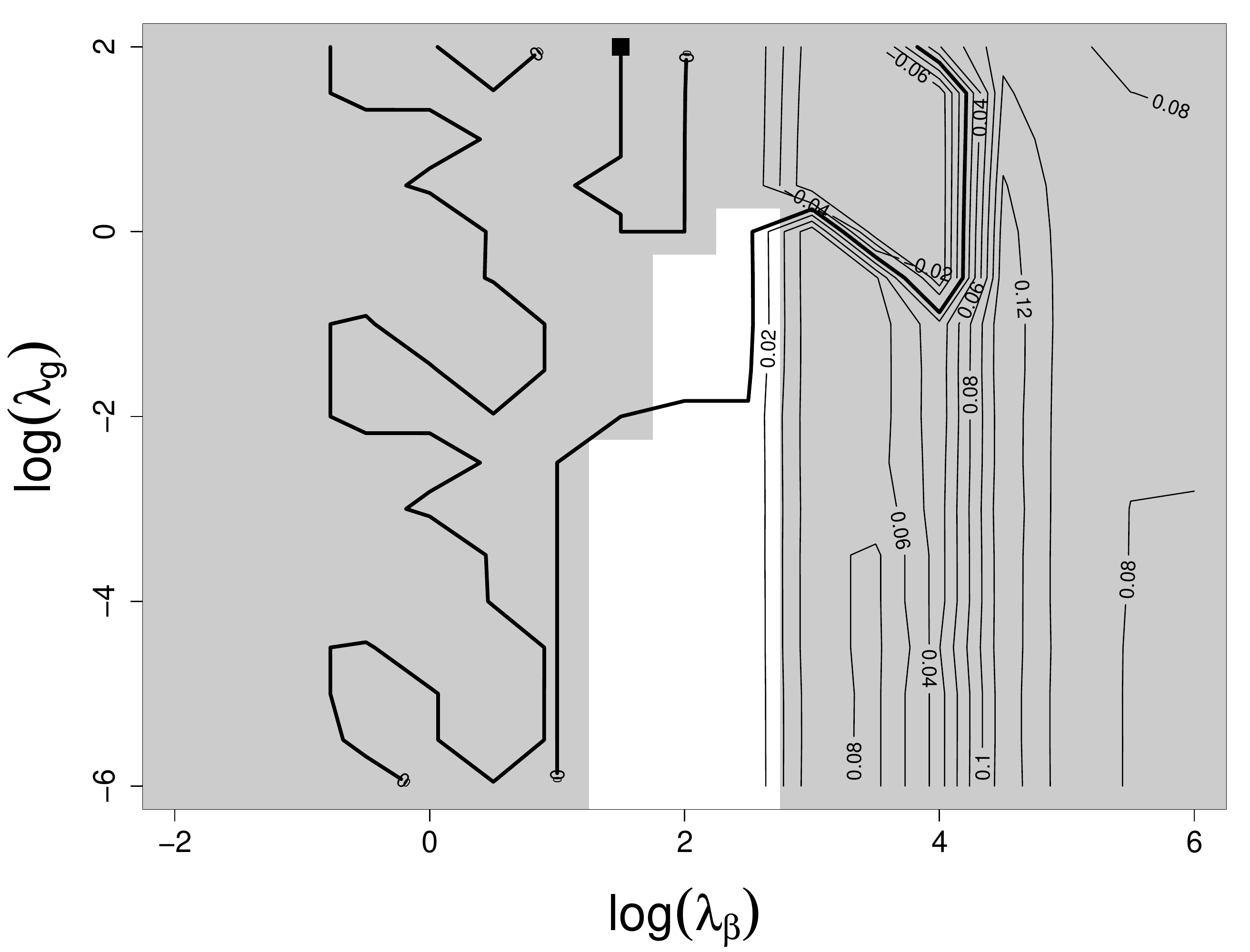}
\vspace{-2mm}
\caption{Diagnostic plots for \emph{Diacyclops Thomasi}, $n=1301$. The region of significance includes negative values (top) and positive values (bottom) with a range of [-0.079, 0.1362]. See dicussion in Section \ref{sec:real} for this species. }  \label{fig:20302}
\label{20302}
 \end{figure}

\begin{figure}[H]
\centering
\includegraphics[height =0.45\textwidth,angle=270]{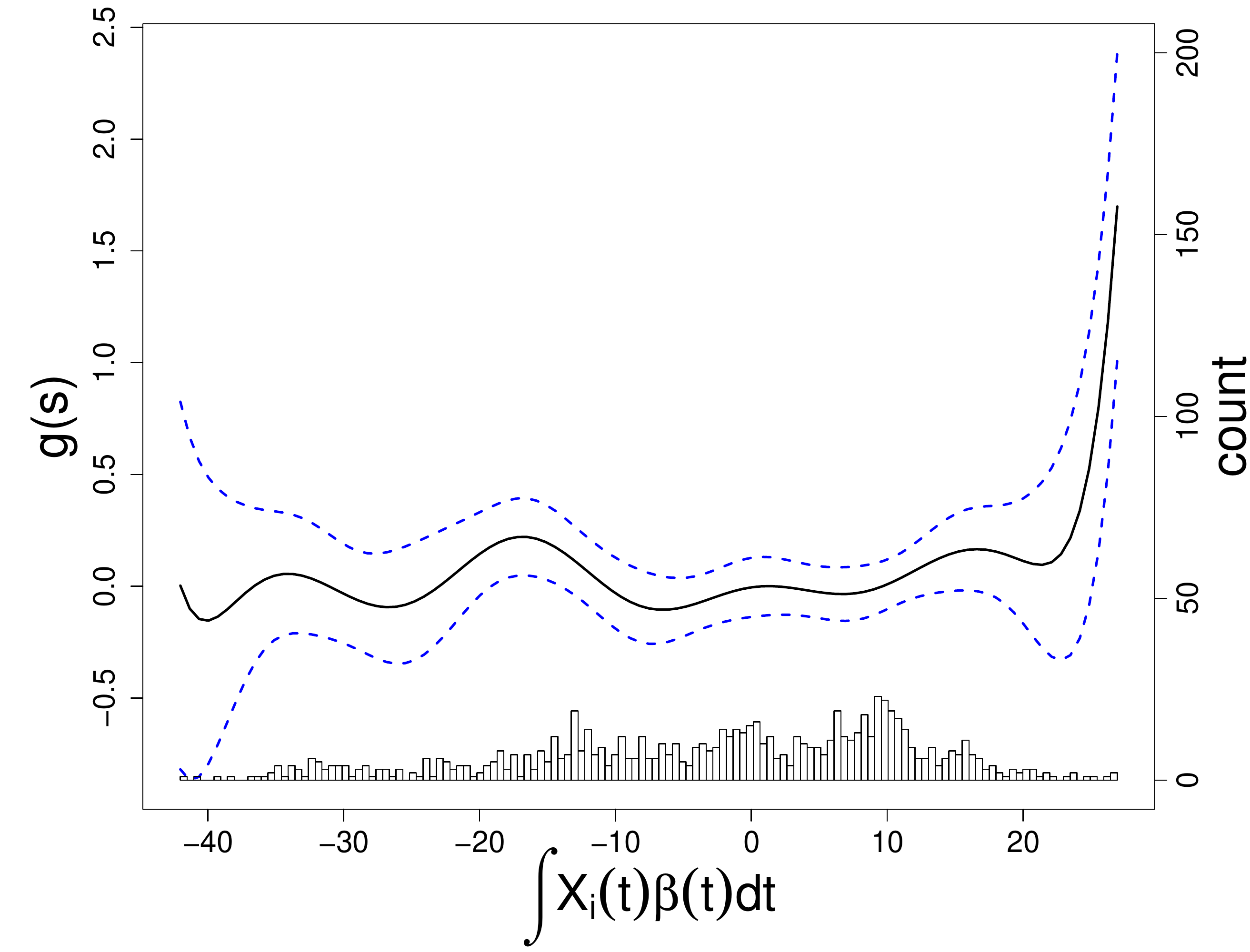} 
\includegraphics[height =0.45\textwidth,angle=270]{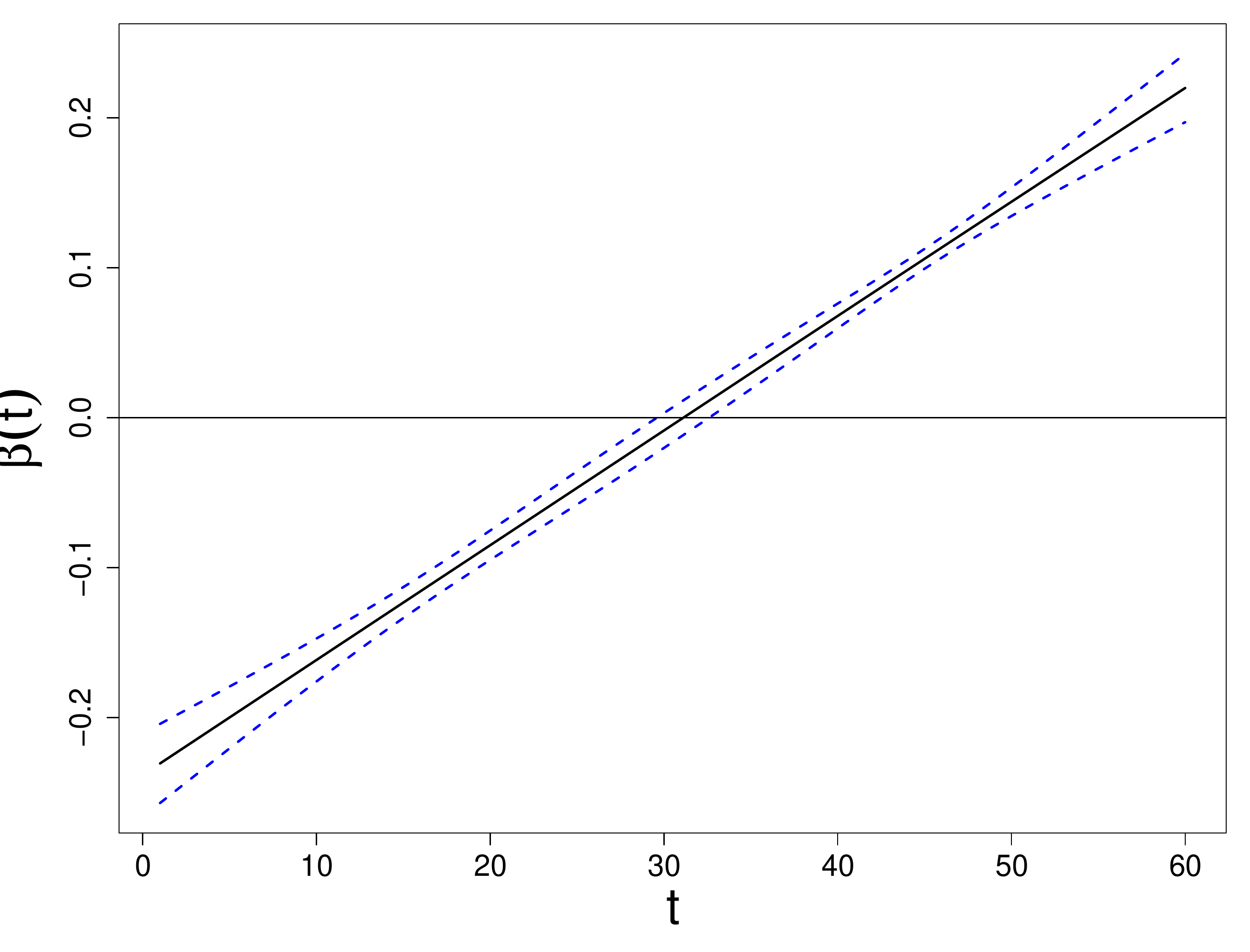} 
\includegraphics[height =0.45\textwidth,angle=270]{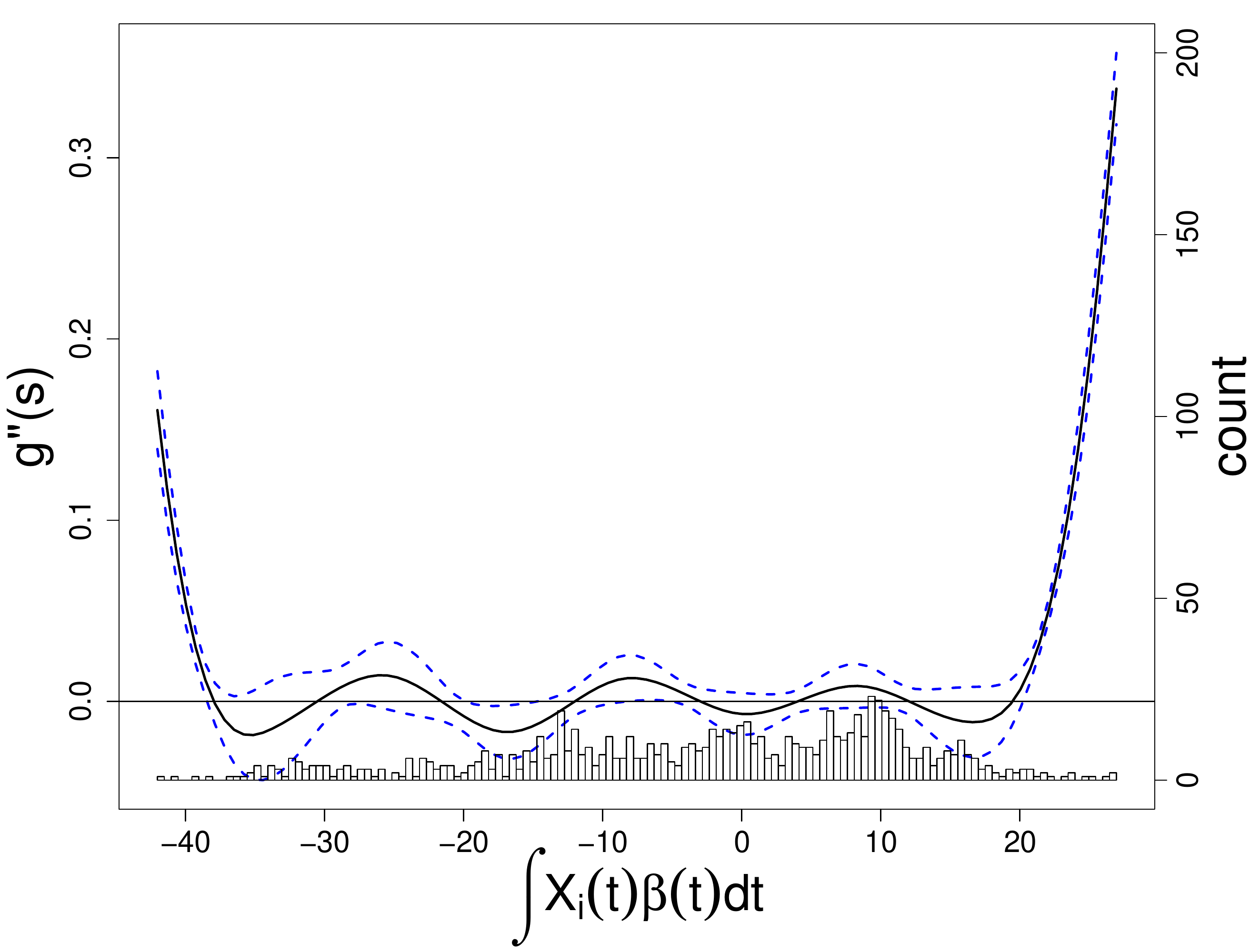}
\includegraphics[height =0.45\textwidth,angle=270]{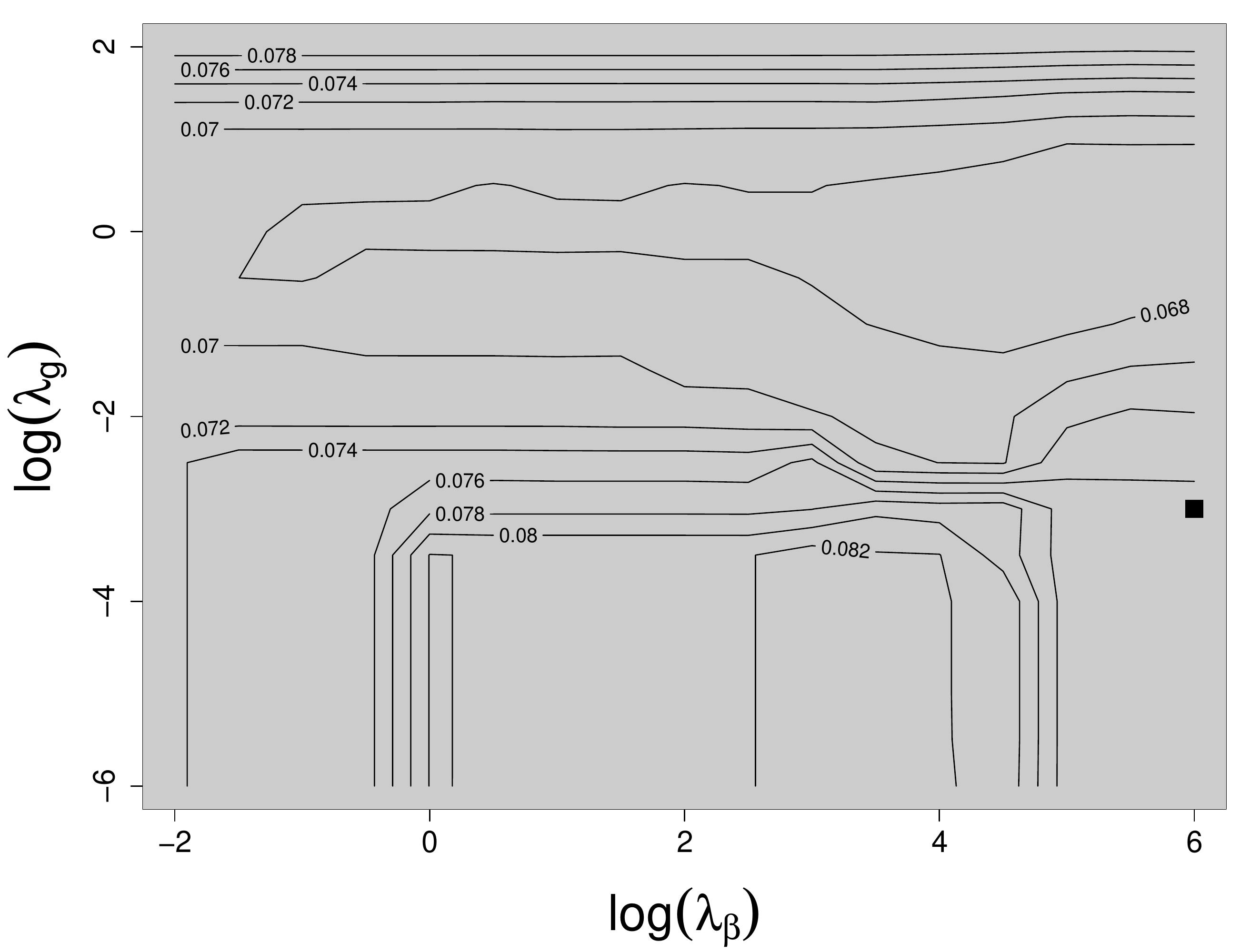}
\caption{Plots for \emph{Gastropus Stylifer}, $n = 893$, $\delta$ is estimated to be in the range [0.0663, 0.0830] but is nowhere significant. }
\label{61502}
\end{figure}

\begin{figure}[H]
\centering
\includegraphics[height =0.45\textwidth,angle=270]{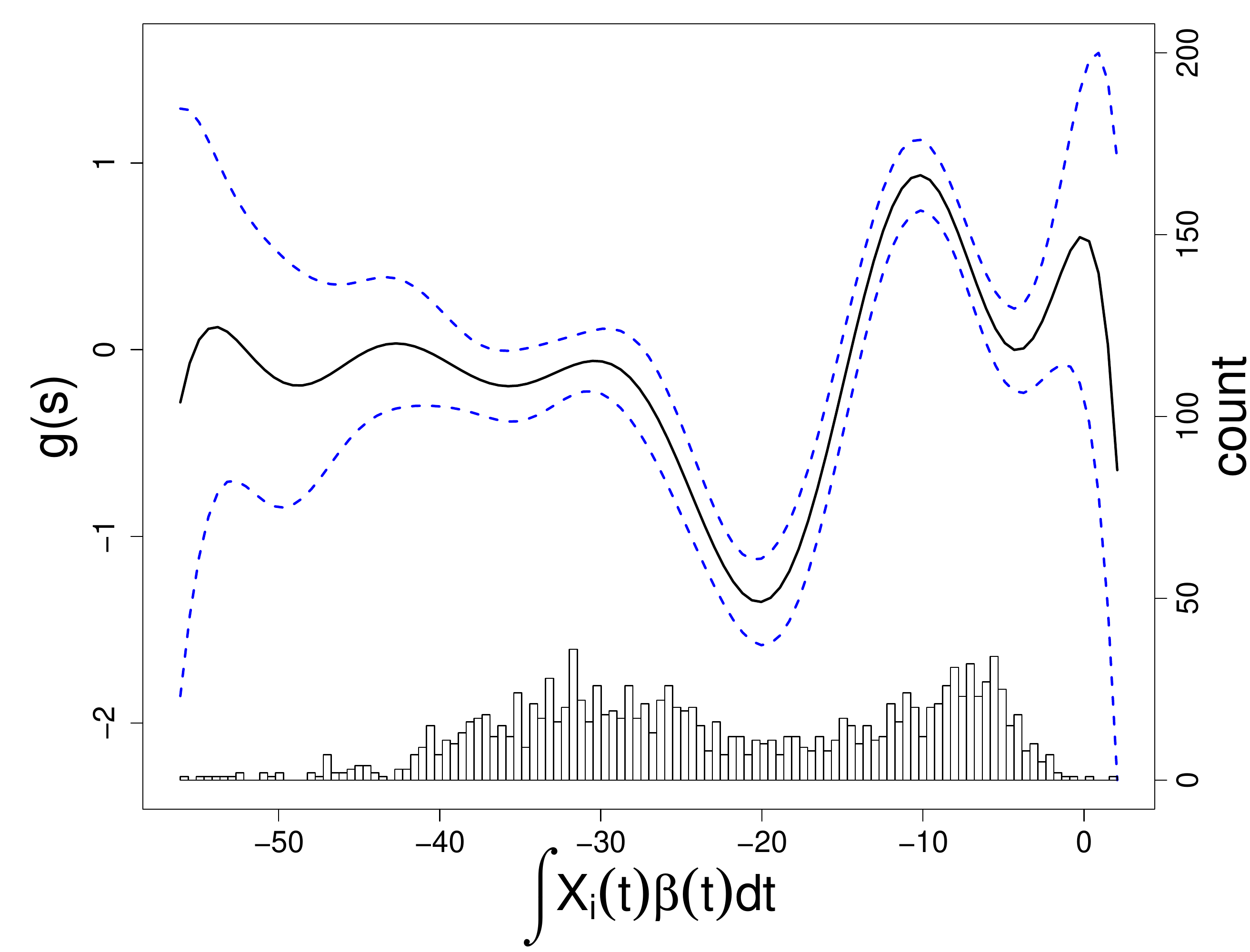} 
\includegraphics[height =0.45\textwidth,angle=270]{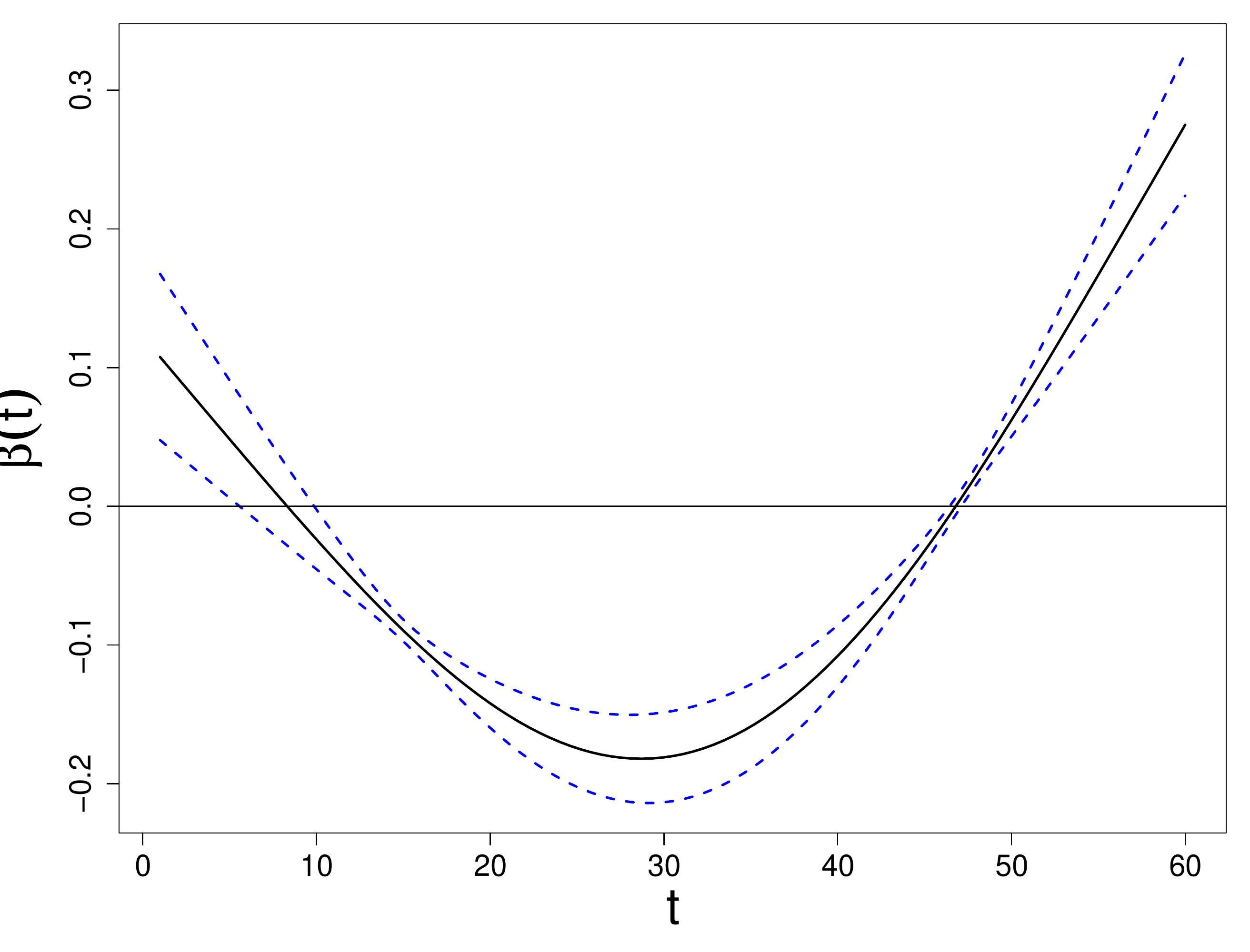} 
\includegraphics[height =0.45\textwidth,angle=270]{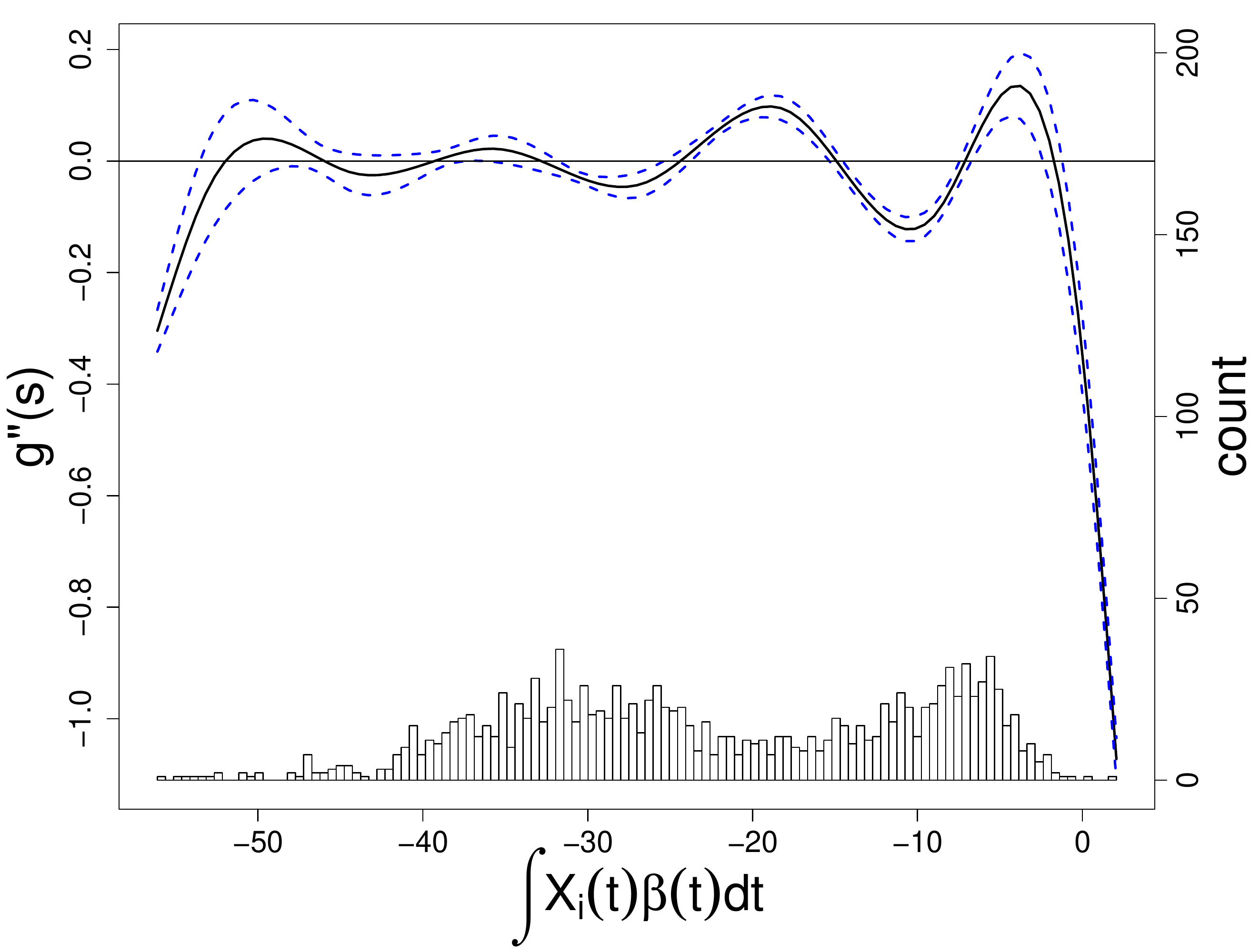}
\includegraphics[height =0.45\textwidth,angle=270]{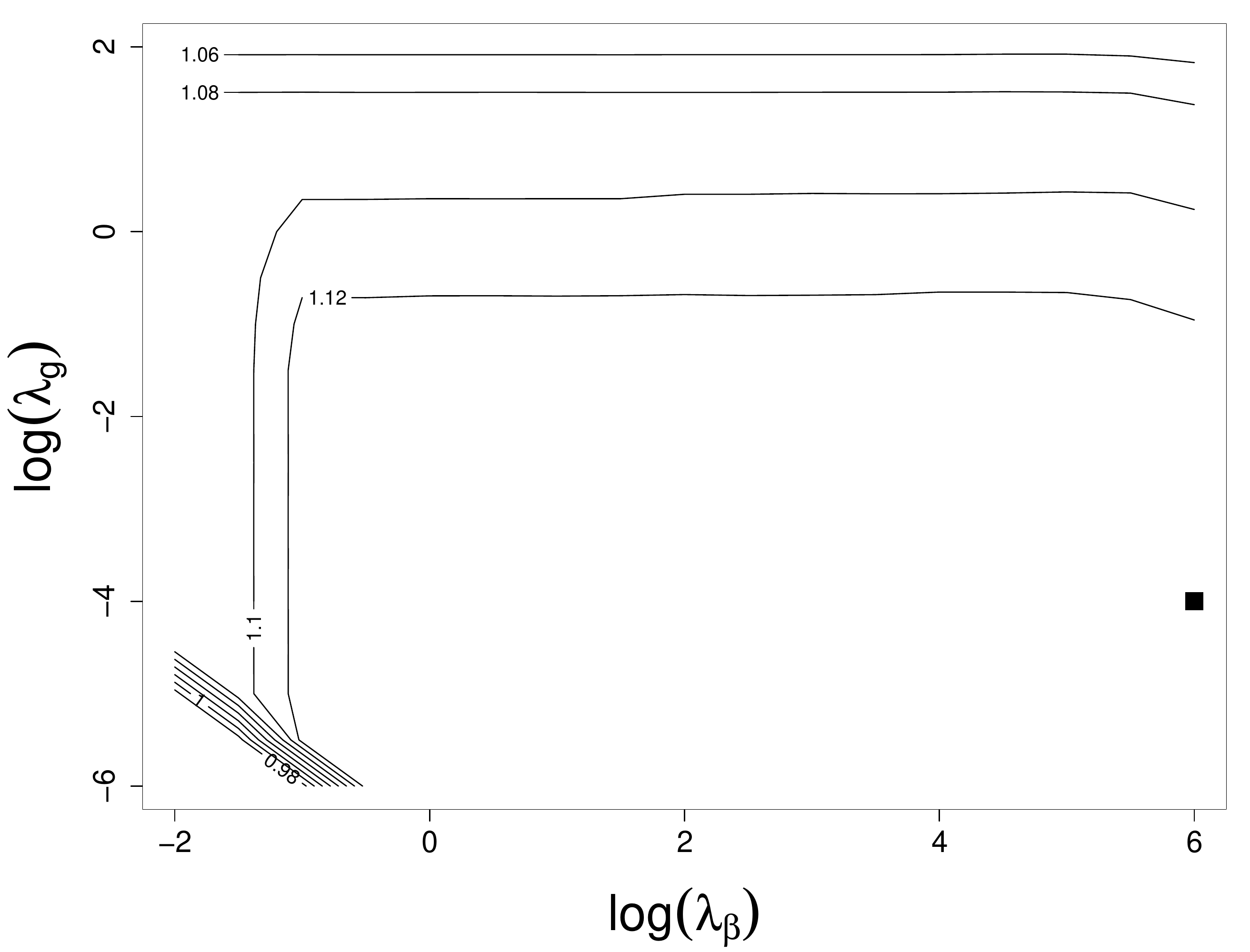}
\caption{Plots for \emph{Kellicottia Longispina}, $n = 1347$. $\delta$ is positive and significant everywhere with values in [0.970, 1.132], but note that the largest values occur at the smallest values of the smoothing parameters and may be over-estimates. }
\label{61702}
 \end{figure}

\begin{figure}[H]
\centering
\includegraphics[height =0.45\textwidth,angle=270,]{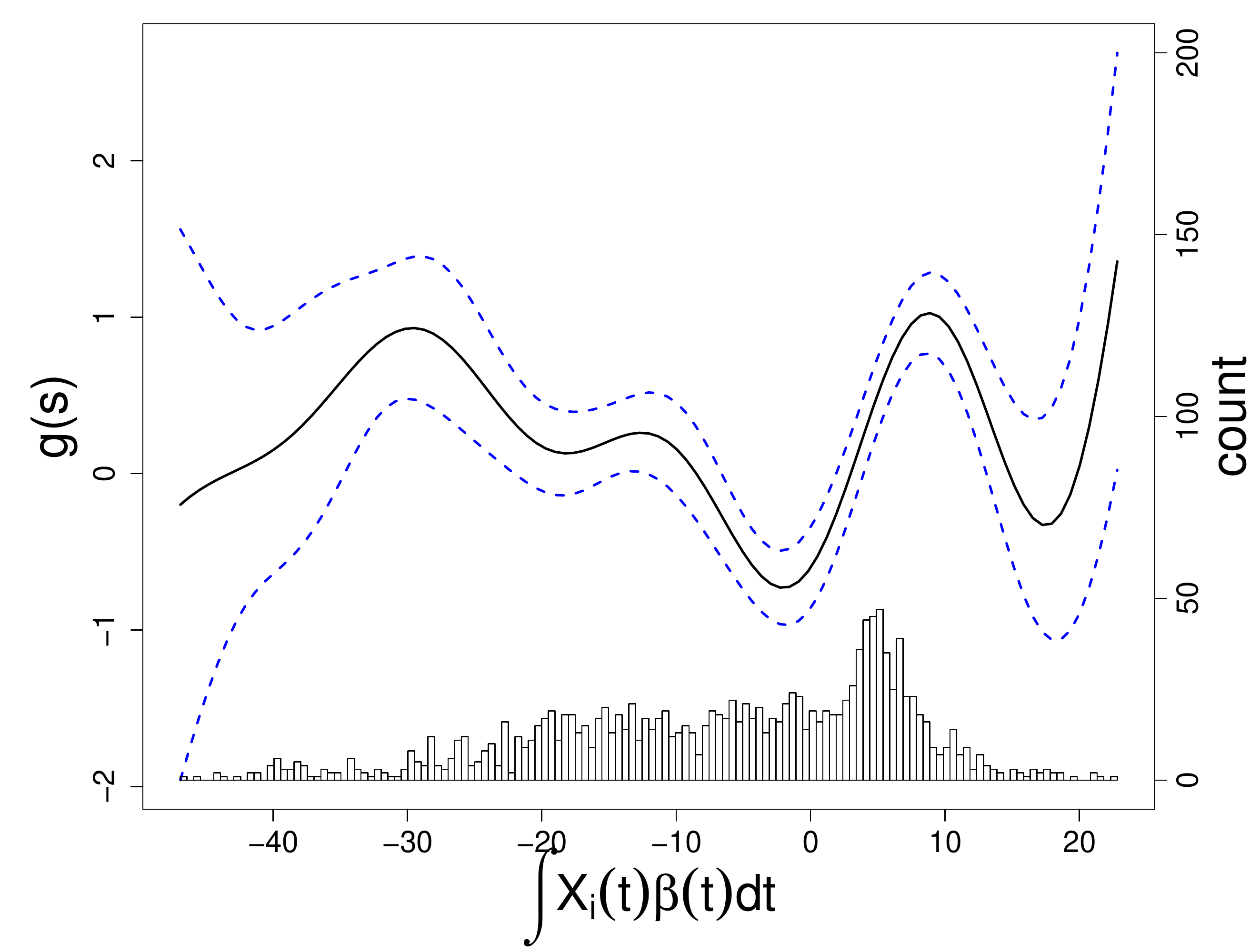} 
\includegraphics[height =0.45\textwidth,angle=270,]{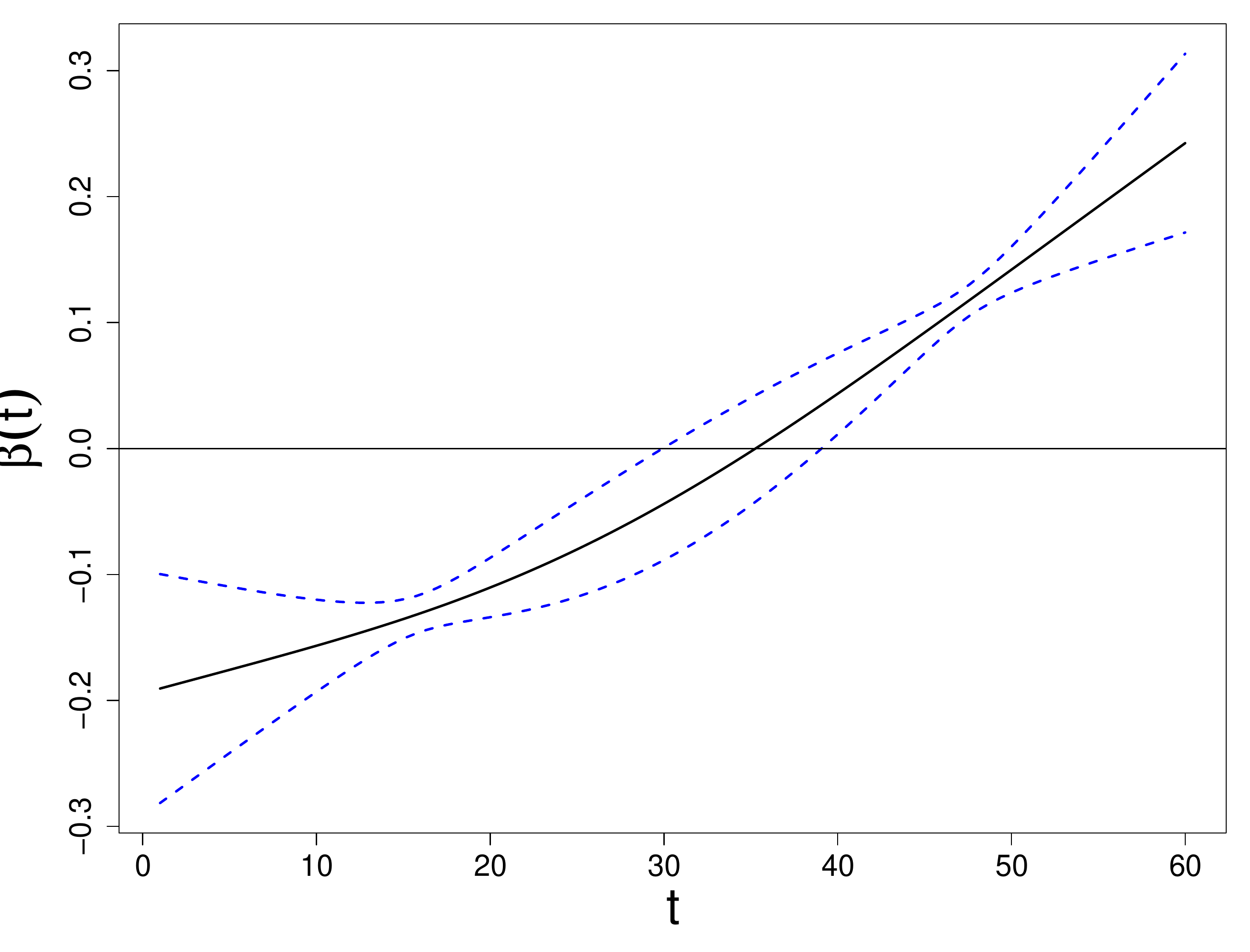} 
\includegraphics[height =0.45\textwidth,angle=270]{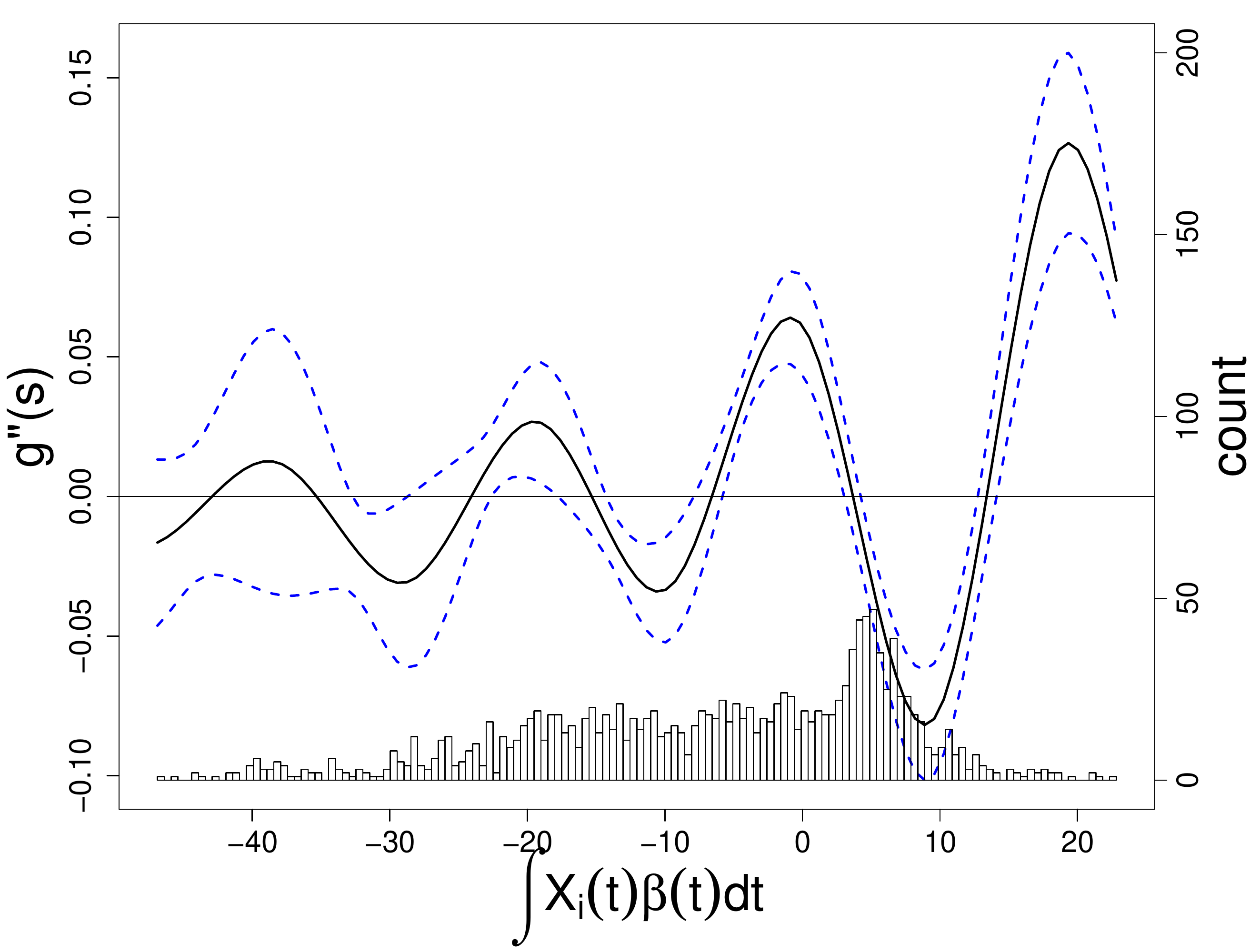}
\includegraphics[height =0.45\textwidth,angle=270]{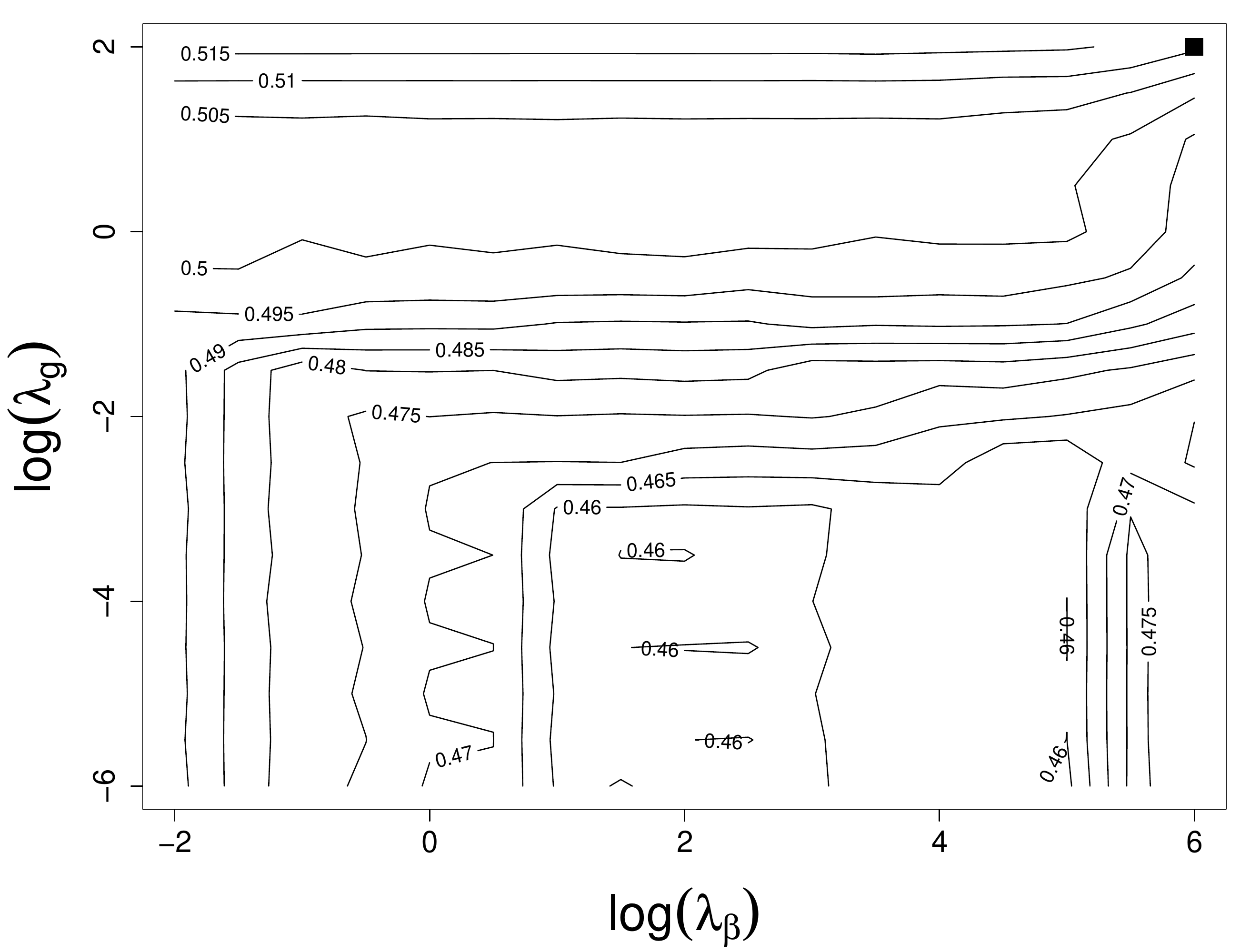}
\caption{Plot for \emph{Keratella Cochlearis}, $n = 1467$. $\delta$ significant everywhere and in the range [0.458, 0.516].  }
\label{61801}
 \end{figure}

\begin{figure}[H]
\centering
\includegraphics[height=0.45\textwidth,angle=270]{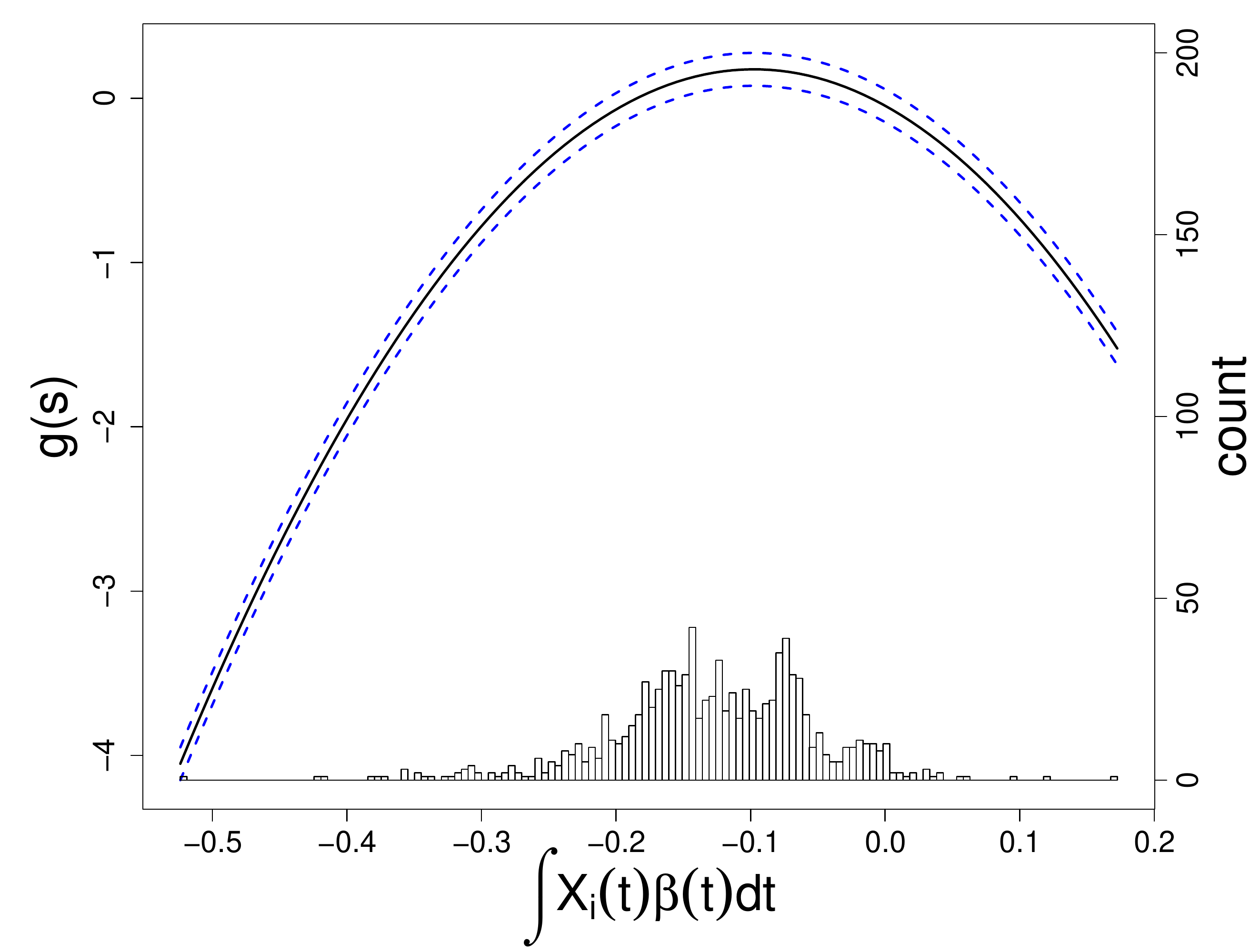} 
\includegraphics[height=0.45\textwidth,angle=270]{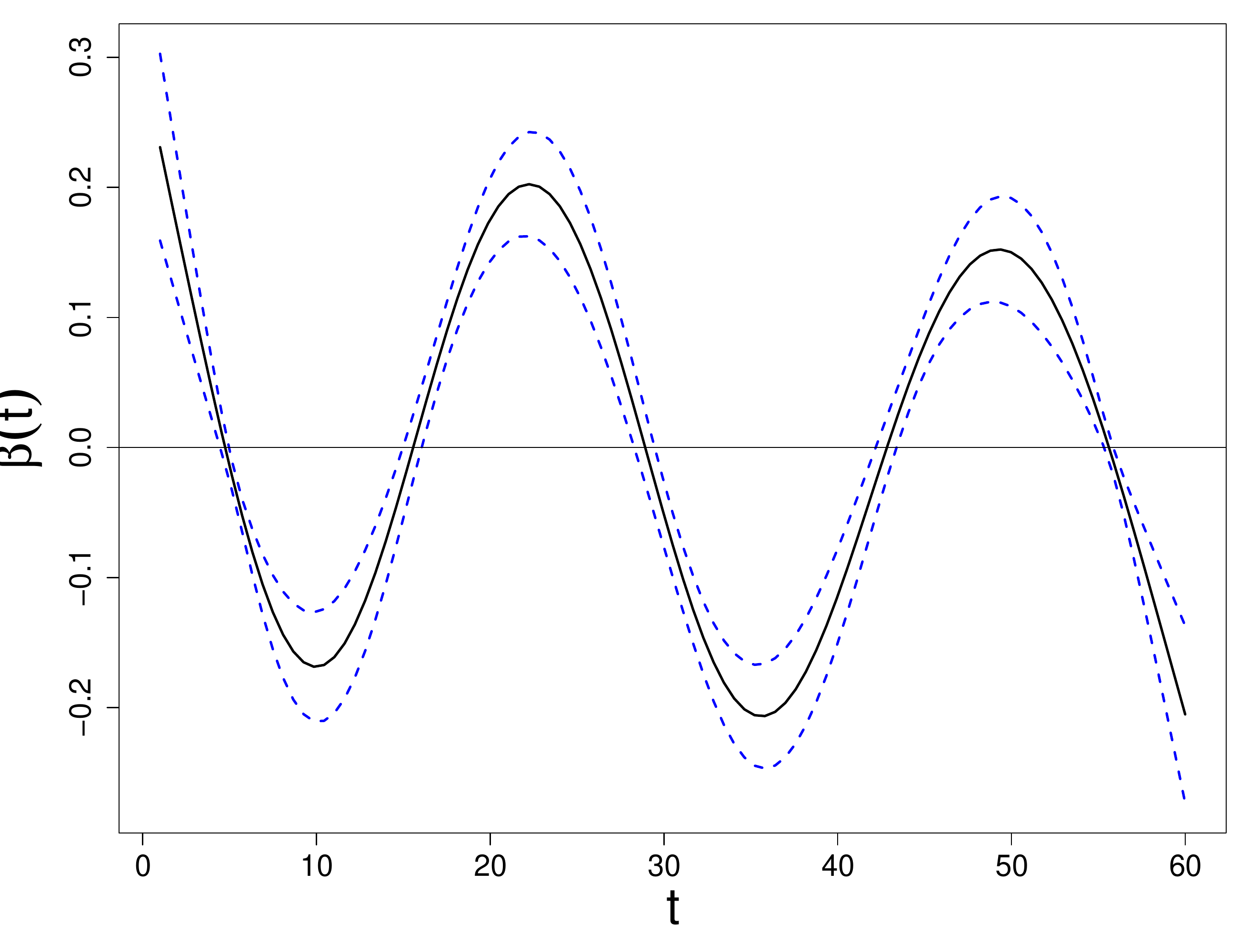} 
\includegraphics[height=0.45\textwidth,angle=270]{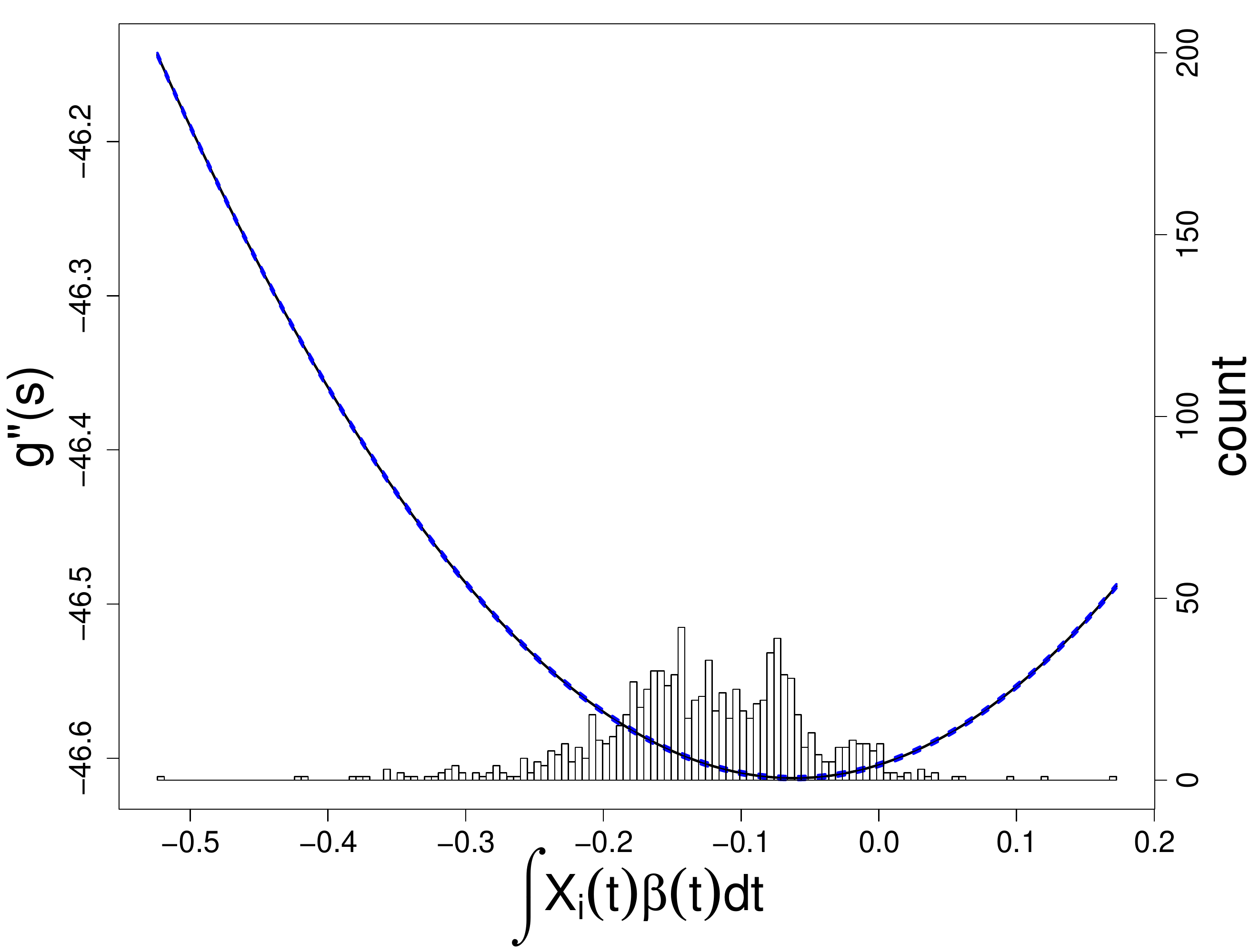}
\includegraphics[height=0.45\textwidth,angle=270]{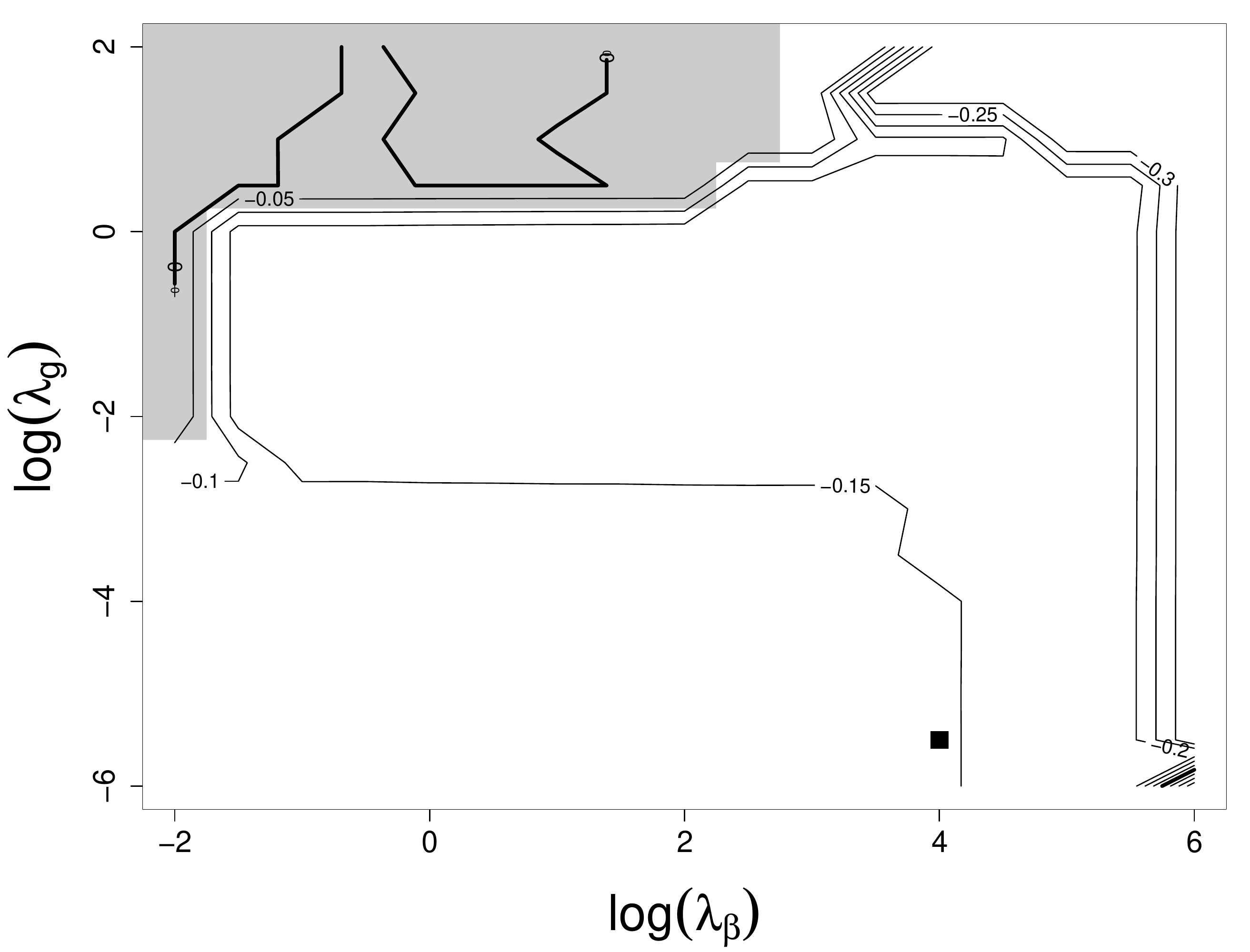}
\caption{Plot for \emph{Keratella Earlinae}, $n = 946$. $\delta$ takes values in the range [-0.348, 0.191]. but note that negative values only occur in the bottom right corner of the contour plot.   }
\label{61804}
 \end{figure}

\begin{figure}[H]
\centering
\includegraphics[height =0.45\textwidth,angle=270]{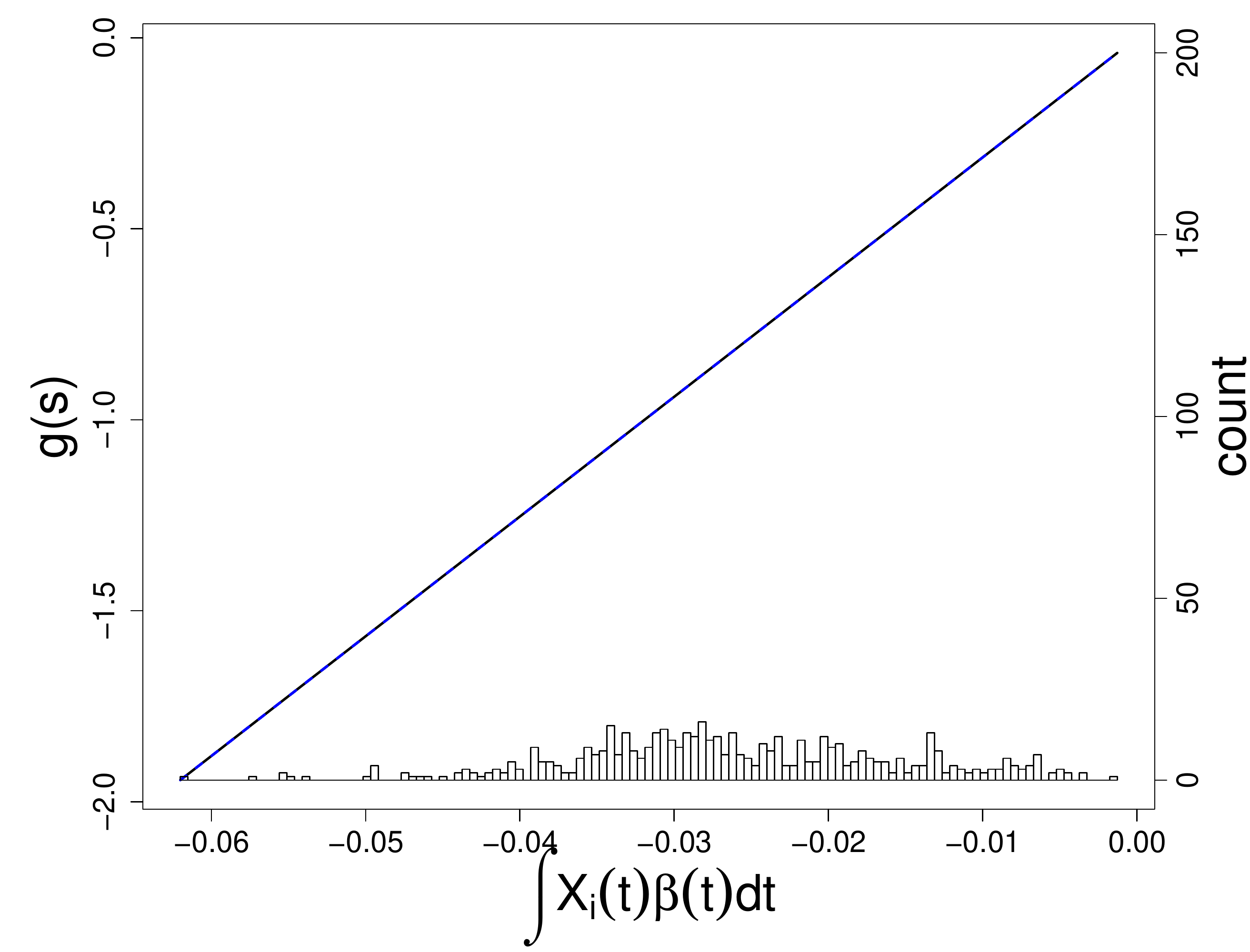} 
\includegraphics[height =0.45\textwidth,angle=270]{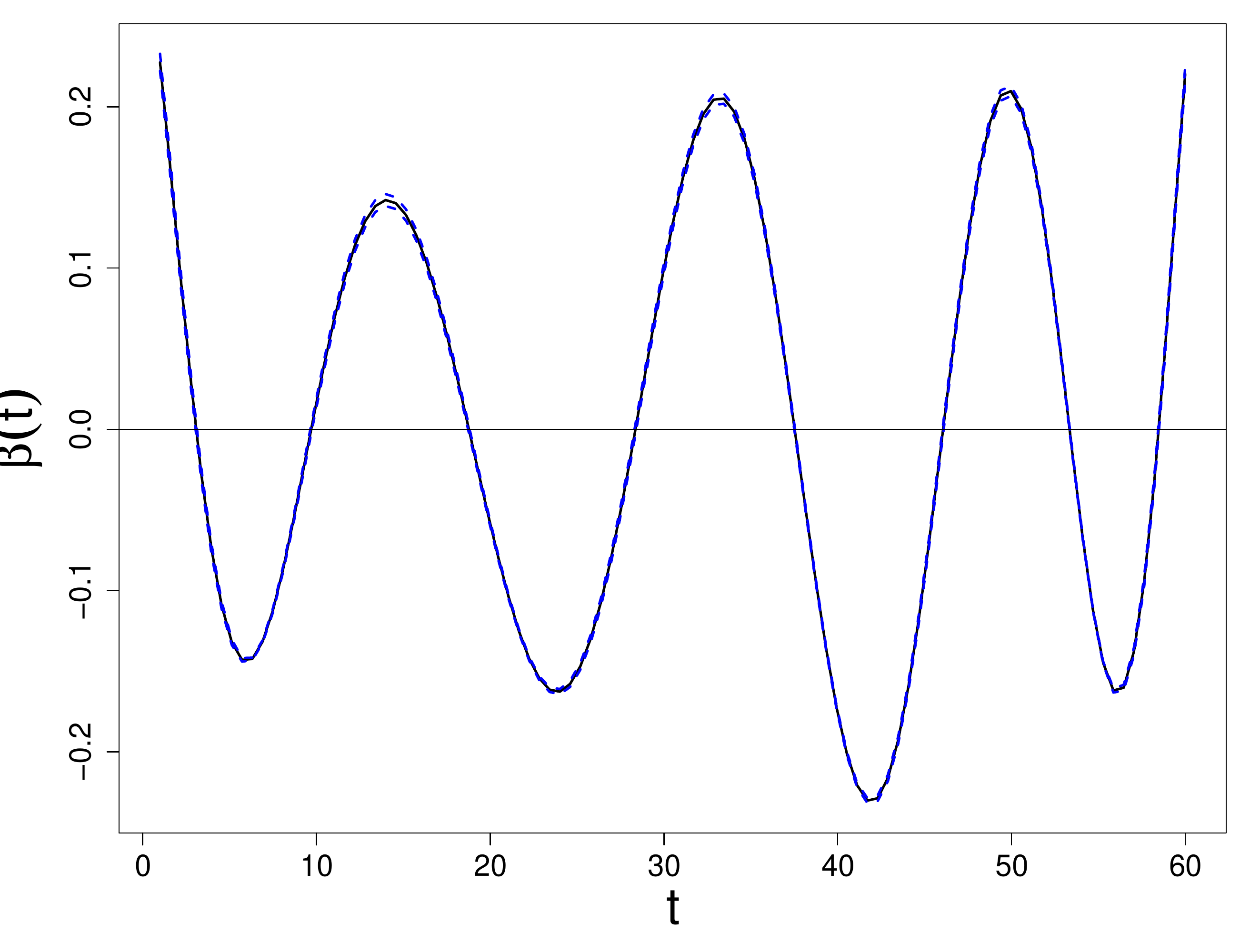} 
\includegraphics[height =0.45\textwidth,angle=270]{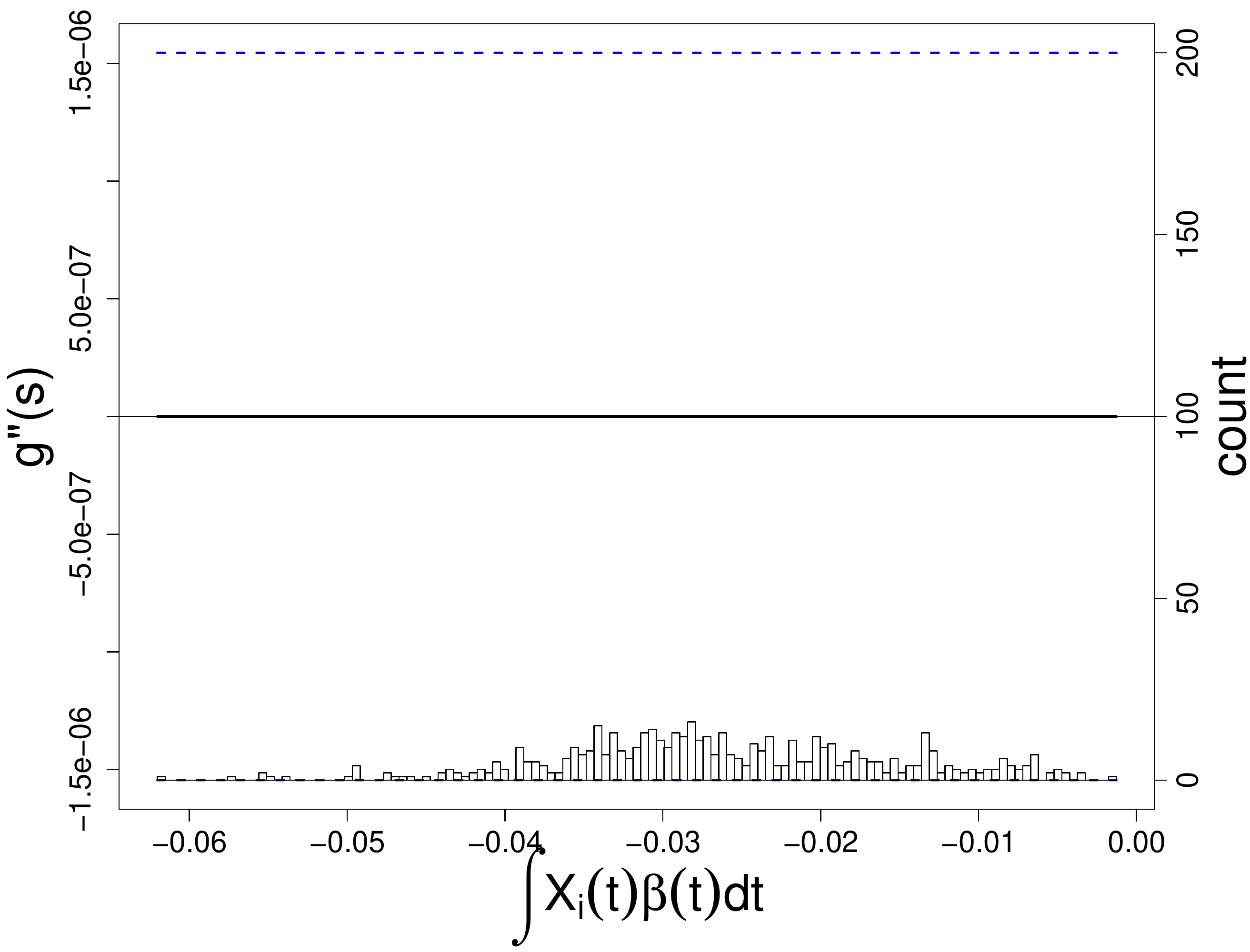}
\includegraphics[height =0.45\textwidth,angle=270]{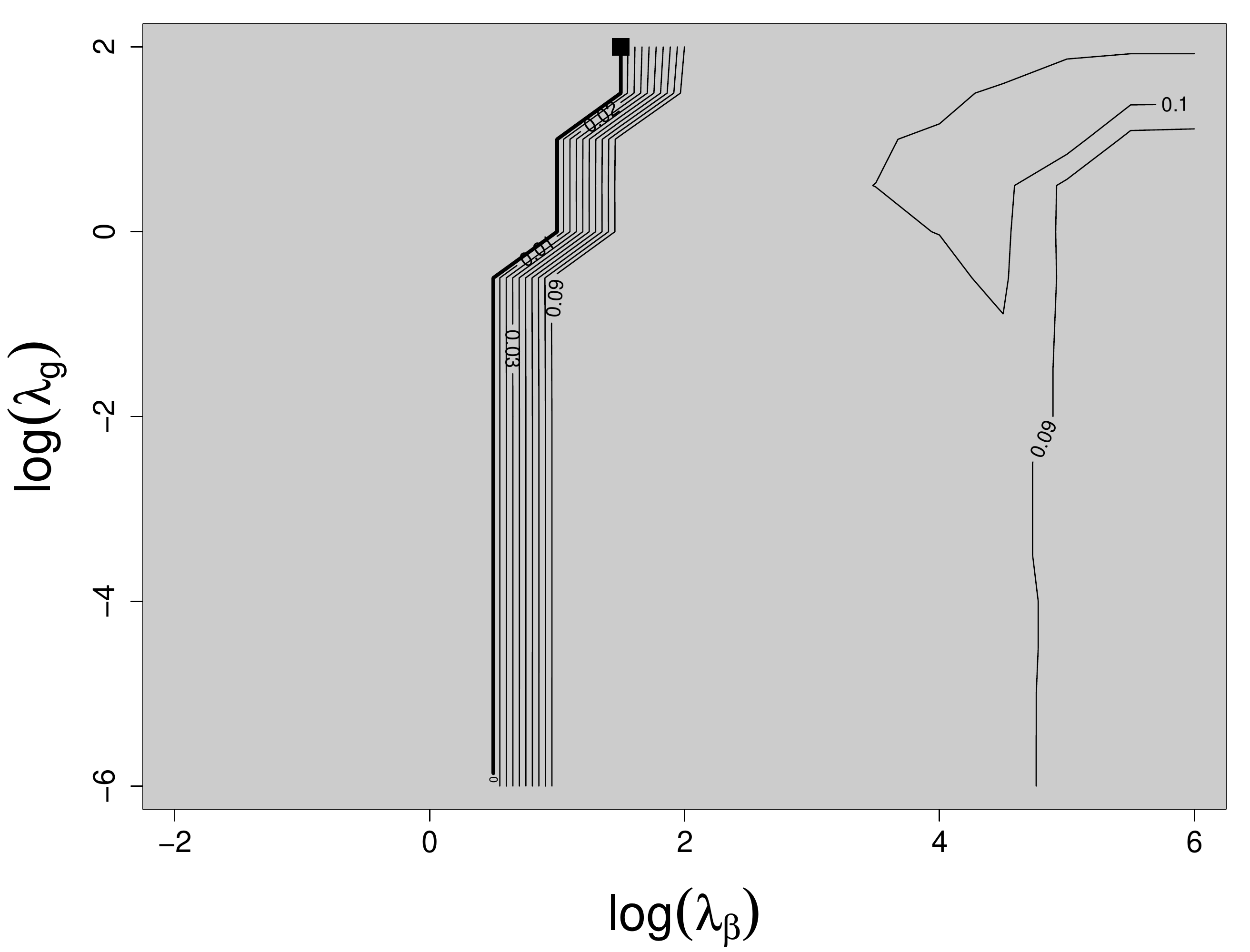}
\caption{Plot for \emph{Keratella Quadrata}, $n = 521$. $\delta$ is estimated in the range [0, 0.1] and is nowhere significant.  }
\label{61806}
\end{figure}

\begin{figure}[H]
\centering
\includegraphics[height =0.45\textwidth,angle=270]{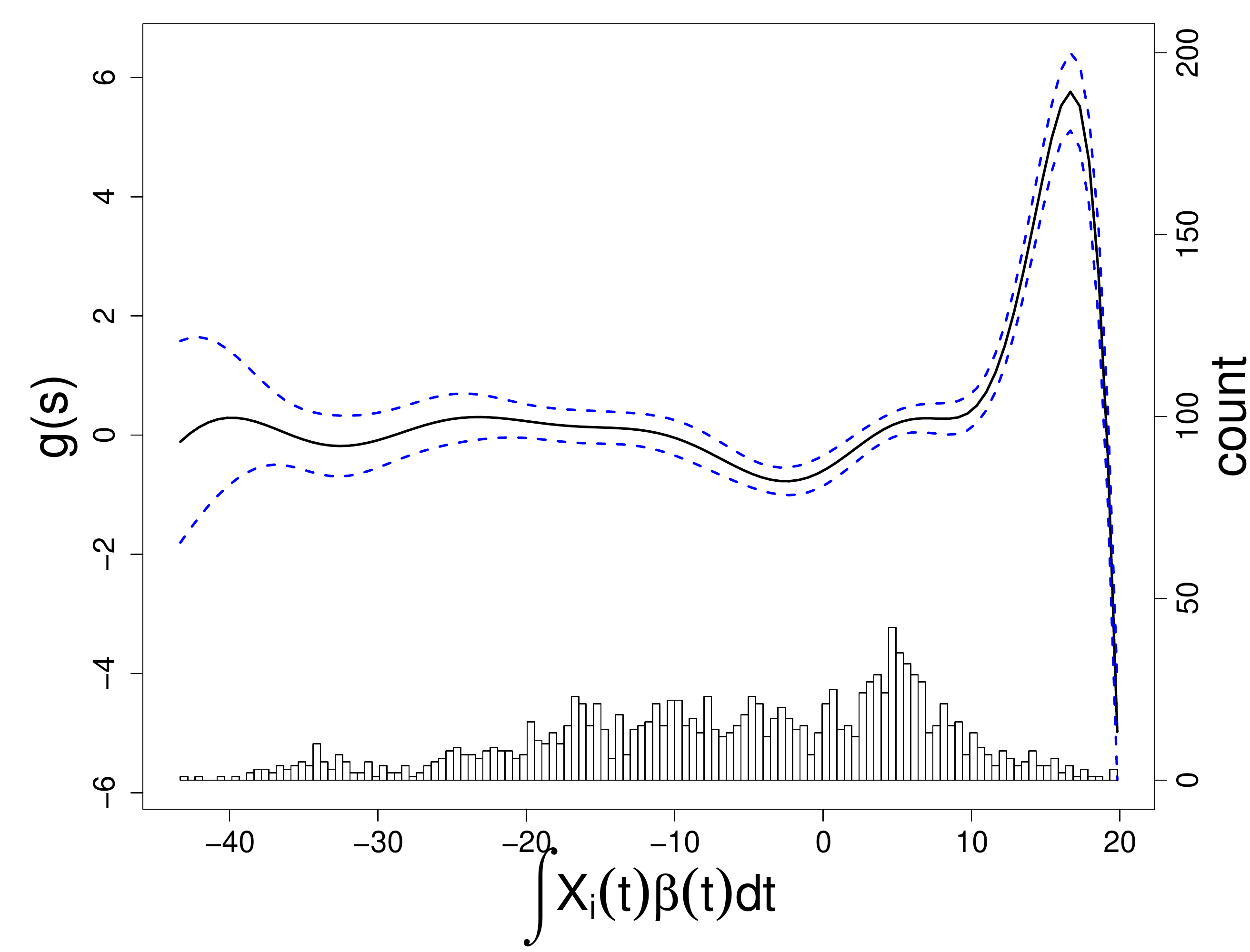} 
\includegraphics[height =0.45\textwidth,angle=270]{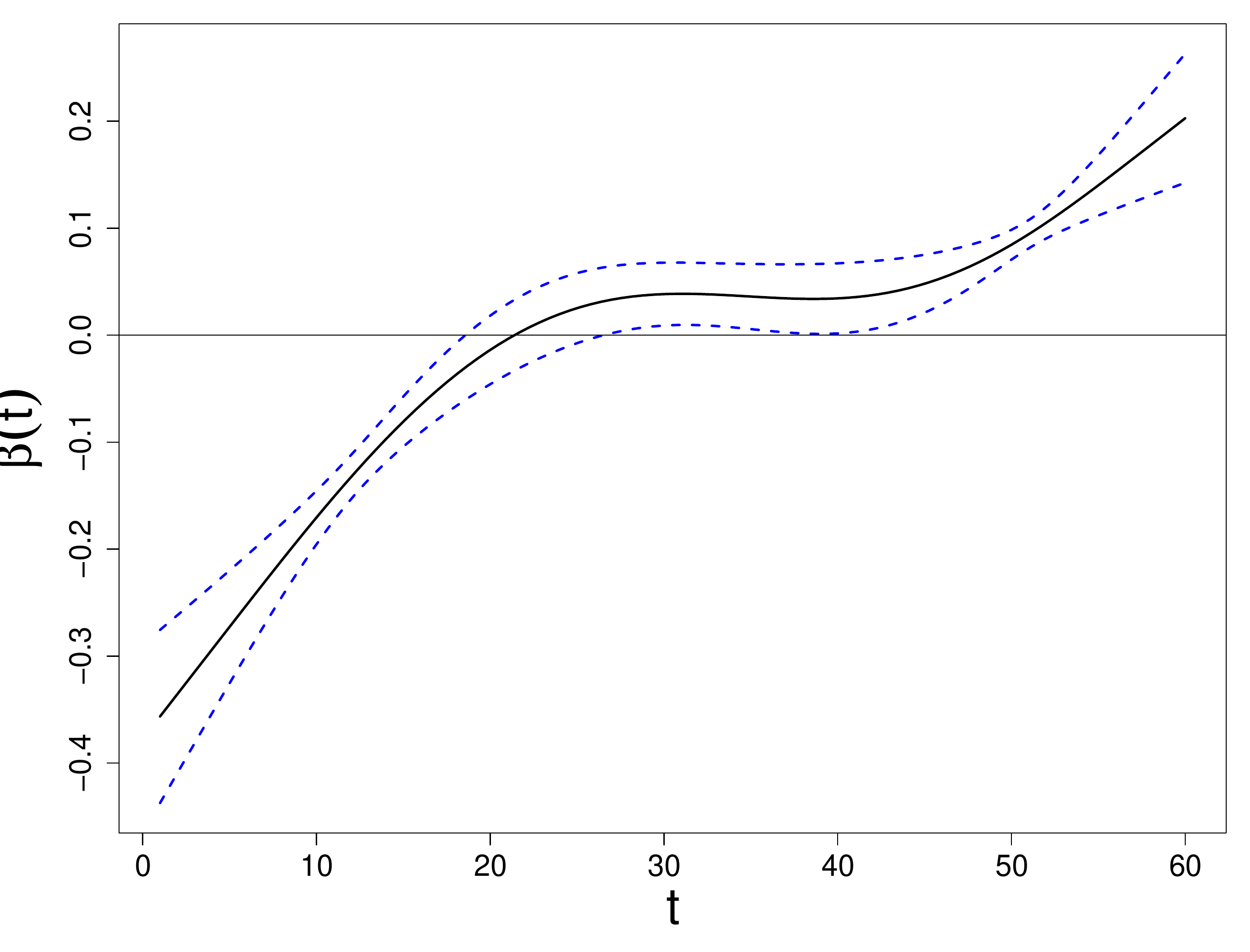} 
\includegraphics[height =0.45\textwidth,angle=270]{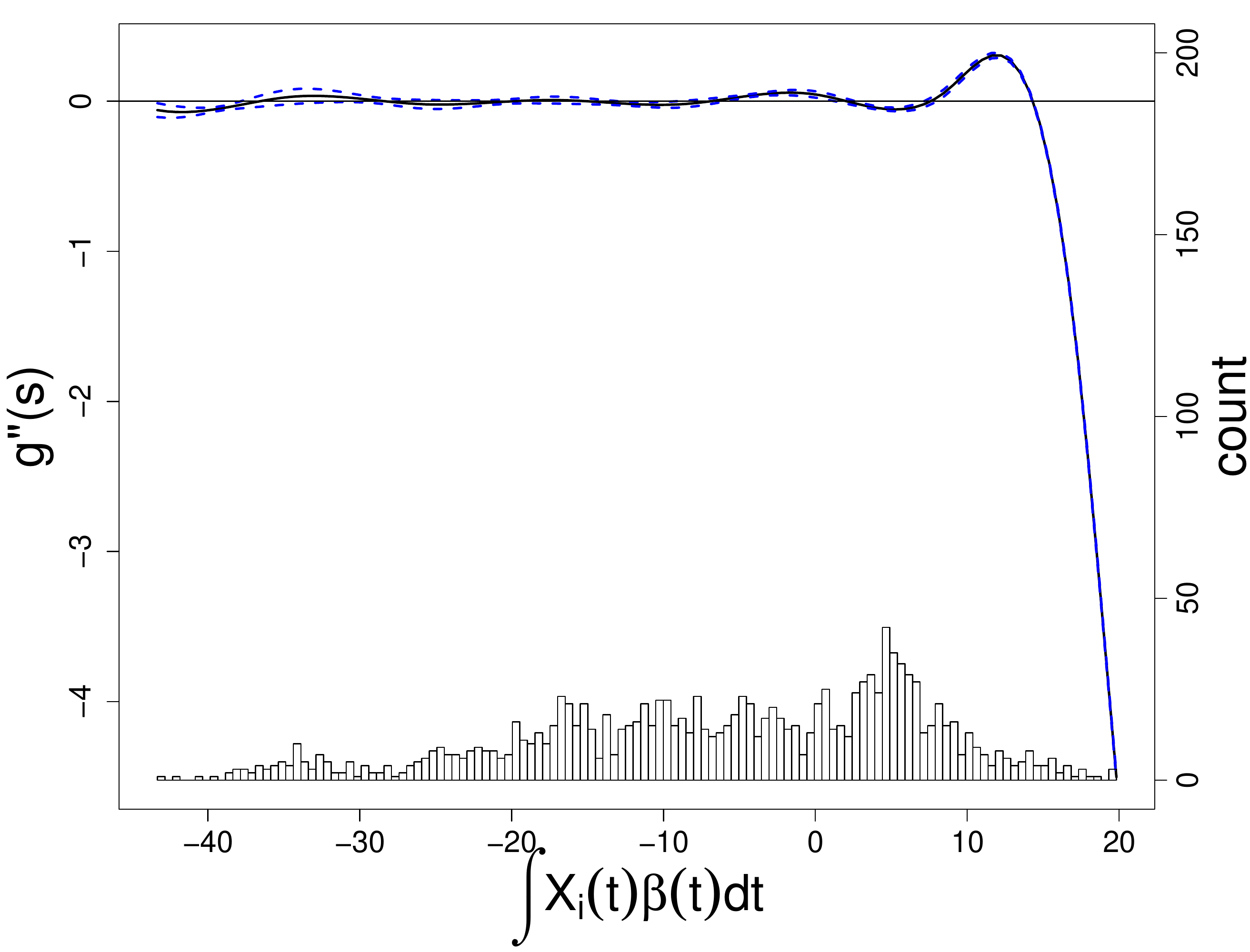}
\includegraphics[height =0.45\textwidth,angle=270]{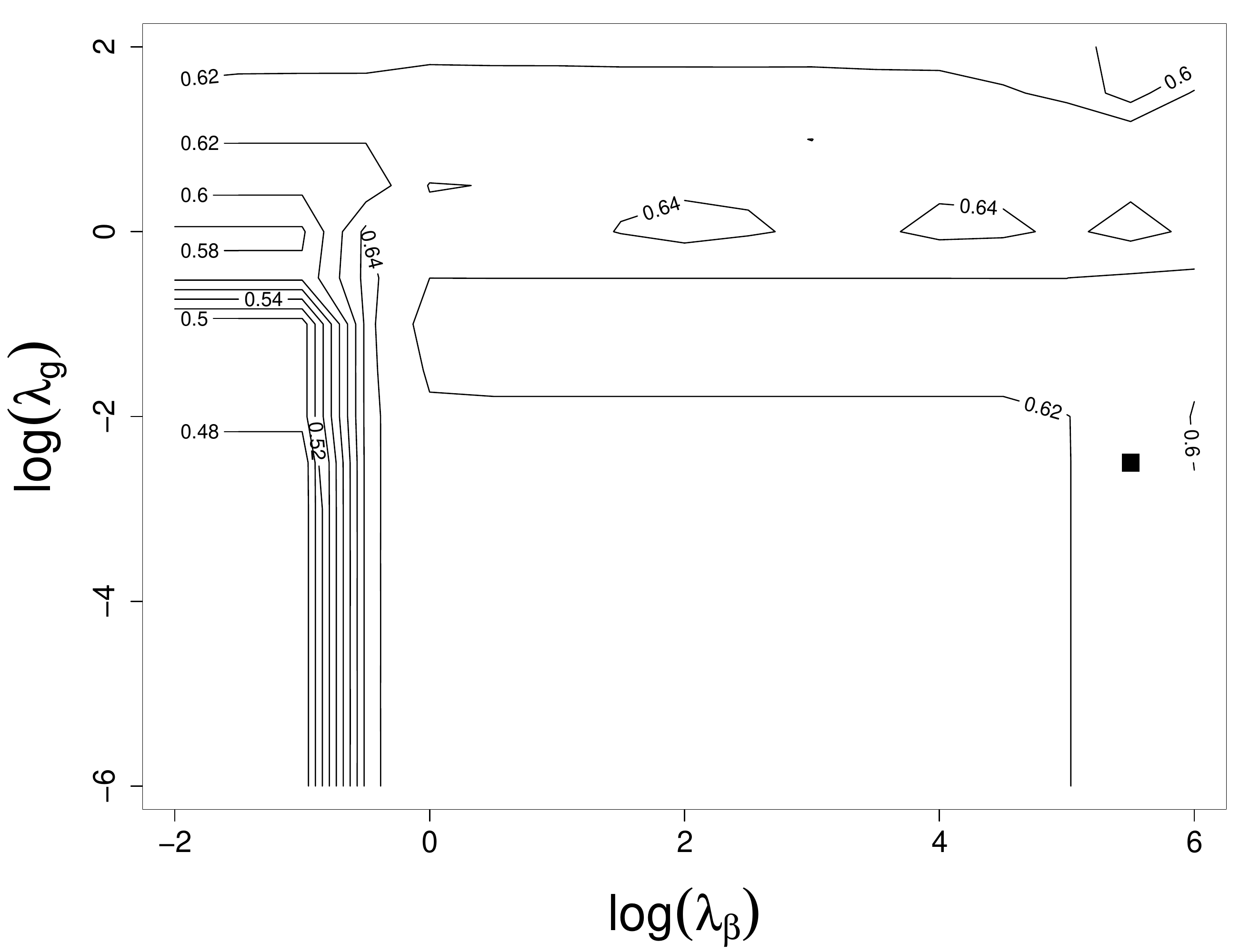}
\caption{Plot for \emph{Polyarthra Remata}, $n = 1347$.  $\delta$ takes values in the range [0.462, 0.647]. }
\label{63004}
 \end{figure}

\begin{figure}[H]
\centering
\includegraphics[height =0.45\textwidth,angle=270]{g_63005_diff-eps-converted-to.pdf} 
\includegraphics[height =0.45\textwidth,angle=270]{beta_63005_diff-eps-converted-to.pdf} 
\includegraphics[height =0.45\textwidth,angle=270]{d2g_63005_diff-eps-converted-to.pdf}
\includegraphics[height =0.45\textwidth,angle=270]{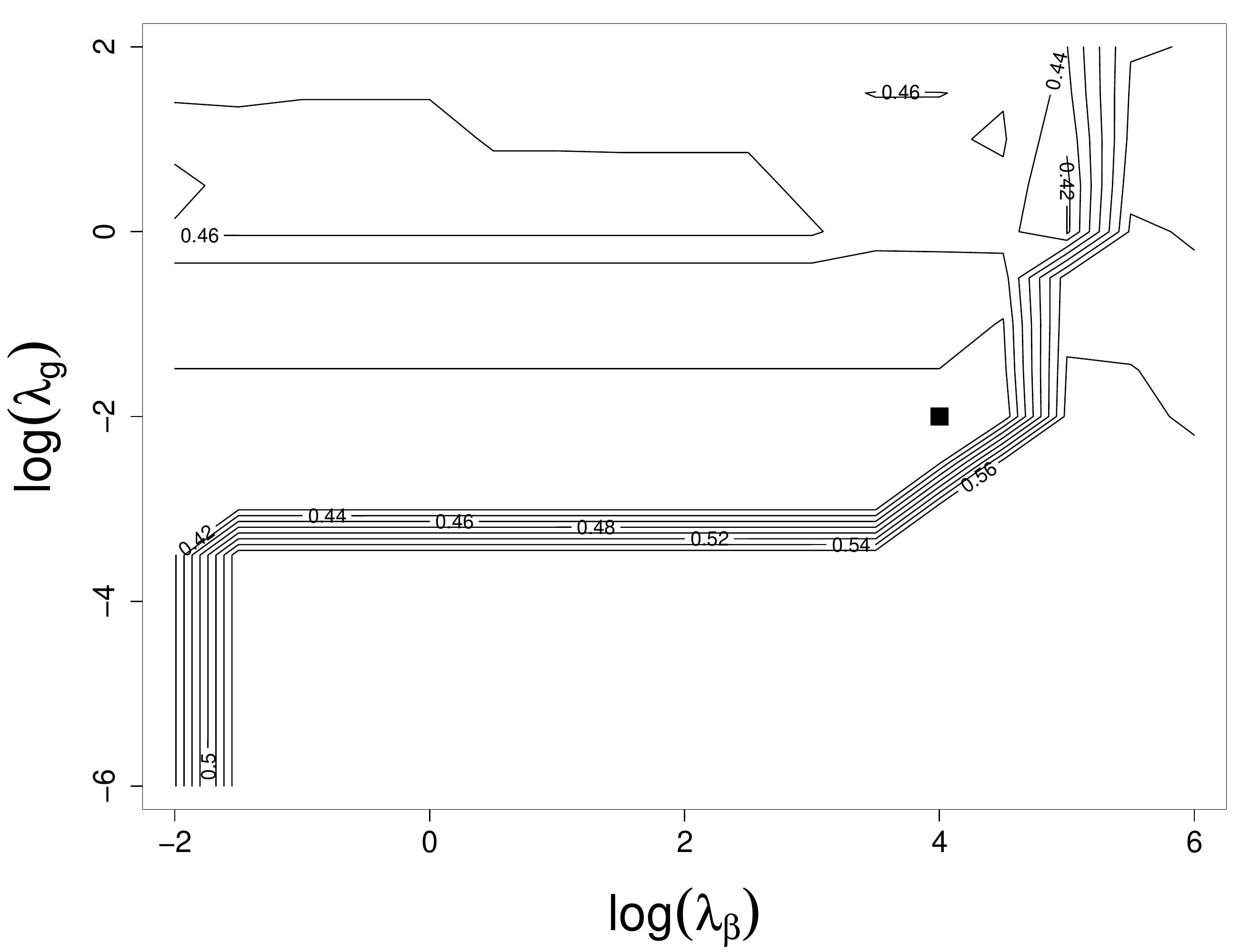}
\caption{Plot for \emph{Polyarthra Vulgaris}, $n = 1577$. $\delta$ takes values in the range [0.4023, 0.5763] and is signifcant at all smoothing parameters. } \label{fig:63005}
\label{63005}
 \end{figure}

Plots \ref{fig:20302} through \ref{fig:63005} repeat the top four panels  in Figure \ref{fig:63005_ex} for each species from the copepod study. For each species,  we provide plots of $g(s)$ (top left), $\beta(t)$ (top right), and $g''(s)$ (bottom left) along with pointwise confidence intervals.  All these plots are given at the values which minimize GCV. The bottom right of each figure plots $\delta$ as a function of both $\lambda_g$ and $\lambda_{\beta}$. Regions where $t_{\lambda_g,\lambda_\beta}$ exceed the critical value are indicated by a white background and a black square gives the minimizing value of GCV.



\end{document}